%% file: main.tex
\newcommand{\markgrey}[1]{\color{gray}#1\color{black}}

\documentclass[twocolumn]{aastex63}
\usepackage{epsfig}
\usepackage{natbib}
\usepackage{amsmath}
\usepackage{hyperref}
\usepackage{comment}
\usepackage{multirow,tabularx}
\usepackage[flushleft]{threeparttable}
\usepackage{amssymb}

 \usepackage{lineno}

\usepackage{xcolor}
\defcitealias{Jones_2017_I}{J17}
\defcitealias{Hlozek_2012}{H12}
\defcitealias{Shivvers_2017}{S17}
\defcitealias{2019MNRAS.485.1171K}{K19}
\defcitealias{Vincenzi_2019}{V19}
\defcitealias{Vincenzi_2020}{V21}
\defcitealias{P21_dust2dust}{P21-dust}
\defcitealias{BS20}{BS21}
\defcitealias{W22_x1age}{W22}
\defcitealias{DES-CC}{DES-nonIa}
\def\snn{SuperNNova}
\def\BinInd{\zeta}
\def\snana{\texttt{SNANA}}
\def\DESFiveYr{DES-SN5YR}

\def\numtot{1829}
\def\numdeshd{1635} 
\def\mindesz{0.10}
\def\maxdesz{1.13}
\def\numdesia{1499} 
\def\numlowz{194}

\def\desomflcdm{0.017}
\def\desomfwcdm{0.082}
\def\deswfwcdm{0.152}
\def\deswfwcdmstatonly{0.132}

\def\desplanckwfwcdm{0.036}
\def\desplanckwfwcdmstatonly{0.032}

\newcommand{\PBEAMS}{\mathcal{P}_{\mathrm{BEAMS(Ia)},i}}

\begin{document}

\title{The Dark Energy Survey Supernova Program: Cosmological Analysis and Systematic Uncertainties}

\def\andname{}

\author{
\parbox{\linewidth}{\centering
M.~Vincenzi$^{1,2}$,
D.~Brout$^{3,4,5}$,
P.~Armstrong$^{6}$,
B.~Popovic$^{7}$,
G.~Taylor$^{6}$,
M.~Acevedo$^{8}$,
R.~Camilleri$^{9}$,
R.~Chen$^{1}$,
T.~M.~Davis$^{9}$,
S.~R.~Hinton$^{9}$,
L.~Kelsey$^{10}$,
R.~Kessler$^{11,12}$,
J.~Lee$^{8}$,
C.~Lidman$^{13,6}$,
A.~M\"oller$^{14}$,
H.~Qu$^{8}$,
M.~Sako$^{8}$,
B.~Sanchez$^{15}$,
D.~Scolnic$^{1}$,
M.~Smith$^{16}$,
M.~Sullivan$^{16}$,
P.~Wiseman$^{16}$,
J.~Asorey$^{17}$,
B.~A.~Bassett$^{18,19}$,
D.~Carollo$^{20}$,
A.~Carr$^{9}$,
R.~J.~Foley$^{21}$,
C.~Frohmaier$^{16}$,
L.~Galbany$^{22,23}$,
K.~Glazebrook$^{14}$,
O.~Graur$^{10}$,
E.~Kovacs$^{24}$,
K.~Kuehn$^{25,26}$,
U.~Malik$^{6}$,
R.~C.~Nichol$^{27}$,
B.~Rose$^{28,1}$,
B.~E.~Tucker$^{6}$,
M.~Toy$^{16}$,
D.~L.~Tucker$^{29}$,
F.~Yuan$^{6}$,
T.~M.~C.~Abbott$^{30}$,
M.~Aguena$^{31}$,
O.~Alves$^{32}$,
F.~Andrade-Oliveira$^{32}$,
J.~Annis$^{29}$,
D.~Bacon$^{10}$,
K.~Bechtol$^{33}$,
G.~M.~Bernstein$^{8}$,
D.~Brooks$^{34}$,
D.~L.~Burke$^{35,36}$,
A.~Carnero~Rosell$^{37,31,38}$,
J.~Carretero$^{39}$,
F.~J.~Castander$^{23,22}$,
C.~Conselice$^{40,41}$,
L.~N.~da Costa$^{31}$,
M.~E.~S.~Pereira$^{42}$,
S.~Desai$^{43}$,
H.~T.~Diehl$^{29}$,
P.~Doel$^{34}$,
I.~Ferrero$^{44}$,
B.~Flaugher$^{29}$,
D.~Friedel$^{45}$,
J.~Frieman$^{29,12}$,
J.~Garc\'ia-Bellido$^{46}$,
M.~Gatti$^{8}$,
G.~Giannini$^{39,12}$,
D.~Gruen$^{47}$,
R.~A.~Gruendl$^{45,48}$,
D.~L.~Hollowood$^{21}$,
K.~Honscheid$^{49,50}$,
D.~Huterer$^{32}$,
D.~J.~James$^{5}$,
N.~Kuropatkin$^{29}$,
O.~Lahav$^{34}$,
S.~Lee$^{51}$,
H.~Lin$^{29}$,
J.~L.~Marshall$^{52}$,
J. Mena-Fern{\'a}ndez$^{53}$,
F.~Menanteau$^{45,48}$,
R.~Miquel$^{54,39}$,
A.~Palmese$^{55}$,
A.~Pieres$^{31,56}$,
A.~A.~Plazas~Malag\'on$^{35,36}$,
A.~Porredon$^{57}$,
A.~K.~Romer$^{58}$,
A.~Roodman$^{35,36}$,
E.~Sanchez$^{59}$,
D.~Sanchez Cid$^{59}$,
M.~Schubnell$^{32}$,
I.~Sevilla-Noarbe$^{59}$,
E.~Suchyta$^{60}$,
M.~E.~C.~Swanson$^{45}$,
G.~Tarle$^{32}$,
C.~To$^{49}$,
A.~R.~Walker$^{30}$,
N.~Weaverdyck$^{32,61}$,
and M.~Yamamoto$^{1}$
\\ \vspace{0.2cm} (DES Collaboration) \\
}}
\email{$\star$ maria.vincenzi@duke.edu}
\submitjournal{(to be submitted to) The Astrophysical Journal}

\begin{abstract}
We present the full Hubble diagram of photometrically-classified Type Ia supernovae (SNe Ia) from the Dark Energy Survey supernova program (DES-SN). DES-SN discovered more than 20,000 SN candidates and obtained spectroscopic redshifts of 7,000 host galaxies. Based on the light-curve quality, we select \numdeshd\ photometrically-identified SNe Ia with spectroscopic redshift \mindesz\ $< z <$\maxdesz, which is the largest sample of supernovae from any single survey and increases the number of known $z>0.5$ supernovae by a factor of five.  
In a companion paper, we present cosmological results of the DES-SN sample combined with \numlowz\ spectroscopically-classified SNe Ia at low redshift as an anchor for cosmological fits. Here we present extensive modeling of this combined sample and validate the entire analysis pipeline used to derive distances. We show that the statistical and systematic uncertainties on cosmological parameters are $\sigma_{\Omega_M,{\rm stat+sys}}^{\Lambda{\rm CDM}}=$\desomflcdm\ in a flat $\Lambda$CDM model, and $(\sigma_{\Omega_M},\sigma_w)_{\rm stat+sys}^{w{\rm CDM}}=($\desomfwcdm, \deswfwcdm$)$ in a flat $w$CDM model.  Combining the DES SN data to the highly complementary CMB measurements by \citet{collaboration2018planck} reduces by a factor of 4 uncertainties on cosmological parameters. In all cases, statistical uncertainties dominate over systematics.
We show that uncertainties due to photometric classification make up less than 10\% of the total systematic uncertainty budget. This result sets the stage for the next generation of SN cosmology surveys such as the Vera C. Rubin Observatory’s Legacy Survey of Space and Time.

\reportnum{DES-2023-0804}
\reportnum{FERMILAB-PUB-23-693-PPD}

\end{abstract}


\section{Introduction}

The modern understanding of the physical evolution of our universe comes from a number of cosmological probes that can constrain the Universe's expansion history and growth of structure.  The Dark Energy Survey (DES) employs multiple probes (Type Ia supernovae, weak lensing, large scale structure, galaxy clusters) to accurately measure a generation-defining picture of the components of the Universe.  In this paper, we present the analysis and distance constraints of the full five years of Type Ia supernovae (SNe Ia) discovered and measured by the DES Supernova program (DES-SN).

SNe Ia are used to make some of the most precise constraints on the nature of dark energy and the expansion history of the Universe from a large span in cosmic history $0\lesssim z \lesssim 2$.  The latest compilations (e.g., Pantheon+: \citealt{PantheonP_sample,PantheonP_cosmo}) include $\sim1500$ distinct SNe; however, they have relied on real-time spectroscopic confirmations of the SNe themselves to be verified as type Ia. There is already an equally large number of SNe discovered for which spectroscopic confirmation was not possible, and recent cosmological analyses have begun to show that the contamination from other types of SNe in the analyses is not debilitating, and not even the largest systematic uncertainty \citep[][ and Popovic et al. in prep.]{2011ApJ...738..162S,2013ApJ...763...88C, Jones_2018_II, Jones_Foundation}. This is, in part, due to the advancement of photometric classification algorithms (e.g., \citealt{2020MNRAS.491.4277M,SCONE_Qu}) that incorporate an improved set of non-Ia spectro-photometric templates and modeling \citep{Vincenzi_2019}. Despite the success of recent photometric analyses \citep{Jones_2018_II, Jones_Foundation}, these samples have not received the same level of use in the broader cosmological community of combined-probe analyses.  The onus has remained on the SN community to demonstrate that the accuracy of photometric SN analyses is not a limitation. This work presents an opportunity to set the stage for future analyses of orders of magnitude larger SN samples, such as the Rubin Observatory Legacy Survey of Space and Time \citep[LSST;][]{2019ApJ...873..111I} and the Nancy Grace Roman Space Telescope \citep[Roman; ][]{2018ApJ...867...23H, 2021arXiv211103081R}, where photometric classification represents the only viable path to fully exploit the statistical power of these surveys.  

The analysis presented here is of the full 5-year photometrically classified set of SNe from DES and additional external samples of spectroscopically confirmed low-$z$ SNe (the \DESFiveYr\ analysis). This work was preceded by the analysis of the first 3 years of spectroscopically classified DES SNe Ia (DES-SN3YR: \citealt{DES_abbott}, also including external low-$z$ SN samples). The DES-SN3YR sample included 207 SNe Ia from DES and was critical in the development and motivation of analyses leading up to the work presented here. This includes photometry and calibration \citep{2018AJ....155...41B,DES_SMP,DES_chrom}, survey and SN Ia population modeling \citep{DES_biascor,2021ApJ...913...49P}, understanding and modeling of the `mass/dust step' \citep{Sullivan_2010, Lampeitl_2010, Smith_2018,DES:2020lzr,BS20,P21_dust2dust,W22_x1age,2022arXiv221114291D, DES:2022hav, DES:2022tjd, 2023MNRAS.518.1985M}, estimates and treatment of systematic uncertainties \citep{DES_syst,BinningSinning}, and the automation of the analysis pipeline \citep{Hinton2020}.  

In this work, we also introduce a number of new supporting analyses that are part of the full \DESFiveYr\ suite of papers.   
Besides photometric classification \citep{Vincenzi_2020,Vincenzi_2021,2022MNRAS.514.5159M}, the biggest differences and improvements in the methodology of  \DESFiveYr\ compared to the DES-SN3YR analysis are due to: 
\begin{itemize}
\item Upgrading to the SALT3 light-curve model from the SALT2 model \citep{SALT3, 2023MNRAS.520.5209T}. 
\item Improved modeling of SN Ia intrinsic scatter using host dust-based models \citep{P21_dust2dust}, focusing on modeling correlations between SN Ia properties and both host mass and host color \citep[rest-frame $u-r$, following][]{2023MNRAS.519.3046K} and modeling the SN Ia progenitor age distribution \citep{W22_x1age}, thus significantly improving previous analyses that neglect modeling of host properties; \citep{Scolnic_2016}. 
\item Modeling of the host spectroscopic redshift efficiency  \citep{Vincenzi_2020} instead of efficiency of spectroscopic typing as in the DES-SN3YR spectroscopic sample \citep{DES_spec} and modeling the fraction of host mis-matched SNe \citep{Qu_hostMismatch}.
\item Updated DES internal calibration \citep{Sevilla_Noarbe_2021} and cross-calibration \citep{Fragilistic}.
\item The inclusion of wavelength-dependent atmospheric effects on photometry \citep*{DES5YR-DCR} in addition to the chromatic corrections used in the DESSN-3YR analysis \citep{DES_chrom}.
\item Using an un-binned systematic covariance matrix for cosmological constraints \citep{BinningSinning, KV_2022}.  
\item {Improved statistical validation of the methodology \citep{Validation_PA,Camilleri}.}
\end{itemize}

This analysis is the culmination of these works. In this paper, we provide the derivation of the DES-SN5YR distances and the assessment of the impact of systematic uncertainties on distances and cosmological fits. The unblinded cosmological constraints from the techniques established in this analysis are presented by \citet{DES_final_cosmo}.

\begin{figure}
    \includegraphics[width=0.95\linewidth]{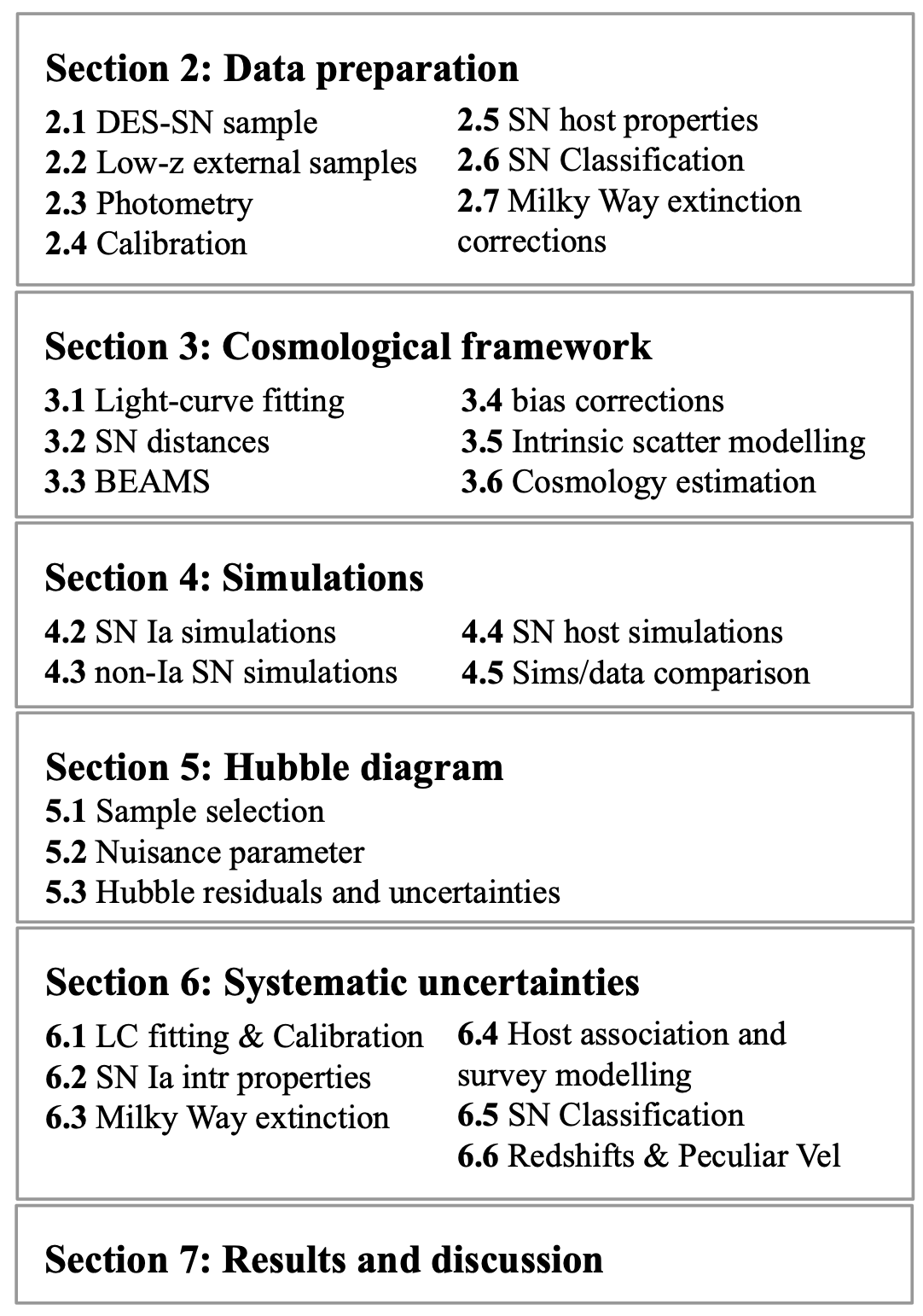}
    \caption{Overview of the paper.}
    \label{fig:overview}
\end{figure}

The structure of this paper is presented in Fig.~\ref{fig:overview}. In Sec.~\ref{sec:data}, we present the DES-SN5YR data set. In Sec.~\ref{sec:cosmo}, we briefly review the cosmological framework implemented in the analysis, with particular attention to how potential contamination from non-Ia SNe is accounted for in the cosmological fitting.  In Sec.~\ref{sec:simulations}, we describe how simulations of DES-SN5YR are built. In Sec.~\ref{sec:HD}, we present the inferred SN distances and our final Hubble diagram. 
In Sec.~\ref{sec:syst_overview}, we discuss the various sources of systematic uncertainties included in our analysis and present our systematic error-budget. Finally, in Sec.~\ref{sec:discussion} and Sec.~\ref{sec:conclusions}, we discuss our results and present our conclusions.

\section{Data}
\label{sec:data}
	\subsection{Dark Energy Survey SN sample}
	The DES-SN program is a five-year survey using the Dark Energy Camera \citep[DECam;][]{2015AJ....150..150F} on the Victor M. Blanco telescope (Cerro Tololo, Chile), covering ten $\sim3$ deg$^2$ fields distributed across the DES footprint \citep[two \lq E\rq\ fields, two \lq S\rq\ fields, three \lq X\rq\ and \lq C\rq\  fields, see][Figure~1 and Table~2]{DES_spec}. Two out of ten fields (\lq X3\rq\ and \lq C3\rq) have been observed to a single-visit depth of 24.5 mag in $r$-band (deep fields), while the remaining eight fields we reach a single-visit depth of 23.5 mag. Only a small fraction of the DES-SN candidates have been spectroscopically followed-up using several spectroscopic facilities. 
 For the majority of the transients, host galaxy redshifts have been collected using the auxiliary Australian DES survey (OzDES), which used the 2dF fibre positioner and AAOmega spectrograph \citep{2004SPIE.5492..410S} on the Anglo-Australian Telescope to collect host galaxy redshifts \citep{lidman2020ozdes}. The SN sample collected by the DES-SN program is the largest and deepest cosmological SN sample from a single telescope to date (see Fig.~\ref{fig:redshift_distr}). \citet{2015AJ....150..172K} and \citet{DES_spec} describe in detail the SN search strategy and spectroscopic follow-up associated with the DES SN program.
	
\subsection{Low redshift samples}
\label{sec:lowz_samples}
We combine the DES-SN sample with various external low redshift ($z<0.1$) SN surveys. These include CfA3 \citep{2009ApJ...700..331H}, CfA4 \citep{2012ApJS..200...12H}, CSP \citep{2017AJ....154..211K} (DR3) and the Foundation SN sample \citep{Foley_Foundation}. These external surveys span a redshift range of $0.01 <z< 0.1$ and provide a lever arm to improve constraints on the dark energy equation of state. For this analysis, we include only low-$z$ SNe above redshift $0.025$ to mitigate the effects of peculiar velocities. Finally, we add a 1 percent error floor in quadrature to the low-$z$ SN photometry (2 percent for Foundation $z$-band), following \citet{scolnic2018} and \citet{Jones_Foundation}.

We don't include other historical low-$z$ SN samples e.g., LOSS \citep{2013MNRAS.433.2240G}, SOUSA\footnote{Light curves available at \url{https://pbrown801.github.io/SOUSA/.}}, or intermediate redshift SN samples, e.g., the SDSS SN sample \citep{2018PASP..130f4002S}. This choice is in order to avoid including a larger number of systematic uncertainties in our analysis (for every survey, we need to take into account for additional systematics related to survey calibration and survey-specific selection effects) and emphasise the contribution of the DES SN program at redshift $z>0.1$.

\subsection{SN Photometry}
	We measure DES-SN photometry using the Scene modeling Photometry \citep[SMP;][]{2013A&A...557A..55A} pipeline presented by \citet{DES_SMP}, which simultaneously models the time-varying SN flux and the time-independent background host-galaxy flux. In comparison to faster difference imaging pipelines, this technique provides more accurate flux and flux uncertainty measurements. 
 \citet{DES5YR_SMP} present a detailed comparison between DES SMP photometry and photometry from difference imaging and demonstrate that the implementation of SMP significantly reduces \textit{(i)} flux uncertainties and \textit{(ii)} effects attributed to the so-called surface-brightness anomaly \citep[i.e., unexplained flux scatter increasing with the host galaxy surface brightness at the SN location, ][]{2015AJ....150..172K, DES_biascor}.
	
	In addition, DES-SN photometry is corrected for wavelength-dependent atmospheric effects such as Differential Chromatic Refraction \citep[DCR, ][]{1982PASP...94..715F} and wavelength-dependent ($\lambda$-dependent) seeing, which affect ground-based observations. DCR occurs because the index of refraction of our atmosphere is wavelength-dependent, while $\lambda$-dependent seeing is caused by variations in the atmospheric refractive index due to atmospheric turbulence. These two effects cause a color-dependent mis-modeling of the shape of the PSF (which is reconstructed using stars that are generally redder than the average SN at $z = 0$) and of the position of the SN. \citet*{DES5YR-DCR} describe the methods used to correct DES-SN photometry for such wavelength-dependent atmospheric effects and assess their impact on DES-SN distance estimation and cosmological results. We do not include wavelength-dependent atmospheric corrections for external low-$z$ samples.

	\subsection{Calibration}
	
Accurate photometric calibration of DECam filters and inter-survey calibration is essential in SN cosmology to estimate SN brightnesses at different redshifts and when combining SNe from different surveys. For the DES-SN sample, calibration is performed in two stages. 

First, DES images are internally calibrated using a catalog of 17 million tertiary standard stars within the DES footprint built using the Forward Global Calibration Method (FGCM) as conceived by \citet[][]{Stubbs_2006} and as implemented in DES by \citet[][]{2018AJ....155...41B}. Not only does this method provide accurate ($\sim1\%$) absolute calibration, but it also provides excellent all-sky uniformity of $<3$mmag for DES \citep{Sevilla_Noarbe_2021,rykoff23}. 

The FGCM tertiary standard star catalog provided in \citet[][]{2018AJ....155...41B} was utilized in the preliminary DES-SN3YR cosmological analysis. The FGCM catalog was updated in the period between DES-SN3YR and DES-SN5YR and here we use the \texttt{Y3GOLD} stellar catalogs as presented in Appendix 3 of \cite{Sevilla_Noarbe_2021}. The improvements are summarized as follows: \textit{(i)} improved aperture photometry corrections, \textit{(ii)} an update to the publicly released DES Y3A2 Standard Bandpass \citep[see][]{Sevilla_Noarbe_2021}, \textit{(iii)} improved uniformity in years following the bad weather of year 3, \textit{(iv)} improved astrometric solution utilizing longer temporal baseline, and \textit{(v)} other technical and practical improvements.

\begin{deluxetable*}{lcccc}
\centering
\tablecolumns{5}
\tablewidth{10pc}
\tablecaption{DECam AB Offsets and Uncertainties}
\tablehead {
\colhead {\textbf{Ref}}  &
\colhead {$g$} &
\colhead {$r$} &
\colhead {$i$} &
\colhead {$z$} 
}
\startdata
    \vspace{-2mm}\\
    \cite{rykoff23} & $+0.001\pm0.011$ & $-0.003\pm0.011$ & $-0.001\pm0.011$ & $+0.002\pm0.012$\\
    \cite{Fragilistic} & $+0.002\pm0.006$ & $-0.009\pm0.006$ & $-0.007\pm0.006$ & $+0.006\pm0.006$\vspace{-2mm}\\
    \label{tab:newcal}
\enddata
\end{deluxetable*}

Second, the tertiary standard stars are calibrated to primary standard stars to place them on the AB system. Within DES, AB offsets were calculated to the HST Caslpec standard star C26202 in \citet[][]{rykoff23}, given in Table~\ref{tab:newcal}. However, because SNIa cosmology analyses combine multiple surveys to cover both low redshift and high redshift to obtain competitive cosmological constraints, here we utilize the calibration of \citet[][Supercal-Fragilistic]{Fragilistic} which is an improvement on the \citet[][Supercal]{Supercal} method. This method consists of simultaneously cross-calibrating the FGCM catalog with the AB calibrated stellar catalogs from numerous other wide-field surveys (e.g., PS1, SDSS, SNLS). Supercal-Fragilistic use priors on each modern survey's published AB calibrations and re-fit for a new solution that minimizes the differences between surveys and mitigates potential systematic errors.
Supercal-Fragilistic find similar offsets as those found in \citet[][]{rykoff23}, but of larger magnitude (see Table~\ref{tab:newcal}); though these offsets are consistent with each other given that the external data used to perform the calibration is largely independent. In this work we have chosen to adopt the offsets from Supercal-Fragilistic because: \textit{(i)} the low-$z$ samples also analyzed in this work have been calibrated in Supercal-Fragilistic, \textit{(ii)} this includes covariance between DECam filters and low-$z$ filters (utilized in our distance likelihood Eq.~\ref{eq:tripp}), \textit{(iii)} Supercal-Fragilistic provides the mechanism to create multiple realizations of inter-filter correlated calibrations from the Supercal-Fragilistic covariance matrix with the other low-$z$ samples, \textit{(iv)} Supercal-Fragilistic obtain smaller uncertainties due to the utilization of more external data. This change results in a $\sim$5~mmag color correction for $g-z$ ($\sim$3~mmag for $g-i$) relative to what was used in DES-SN3YR.

The AB offset uncertainty reported in the C26202-based analysis of \citet{rykoff23} is $\sim$0.011 mags. The reported DES5YR uncertainties (stat+syst) in Supercal-Fragilistic covariance are roughly half of the size on the diagonal (6 mmag), which is the result of leveraging the cross-calibration of multiple surveys utilizing multiple primary standard stars. The full Supercal-Fragilistic covariance\footnote{\url{https://github.com/PantheonPlusSH0ES/DataRelease}} is used to determine the effects of correlated systematic uncertainties in both light-curve fitting and in SALT3 model training. Systematic uncertainties due to absolute calibration of the DECam and low-$z$ filters are discussed in Section~\ref{sec:syst_overview}.

\subsection{Host galaxy association, redshifts and host properties}
\label{sec:data_host}
\begin{figure}
    \includegraphics[width=\linewidth]{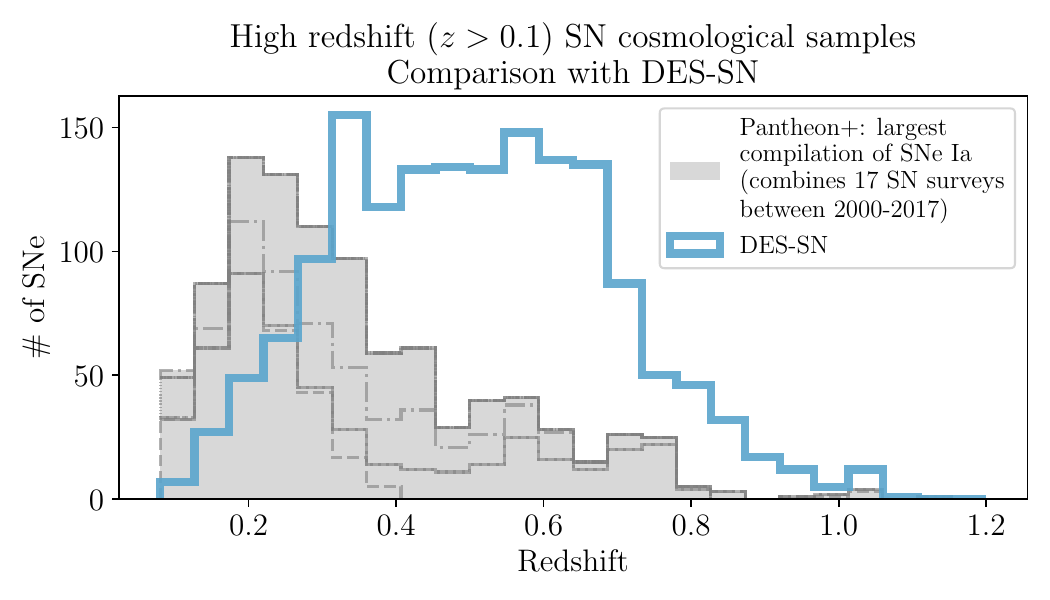}
    \caption{Redshift distribution of the DES-SN sample compared to the Pantheon+ compilation. DES-SN is the largest and deepest sample of SN~Ia to date. The dashed lines represent the different surveys' redshift distributions contributing to the total Pantheon+ compilation. The Pantheon+ and the DES-SN samples presented in this paper have a small overlap of 146 SNe \citep[previously published by][]{DES_SMP}).}
    \label{fig:redshift_distr}
\end{figure}

For each SN, we identify the host galaxy using the Directional Light Radius (DLR) method presented by \citet{2006ApJ...648..868S, Gupta_2016}. We define as \lq hostless\rq\ SNe for which no galaxy is detected with DLR$<4$. The galaxies identified as likely hosts of DES transients are targeted using the AAOmega spectrograph on the 3.9-m Anglo-Australian Telescope (AAT) as part of the OzDES programme \citep{2015MNRAS.452.3047Y, 2017MNRAS.472..273C, lidman2020ozdes}.
A full description of the different sources of redshifts used in our sample and the host spectroscopic redshift efficiency for the DES-SN sample are presented by \citet{Vincenzi_2020} and \citet{DES5YR_SMP}.

To characterize SN host galaxies, we mainly focus on two global host galaxy properties: stellar mass ($M_{\star}$) and rest-frame $u-r$ color. These are the two properties we can most reliably estimate given the limited broad-band photometry available for our SN hosts. For DES SN hosts, these galaxy properties are measured using DES broad-band photometry and, when available, $u$-band and $JHK$ photometry from external surveys \citep{DES_deepstacks, 2022MNRAS.509.3547H}. We use the galaxy Spectral Energy Distribution fitting code by \citet{2010MNRAS.406..782S} and the P\'EGASE2  galaxy spectral templates \citep{1997A&A...326..950F, 2002A&A...386..446L}, assuming a \citet{2001MNRAS.322..231K} initial mass function. In the DES SN cosmological sample used in this work (see Sec.~\ref{sec:HD}), we find that 68 per cent of the DES SN hosts are assigned to high stellar mass ($>10^{10} M_{\odot}$).

For consistency, we remeasure stellar masses and restframe $u-r$ colors of the low-$z$ SN host galaxies using the same code and initial mass function used for the DES SN hosts. We use optical and UV photometry\footnote{We use UV photometry from GALEX \citep{2017ApJS..230...24B} and SDSS ($u$ band). } to ensure the same rest-frame wavelength coverage used for the DES SN hosts. We find that a significant fraction of low-$z$ hosts previously assigned a stellar mass lower than $10^{10} M_{\odot}$ are re-assigned a larger stellar mass ($>10^{10} M_{\odot}$). As a result, the fraction of low-$z$ SNe in high mass hosts is 69 per cent, compared to 59 per cent in previous analysis (we only consider SNe included in the cosmological sample presented in Sec.~\ref{sec:HD}). The details of the host property measurements for DES-SN and for the external low-$z$ samples are presented in \citet{DES5YR_SMP} and \citet{2023MNRAS.519.3046K}.

Spectroscopic redshifts for the low-$z$ sample are incorporated following the revisions presented by \citet{2022PASA...39...46C}. Low-$z$ SN redshifts require additional corrections for peculiar velocities (that are negligible for high redshift DES SNe). The nominal peculiar velocities used for this analysis were determined by \citet{2022ApJ...938..112P} and are based on 2M++ density fields \citep{2015MNRAS.450..317C} with global parameters found in \citet{2020MNRAS.497.1275S}, combined with group velocities estimated by \citet{2015AJ....149..171T}. We consider uncertainties on peculiar velocity estimates to be 250 km~s$^{-1}$ \citep{scolnic2018}.

\subsection{Non-Ia Classification}
\label{sec:classifiers}
In our baseline approach, we classify the DES-SN sample using the open-source algorithm SuperNNova \citep[][]{2020MNRAS.491.4277M},\footnote{\url{https://github.com/supernnova/SuperNNova}} a photometric SN classifier based on recurrent neural networks. \snn\ is trained to classify different types of transients using photometric data only (i.e., fluxes and flux uncertainties in different filters) and, optionally, redshift information. It does not rely on feature extraction or light-curve fitting and it uses machine learning techniques, e.g., recurrent neural networks.
This is the first SN cosmological analysis that exploits machine learning techniques for classification. 

The training of \snn\ (and most classification algorithms based on machine-learning) requires large ($>100,000$) and representative samples of SN light-curves \citep{2020MNRAS.491.4277M}. For this reason, the subset of spectroscopically confirmed DES SNe is not suitable as a training sample, and simulations are used instead. We train \snn\ on realistic simulations of DES-like light-curves, built following the approach described in Sec.~\ref{sec:simulations} and by \citet{Vincenzi_2020}.
Simulations include SNe Ia, peculiar SNe Ia and core-collapse SNe generated using the core-collapse template library by \citet{2019PASP..131i4501K, Vincenzi_2019}. A detailed analysis of training methods and performances of \snn\ in the context of the DES-SN5YR analysis is presented by \citet{Vincenzi_2021, 2022MNRAS.514.5159M}.

Using \snn, we estimate for each SN event its probability of being a SN Ia, $P_{\mathrm{Ia}}$. These probabilities are then incorporated in the cosmological analysis as described in Sec.~\ref{sec:BEAMS}.

\subsection{Milky Way Extinction corrections}
\label{sec:MW}
Milky Way extinction corrections are applied to the light-curve fitting model. In the analysis, we exclude SNe with Milky Way reddening larger than 0.25. The 10 DES SN fields have been specifically chosen in low Milky Way dust extinction regions (median reddening $E(B-V)_{\mathrm{MW}}<0.02$); however, significant differences are observed from field to field (average $E(B-V)_{\mathrm{MW}}$ is $<0.01$ in E,C fields, $\sim0.02$ in X fields and $\sim0.03$ in S fields). SNe in the low-$z$ SN samples have a median Milky Way extinction of 0.04 (low redshift SN surveys generally require large sky coverage, therefore higher Milky Way dust extinction regions cannot  be avoided).

For each SN, we estimate $E(B-V)_{\mathrm{MW}}$ using the Milky Way reddening maps presented by \citet{Schlafly2010} and use the reddening law by \citet{Fitzpatrick99}, with $R_V=3.1$.

\section{Cosmological analysis framework}
\label{sec:cosmo}
In this section, we give an overview of the cosmological framework used to go from observed light-curves to SN distances and cosmological fitting. SN distances are obtained after light-curve fitting (Sec.~\ref{sec:LCfitting}) using the BBC framework (\lq BEAMS with Bias Corrections\rq\, Sec.~\ref{sec:BEAMS}, \ref{sec:bias_corrections} and \ref{sec:dist_uncertainties}). In Table~\ref{tab:BBCinput_output}, we present a schematic overview of the inputs and (intermediate and final) outputs of the BBC framework. 

\subsection{Light-curve fitting}
\label{sec:LCfitting}
The first step to measure SN Ia distances is to perform light-curve fitting of the multi-band photometry observed for each SN. This step is necessary to standardize SN brightnesses.
In this analysis, we perform light-curve fitting using the SALT3 model framework \citep{SALT3}. The SALT3 model is defined by five SN-dependent parameters: redshift $z$, day of peak brightness ($t_{\rm peak}$), stretch $x_1$, color $c$ and an amplitude term $m_x$ or $x_0$ (where $m_x=-2.5 \mathrm{ log}_{10} (x_0)$ and it is approximately the B-band peak brightness of the SN). In the SALT3 fitting process, the best fit values and uncertainties of each parameter are determined in order to measure SN distances (see Sec.~\ref{sec:measure_distances}). In our analysis, SN redshifts are known with high accuracy (spectroscopic redshifts included in our sample have an uncertainty of $\sim10^{-4}$, see Sec.~\ref{sec:data_host}), therefore the parameter $z$ is fixed in the light-curve fitting. 

The SALT3 model is based on the widely used SALT2 model \citep{Guy07}; however, it provides improved estimation of uncertainties as a function of phase and color and an extended central passband wavelength range 2800-8000\AA\ (compared to a range of 2800-7000\AA\ in SALT2). 
The SALT3 model is trained on a compilation of 1083 SNe with 1207 spectra presented by \citet{SALT3}. \citet{2023MNRAS.520.5209T} showed that there is negligible difference on SN cosmology results from the choice of SALT2 or SALT3 model, when the models are trained on the same input data. In this analysis, we train our own SALT3 model \texttt{SALT3.DES5YR}.

Following \citet{2023MNRAS.520.5209T}, we train the \texttt{SALT3.DES5YR} model using the sample from \citep{SALT3}, based on the calibration values presented by \citet{Fragilistic}. We choose to remove any observer-frame U-band training data from the training set, as calibration of the near UV ground-based data is challenging (the atmospheric extinction is  larger and variable from site to site and with airmass, the filter's characterization historically poorer, and the cross-calibration approach presented by \citet{Fragilistic} is not applicable). This affects 97 SNe in the training sample, from CfA and miscellaneous low-$z$ samples.\footnote{Our training sample still includes $u$-band data from the more recent SDSS and CSP SN surveys, for which the filter transmissions have been measured and well characterized. For the older CfA U-band data, we do not have measured filter transmissions. This has caused several calibration issues, as highlighted by various cosmological analyses \citep{2011ApJ...737..102S, PantheonP_cosmo}.} 

Given the difficulty of training SN Ia in the UV where SN flux is low and good quality SN data are limited, we avoid the far UV and use passbands whose central wavelength ($\lambda_b$) satisfies $3500$~\AA$< (\lambda_b)/(1+z) < 8000$~\AA. The lack of good rest-frame UV band modeling is an important limitation for our analysis because the DES-SN survey aims to probe the high redshift SNe (where observer-frame optical is rest-frame UV).

\subsection{Measuring SN Ia distances}
\label{sec:measure_distances}
SN Ia distance moduli, $\mu_\mathrm{obs}$, are defined as  \citep[e.g.,][]{1998A&A...331..815T, 2006A&A...447...31A}
\begin{equation}
    \mu_\mathrm{obs} = m_x + \alpha x_1 - \beta c + \gamma G_{\rm host} - \mathcal{M} - \Delta\mu_\mathrm{bias},
    \label{eq:tripp}
\end{equation}
where $m_x$, $x_1$ and $c$ are the SALT3 light-curve parameters discussed in Sec.~\ref{sec:LCfitting}. 
The global nuisance parameters $\alpha$ and $\beta$ set the amplitude of the stretch-luminosity and color-luminosity corrections, and $\mathcal{M}$ is the absolute magnitude of a SN Ia with $x_1=0$ and $c=0$. 
The fourth term of the equation, $\gamma G_{\rm host}$, encapsulates any residual dependency between SN Ia luminosities and their host galaxy properties. 
This dependency is modelled as a step function, $\gamma G_{\rm host}$, defined as
\begin{equation}
\gamma G_{\rm host} = \begin{cases}
+\gamma/2 & \quad \text{if } P > P_{\text{step}}, \\
-\gamma/2  & \quad \text{otherwise,}
\end{cases}
\label{eq:std_gamma}
\end{equation}
where $P$ is a SN host galaxy property, usually stellar mass $M_{\star}$, $\gamma$ is the size of the residual \lq step\rq\ and $P_{\text{step}}$ is the threshold at which the step is measured, usually fixed to $10^{10}M_{\odot}$ for stellar mass. 

Many cosmological analyses have shown that SNe Ia observed in high mass galaxies ($M_{*}>10^{10}$\,$M_{\odot}$) are approximately 0.07 mag brighter than SNe in lower mass galaxies after color and light curve stretch corrections \citep[][]{Sullivan_2010, Lampeitl_2010, DES_massstep, 2021MNRAS.501.4861K}. Recently, \citet{BS20} and \citet{2023MNRAS.519.3046K} highlighted that this so-called \lq mass-step\rq\ is highly color dependent: smaller (or negligible) for blue SNe and significant ($>0.1$ mag) for redder SNe.

\citet{BS20} (hereafter \citetalias{BS20}), supported by the work of \citet{2020ARA&A..58..529S}, propose that dust is the underlying cause of the SN mass-step and show how different $R_V$'s in high and low mass host galaxies could explain the observed brightness step. 

In addition, \citet{2023MNRAS.519.3046K} analyzed the DES-SN sample and measured the SN brightness step between SNe found in intrinsically blue and intrinsically red host galaxies (a \lq color-step\rq, rather than a mass-step). \citet{2023MNRAS.519.3046K} find a significant color-step even after corrections for the dust-driven mass-step, and suggest that either stellar mass is not the optimal proxy to describe a dust-driven brightness step, or that other astrophysical factors (e.g., SN progenitor physics) might play an important role in explaining the dependence of SN Ia luminosities on their host galaxies. For this reason, in our analysis we implement Eq.~\ref{eq:std_gamma} \textit{either} assuming $P=M_{\star}$, $P_{\text{step}}=10^{10}M_{\odot}$ (the mass-step, $\gamma_{M_{\star}}$) \textit{or} $P=(u-r)$, $P_{\text{step}}=1$ (the color-step, $\gamma_{u-r}$) (see also Sec.~\ref{sec:sim_ia_plus_host}). 

The parameters $\alpha$, $\beta$ and $\gamma$ are determined prior to, and independently of, the cosmological parameters, using the approach presented by \citet[][see Sec.~\ref{sec:BEAMS}]{2011ApJ...740...72M}. The term $\beta$ is treated as the observed, effective $\beta$, i.e., we do not fit separately for an intrinsic $\beta$ and a \lq dust $\beta$\rq\ (or $R_V$) parameter (see further discussion in Sec.~\ref{sec:bias_corrections}). We note that without a calibrated absolute distance scale, $\mathcal{M}_B$ is degenerate with the cosmological parameter $H_0$ and therefore is not addressed in this work.
Finally, corrections for biases resulting from selection effects and analysis choices are applied in the $\Delta\mu_\mathrm{bias}$ term of Eq.~\ref{eq:tripp}. These selection effects are determined from accurate simulations of the survey and using models of the residual scatter. We discuss the modeling and implementation of bias corrections in Sec.~\ref{sec:bias_corrections}.

\begin{deluxetable}{p{2.5cm}p{4.5cm}}
\centering
\tablecolumns{2}
\tablewidth{20pc}
\tablecaption{BBC input and outputs chart.}
\tablehead {
\colhead {\textbf{BBC}}  &
\colhead {Variants} }
\startdata
    \textbf{Inputs}& S3 fit parameters ($^*$), $\sigma_{\rm S3fit}$, $P_{\rm Ia}$, biasCor simulations \vspace{2mm}\\
    \textbf{Intermediate outputs} & $\mu_{\rm bias}$ (Eq.~\ref{eq:bias_corrections}), \newline $\mathcal{P}_{\rm BEAMS(Ia)}$ (Eq.~\ref{eq:PBIa}), \newline $M_{\zeta}$ (Eq.~\ref{eq:likelihoodsIa}), $\sigma_{\mathrm{floor}}$ (Eq.~\ref{eq:sig_floor})\vspace{2mm}\\
    \textbf{Outputs}\newline (for cosmology) & $\mu_{\rm obs}$, $\sigma_{\mu,\mathrm{unbin}}$ (Eq.~\ref{eq:sig1} and \ref{eq:sig_renorm}), \newline $\alpha$, $\beta$, $\gamma$, $\sigma_{\rm gray}$\vspace{2mm}\\
\enddata
    \label{tab:BBCinput_output}
    \begin{tablenotes}\footnotesize
    \item ($^*$) SALT3 fitted parameters $m_x$, $x_1$, $c$.
    \end{tablenotes}\footnotesize
\end{deluxetable}

\subsection{The BEAMS approach}
\label{sec:BEAMS}
Photometric SN samples require the application of photometric classification algorithms to determine the SN types. 
We incorporate type Ia classification \textit{probabilities} in the cosmological fits using the \lq Bayesian Estimation Applied to Multiple Species\rq\ framework \citep[BEAMS, presented by][]{2007PhRvD..75j3508K, Kunz_2012, 2012MNRAS.421..913N}.

The BEAMS approach was developed to incorporate SN probabilities and marginalise over contamination from non-Ia SNe while performing a cosmological fit. \citet{Kessler_2017} extended the BEAMS framework to include modeling and correction of selection effects, and to incorporate the \citet{2011ApJ...740...72M} method of measuring
nuisance parameters independent of cosmological parameters. This extended framework is referred as \lq BEAMS with Bias corrections\rq\ (BBC). 
In the BEAMS approach, the analytical form of the likelihood includes two terms, one that models the SN Ia population, $\mathcal{L}_\mathrm{Ia}$, and the other that models a population of contaminants, $\mathcal{L}_\mathrm{CC}$,
\begin{equation}
    \prod_{i=1}^{N_{\mathrm{SNe}}} \big( \mathcal{L}^i_\text{Ia}+\mathcal{L}^i_\text{CC} \big) .
    \label{eq:SUM}
\end{equation}
The two terms of the likelihood, $\mathcal{L}^i_\text{Ia}$ and $\mathcal{L}^i_\text{CC}$, are defined as
\begin{equation}
  \begin{aligned}
    \mathcal{L}^i_\text{Ia} &=  P^i_{\text{Ia}} \times D_\text{Ia} (z_i, \mu_{\text{obs},i}, \mu_{\text{ref}})\\
    \mathcal{L}^i_\text{CC} &= (1-P^i_{\text{Ia}}) \times D_{\text{CC}}(z_i, \mu_{\text{obs},i}, \mu_{\text{ref}}).
  \end{aligned}
  \label{eq:likelihoods}
\end{equation}
The term $D_{\text{Ia}}$ is defined as:
\begin{equation}
  \begin{aligned}
    D_\text{Ia} &=  \text{exp} \biggl(-\frac{\big(\mu_{\text{obs},i} + M_{\BinInd} - \mu_{\text{ref}}(z_i)\big)^2} {\sigma_{\mu,i}^{2}} \biggr),\\
  \end{aligned}
  \label{eq:likelihoodsIa}
\end{equation}
where $\mu_{\text{obs},i}$ is defined in Eq.~\ref{eq:tripp}, $\mu_{\text{ref}}(z_i)$ is the distance modulus of the $i$-th SN as predicted assuming a fixed reference cosmology (e.g., $\Omega_M=0.3$, $w=-1$), and $M_{\BinInd}$ are offsets quantifying by how much observations deviate from the reference cosmology. They \textit{absorb} the cosmological information, enabling a cosmology-independent fit of the SN nuisance parameters \citep[as shown by][]{2011ApJ...740...72M}.\footnote{The estimated SN distances are not dependent on the choice of the reference cosmology \citep[see][]{KV_2022}.} The $M_{\BinInd}$ offsets are calculated for 20 redshift bins ($\BinInd$), equally spaced on a logarithmic scale, and they effectively constitute a \lq binned\rq\ version of the SN Hubble diagram (${z_{\BinInd}}, M_{\BinInd}+\mu_{\text{ref}}(z_{\BinInd})$). However, we emphasise that we only use this binning to determine the nuisance parameters $\alpha, \beta$; we do not use this binned Hubble diagram to fit cosmology as it has been shown to lead to an overestimate of systematic uncertainties \citep[see][]{BinningSinning}. We follow instead the unbinned approach described and validated by \citet{KV_2022}. The distance modulus uncertainties $\sigma_{\mu,i}$ are discussed in Sec.~\ref{sec:dist_uncertainties}.

Finally, $D_\text{CC}$ in Eq.~\ref{eq:likelihoods} is the contaminants likelihood term, and it models the non-Ia SN distance moduli distribution on the Hubble diagram. Core-collapse SNe are not standardized by the SALT3 framework, therefore it is not trivial to model $D_\text{CC}$ analytically. 
In our baseline analysis, we empirically model the term $D_\text{CC}$ using the core-collapse simulations described in Sec.~\ref{sec:simulations} and test alternative approaches in the systematics analysis.

The two likelihood terms in Eq.~\ref{eq:likelihoods} can be used to estimate a \lq BEAMS probability\rq:
\begin{equation}
   \PBEAMS =  \frac{ P^i_{\text{Ia}} D^i_{\text{Ia}}}{  P^i_{\text{Ia}}D^i_{\text{Ia}} + (1- P^i_{\text{Ia}})D^i_{\text{CC} }} \label{eq:PBIa},
\end{equation}
which effectively quantifies the likelihood of a SN of belonging to the Ia population or the contaminants population, given not only its classification probability ($P^i_{\text{Ia}}$), but also its inferred distance modulus and distance modulus uncertainty.
A more detailed description of the BEAMS framework is given by \citet{Kunz_2012, Hlozek_2012}, the BBC \lq binned\rq\ approach is described by \citet{Kessler_2017, Vincenzi_2021} and the BBC \lq unbinned\rq\ approach used in this analysis is presented by \citet{KV_2022}.

\subsection{Bias corrections}
\label{sec:bias_corrections}
All SN surveys are affected by selection effects introduced by their flux-limited nature. These selection effects can introduce systematic biases in cosmological analyses of SN Ia samples, and thus SN Ia distances are usually corrected for such expected biases (Eq.~\ref{eq:tripp}). The corrections are generally estimated using large SN Ia Monte Carlo simulations that accurately model the survey detection efficiency and other potential selection effects \citep{1999AJ....117.1185H, 2010AJ....140..518P,Betoule_2014,DES_biascor, 2021ApJ...913...49P}. 

In our analysis, bias corrections ($\Delta\mu_\mathrm{bias}$) are estimated using the BBC framework and large SN Ia simulations that model the different surveys considered in the analysis. We follow the approach presented in \citet{2021ApJ...913...49P} referred to as \lq BBC4D\rq. In BBC4D, the term $\Delta\mu_\mathrm{bias}$ is modelled as a function of the four observables $z$, $x_1$, $c$, log$M_{\star}$, and it is defined as: 
\begin{equation}
    \Delta\mu_\mathrm{bias} = m_x + \alpha^{\mathrm{true}} x_1 - \beta^{\mathrm{true}} c + \gamma^{\mathrm{true}} G_{\rm host} + \mathcal{M}_B^{\mathrm{true}} -\mu^{\mathrm{true}},
  \label{eq:bias_corrections}
\end{equation}
where $\alpha^{\mathrm{true}}$, $\gamma^{\mathrm{true}}$, $\mathcal{M}_B^{\mathrm{true}}$ and $\mu^{\mathrm{true}}$ are the true simulated values of nuisance parameters, intrinsic SN brightness and distance modulus. The parameter $\beta^{\mathrm{true}}$ technically is not defined when simulating SNe following the dust-based model by \citetalias{BS20}, in comparison to the historical approach of simulating a single $\beta$. In the \citetalias{BS20} model, a forward modeled distribution of intrinsic color-luminosity relations, $\beta_{\rm int}$, and a distribution of dust $R_V$ are combined. For this reason, in the calculation of bias corrections and uncertainties, an effective $\beta$ must be assumed. While the choice of $\beta^{\mathrm{true}}$ in bias corrections has a negligible effect on the inferred cosmology (see discussion in Sec.~\ref{disc:variants}), we set to $\beta^{\rm true}=2.87$. In the \cite{P21_dust2dust} forward modeling process discussed in Sec.~\ref{sec:sim_ia}, this value of $\beta$ is determined to be the effective observed $\beta$.

Using realistic simulations of SN samples, \citet{2021ApJ...913...49P} tested the ability of the BBC4D approach to estimate unbiased SN distances and recover the input $\alpha$, $\beta$, $\gamma$ and SN intrinsic scatter $\sigma_{\rm int}$.

Throughout the analysis, we will also refer to the \lq BBC0D\rq\ approach, i.e. the approach of fixing $\Delta\mu_\mathrm{bias} =0$ and ignoring bias corrections. This approach is not used for cosmology, but it is useful to estimate \textit{raw} SN distances, removing any assumption on SN Ia intrinsic properties and removing the modeling of selection effects.
 
\subsection{SN distance uncertainties and intrinsic scatter}
\label{sec:dist_uncertainties}
Within BBC, distance modulus uncertainties $\sigma_{\mu,i}$ in Eq.~\ref{eq:likelihoodsIa} are described as
\begin{equation}
\begin{split}
        \sigma_{\mu,i}^2 = f(z_i, c_i, M_{*,i})  &\sigma_{\mathrm{S3fit},i}^2 + \sigma_{\mathrm{floor}}^2(z_i, c_i, M_{*,i}) \\
   &+ \sigma_{z,i}^2 + \sigma_{{\rm vpec},i}^2 + \sigma_{\mathrm{lens},i}^2 ,
   \end{split}
\label{eq:sig1}
\end{equation}
 where $\sigma_{\mathrm{S3fit},i}$ is computed from the SALT3 light-curve fit parameters,\footnote{The term $\sigma_{\mathrm{S3fit},i}$ is defined as 
 \begin{equation*}
    \begin{split}
         \sigma_{\mathrm{S3fit},i}= &(\sigma_{m_x})^2 + (\alpha \sigma_{x_1})^2+(\beta \sigma_{c})^2 \\
         & + 2 \alpha C_{m_x,x_1} - 2 \beta C_{m_x,c} + 2 -\alpha \beta C_{x_1,c} .
     \end{split}
 \end{equation*}} $\sigma_{{\rm vpec},i}$ and $\sigma_{z,i}$ are uncertainties associated with estimates of peculiar velocities and spectroscopic redshifts respectively, and $\sigma_{\mathrm{lens},i}$ are uncertainties associated with weak lensing effects due to the large scale structure the SN photons are traveling through.
 The terms $f(z_i, c_i, M_{*,i})$ and $\sigma_{\mathrm{floor}}(z_i, c_i, M_{*,i})$ are survey-specific scaling and additive factors that are estimated from the same simulations used for bias corrections to ensure that the reduced $\chi^2$ in each cell of a $\{z_i, c_i, M_{*,i}\}$ grid is close to unity. The scaling term is introduced to account for Malmquist bias that suppresses fainter SNe. This bias results in naively computed uncertainties that are overestimated, and thus $f<1$ by construction. Conversely, the additive term accounts for any additional scatter beyond the naively computed $\sigma_{\rm S3fit}$. It is the sum in quadrature of a gray term and a term that depends on redshift, color and host mass:
\begin{equation}
         \sigma_{\mathrm{floor}}^2(z_i, c_i, M_{*,i}) = \sigma_{\mathrm{scat}}^2(z_i, c_i, M_{*,i}) + \sigma_{\rm gray}^2. 
         \label{eq:sig_floor}
\end{equation}

The two terms $f(...)$ and $\sigma_{\rm scat(...)}$ are computed from large simulations (also used to estimate bias corrections). If $f < 1$, we set $\sigma_{\rm floor}=0$ to avoid negative covariances; otherwise $f=1$ and the $\sigma_{\rm floor}$ term is added. The term $\sigma_{\rm grey}$ is global (not survey specific), it is fitted within BBC and it enforces the reduced $\chi^2$ of the BBC fit to be equal to one. This term is expected to be zero if the scatter model used in the simulations is accurate. This novel approach of modeling intrinsic scatter has been first introduced by \citet{PantheonP_cosmo}.

In the previous DES-SN3YR analysis, we used only the scaling term, $f$, as described in \citet{Kessler_2017}. With the introduction of the \citet{BS20} intrinsic scatter model, however, we found that $f$ can take on values much larger than unity, leading to pathologically large distance uncertainties.  The $\sigma_{\rm floor}$ alternative avoids overly large uncertainties. When tested on simulations, both methods provide unbiased cosmological results.

Although $\sigma_{\mu,i}$ are used in the BBC fit, they are not suitable for a cosmology fit with an unbinned Hubble diagram containing photometrically classified events. Following \citet{KV_2022}, an unbinned Hubble diagram requires redefining the distance uncertainties,
\begin{equation}
       \sigma_{\mu,i,\mathrm{final}}  = \sigma_{\mu,i} \times S_{\zeta} / \sqrt{\PBEAMS}, 
       \label{eq:sig_renorm}
\end{equation}
where $\PBEAMS$ is the BEAMS probability \citep[see Eq.~\ref{eq:PBIa} and][]{KV_2022} and $S_{\zeta}$ is a scale such that the weighted average uncertainty in each BBC-fitted redshift bin $\zeta$ is equal to the BBC-fitted offset uncertainty, $\sigma_{M_{\zeta}}$.
The average $S_{\zeta}$ value is 1.01. \citet{KV_2022} used this prescription on 50 data-sized DES-SN5YR samples, and showed that the $w$-bias is consistent with zero at a level below 0.01 (including a CMB-like prior).

Finally, the SN distance uncertainties calculated as described in Eq.~\ref{eq:sig1} are renormalized for photometric SN samples. The renormalization is applied to all SNe and it is necessary to ensure that \textit{(i)} likely SN contaminants have inflated distance uncertainties and are downweighted in the cosmological fit and \textit{(ii)} uncertainties on the $\Delta\mu_{\BinInd}$ offsets estimated with BBC for each redshift bin are equal to the weighted average of the distance uncertainties of the SNe in the bin. The formalism related to the renormalization of the SN distance uncertainties is described in detail in \citet{KV_2022}.

\subsection{Covariance matrix and cosmological parameter estimation}
\label{sec:cosmology_fitting}
The output of BBC is a set of SN distances (as well as their uncertainties) corrected for biases from selection effects and contamination. These SN distances are estimated for the nominal analysis and for a set of analysis variants, implemented to quantify systematic uncertainties (see Sec.~\ref{sec:syst_overview}) and to build the uncertainty covariance matrix.

The $N_{\mathrm{SNe}}\times N_{\mathrm{SNe}}$ uncertainty covariance matrix, $\mathbf{C}$ is defined as the sum of a (diagonal) statistical term ($\mathbf{C_{\mathrm{stat}}}$), and a systematic term ($\mathbf{C_{\mathrm{syst}}}$). Following \citet{2011ApJS..192....1C} and \citet[][section 3.8.2]{DES_syst}, we compute the systematic covariance matrix, $C_{\mathrm{syst}}^{ij}$, defined as
\begin{equation}
    C_{\mathrm{syst}}^{ij} = \sum_{S=1}^{N_{\mathrm{syst}}} \left( \Delta \mu_{\mathrm{obs, S}}^i \right) \left(\Delta \mu_{\mathrm{obs, S}}^j\right) W^2_S,
    \label{eq:cov_matrix}
\end{equation}
where, $\Delta \mu_{\mathrm{Ia, S}}$ are the differences in SN distances after changing the systematic parameter $S$, $W_S$ is the scale of the systematic $S$, and the indices $i$ and $j$ are iterated over the $N_{\mathrm{SNe}}$ in the analysis ($i,j=1,...,N_{\mathrm{SNe}}$). 

SN distances and the uncertainty covariance matrix $C$ are used in the final cosmological fit, and the $\chi^2$ of the SN likelihood is defined as
\begin{equation}
    \chi^2 = \Delta\mu^T \cdot {C^{-1}} \cdot \Delta \mu,
\end{equation}
where $\Delta \mu$ is the $N_{\mathrm{SNe}}$-dimensional vector $\{ \mu_{\mathrm{obs},i}-\mu_{\mathrm{theory},i}(\Omega_M, w) \}_{i=1,..,N_{\mathrm{SNe}}}$.

\section{Simulations and Data comparison}
\label{sec:simulations}

\begin{figure*}
    \includegraphics[width=\linewidth]{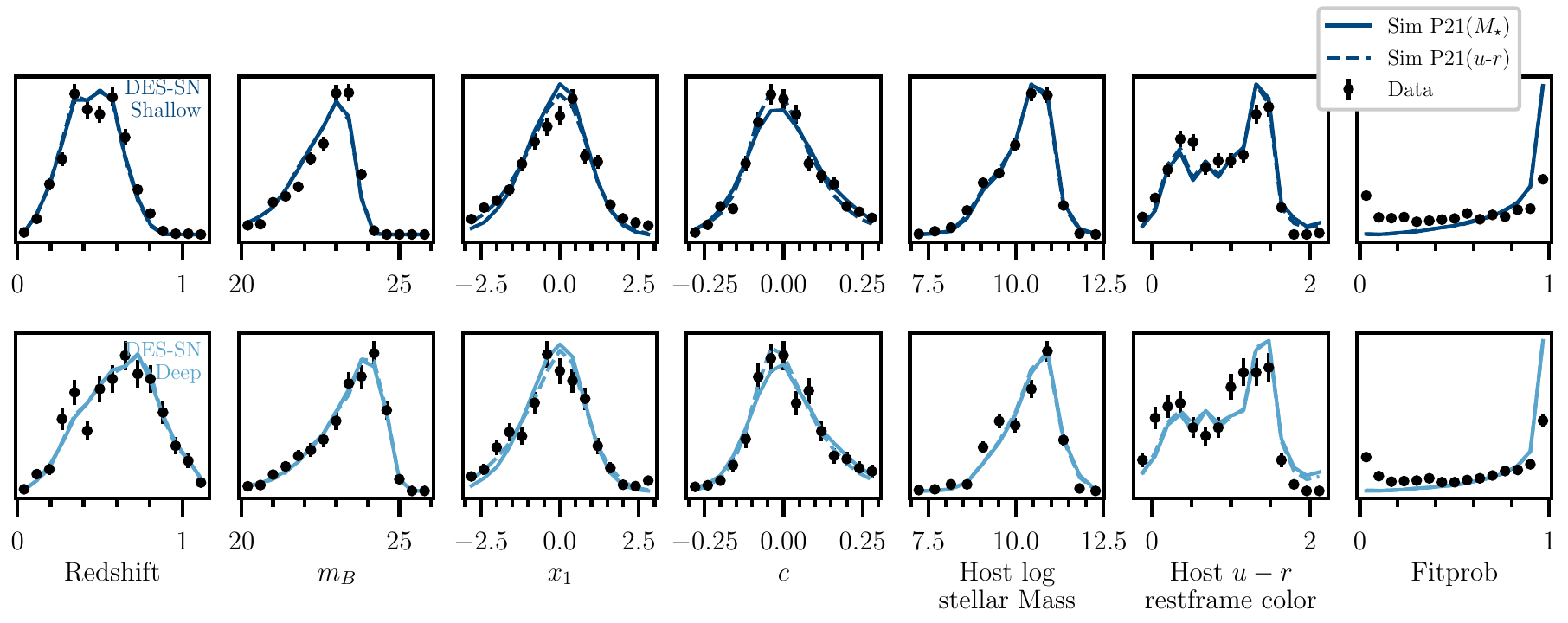}
    \caption{Comparison between observed and simulated SN and host galaxy properties. We present the comparison both for our baseline simulation (where dust and SN-host correlations are modelled as a function of host galaxy stellar mass, P21$(M_{\star})$) and for an alternative simulation where dust and SN-host correlations are modelled as a function of host galaxy restframe $u-r$ color (P21$(u-r)$, see Section \ref{sec:simulations}). We present results for SNe in DES-SN Shallow fields (upper panels) and Deep fields (lower panels) separately.}
   \label{fig:data_sim_comparison_DES}
\end{figure*}

\begin{figure}
    \centering
    \includegraphics[width=0.8\linewidth]{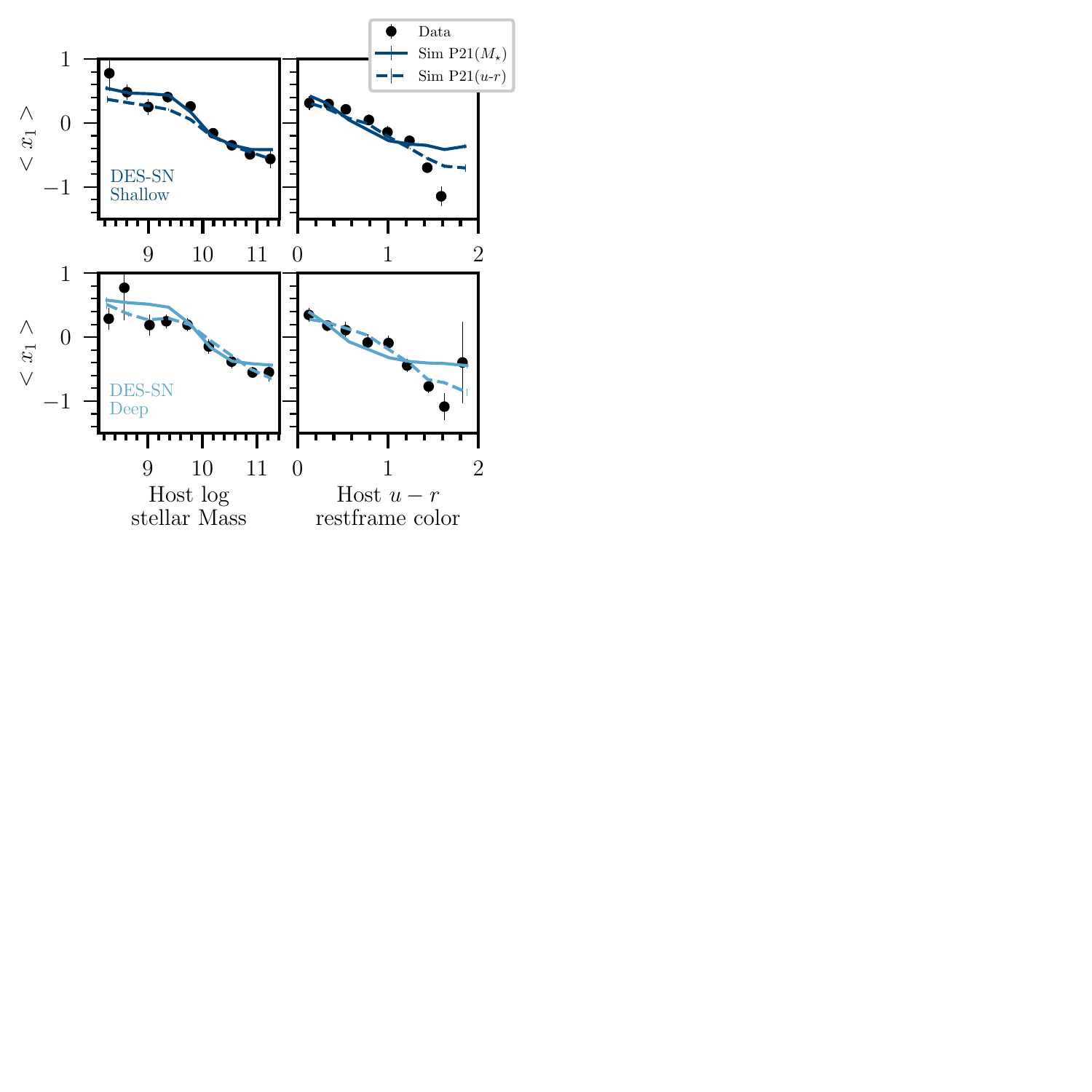}
    \caption{Comparison between observed and simulated correlations between SN stretch ($x_1$), and host galaxy properties (stellar mass on the left, and $u-r$ restframe color on the right). We present results for SNe in DES-SN Shallow (upper panels) and Deep (lower panels) separately. Data (circle markers) are compared to our baseline simulation (modeling SN-host correlations using host galaxy stellar mass, P21$(M_{\star})$, solid lines) and to an alternative simulation where SN-host correlations are modelled using host galaxy restframe $u-r$ color (P21$(u-r)$, dashed line, see Section \ref{sec:simulations}).}
   \label{fig:data_sim_comparison_DES_corr}
\end{figure}

\begin{figure}\centering
    \includegraphics[width=1\linewidth]{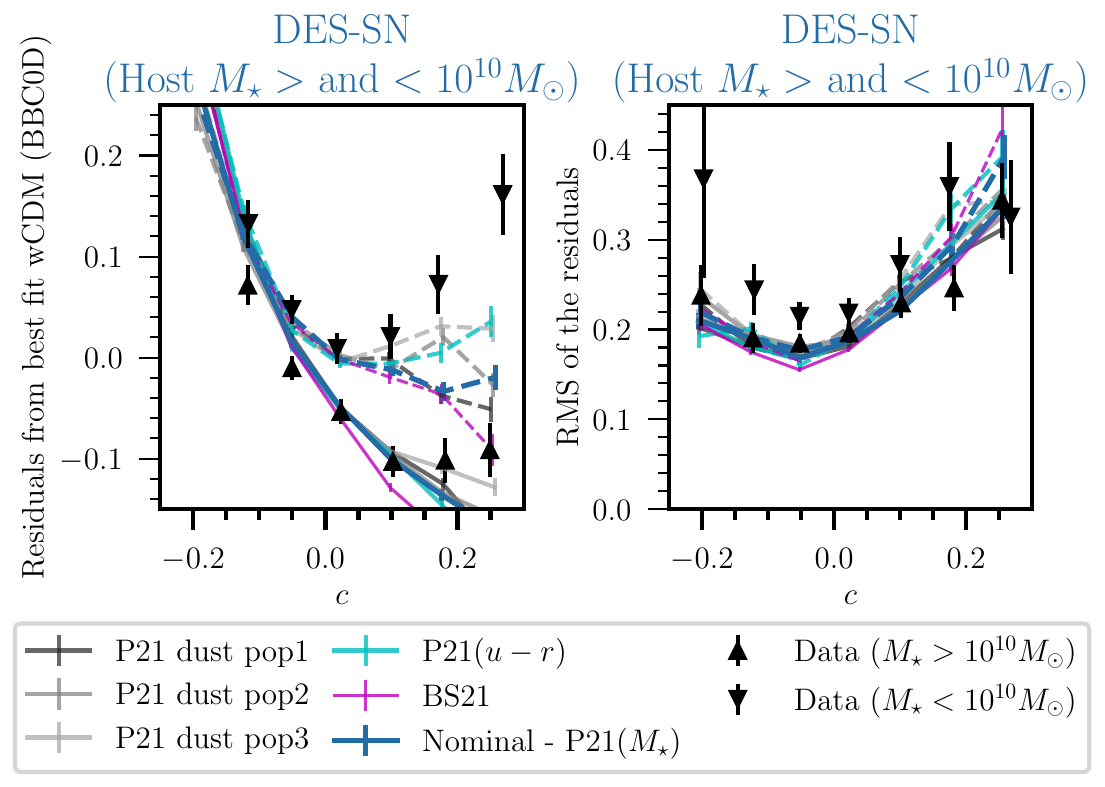} \vspace{2mm}\\
    \includegraphics[width=1\linewidth]{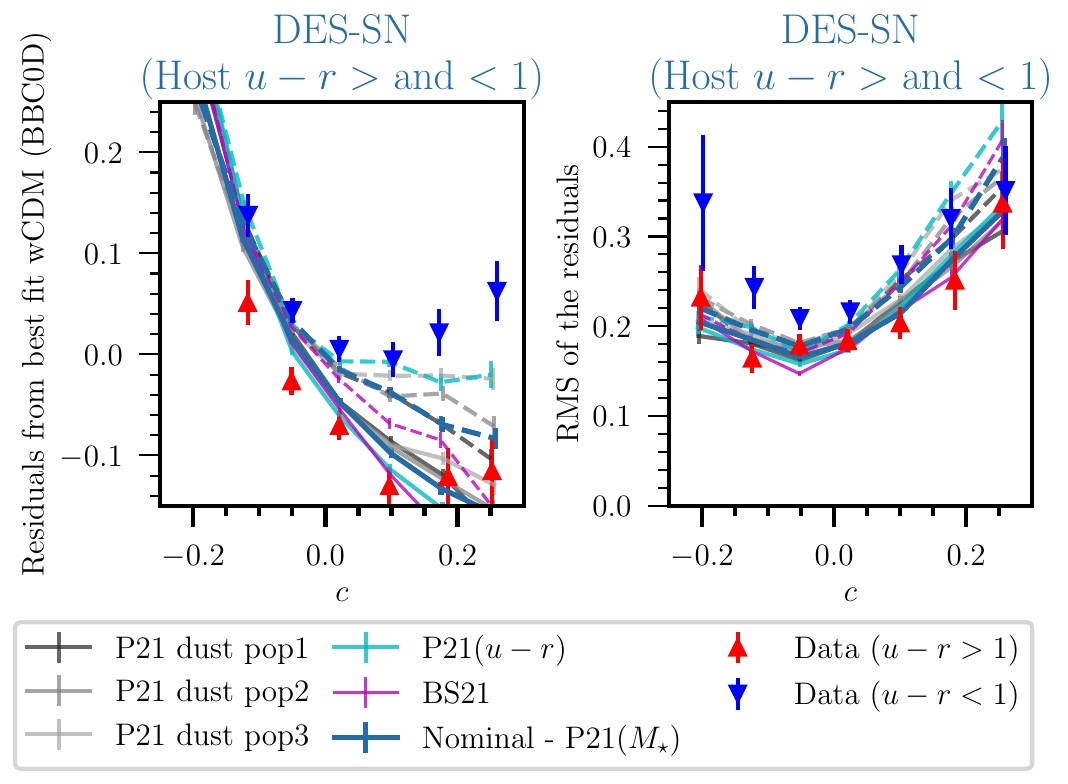}
    \caption{Average Hubble residuals (left) and RMS of the Hubble residuals (right) of the DES-SN sample as a function of SALT3 color $c$ (residuals are measured without applying any bias correction, i.e., in the \lq BBC0D\rq\ approach). The sample is split between high and low host mass galaxies (upper panels) and between intrinsically redder (rest-frame $u-r>1$) and intrinsically bluer (rest-frame $u-r<1$) galaxies (lower panels). Data are presented with upper pointing and down pointing triangles for high and low mass sub-samples respectively. Simulations are presented with solid and dashed lines for high-mass/red and low-mass/blue sub-samples respectively. The models used to generate simulations are described in Sec.~\ref{syst:intr_scatter}. Equivalent plots for Low-$z$ in shown in Fig.~\ref{fig:mu_vs_c_des_lowz}.}
    \label{fig:mu_vs_c_des}
\end{figure}

\subsection{Overview of the simulations}
In this section, we present the set of simulations used in our analysis. These simulations are generated \textit{(i)} to predict the distance biases affecting our SN samples (DES-SN and external low-redshift samples), \textit{(ii)} to train photometric classification algorithms, and \textit{(iii)} to model the core-collapse likelihood in the BBC fit (see Eq.~\ref{eq:likelihoods}). We compare our simulated DES-SN5YR samples with the observed DES-SN5YR sample. The selection criteria used to compile both the observed and simulated DES-SN5YR sample will be discussed in detail in Sec.~\ref{sec:sample_selection}.

Simulations are generated and analysed using the SuperNova ANAlysis software \citep[\snana,][]{Kessler_2009},\footnote{\url{https://github.com/RickKessler/SNANA}} integrated in the \textsc{pippin} pipeline framework \citep{Hinton2020}.\footnote{\url{https://github.com/dessn/Pippin}}

Simulations are built upon the work by \citet{DES_biascor} and \citet{Vincenzi_2020}. \citet{DES_biascor} describes in detail the modeling and simulation of the DES SN photometry and associated uncertainties, DES cadence and observing strategy, and DES detection efficiency and trigger logic to define candidates. \citet{Vincenzi_2020} focuses on the modeling of contamination from non-Ia SNe, simulations of SN host galaxies and characterization of selection effects introduced by the requirement of a spectroscopic redshift from SN host galaxies.

An \snana\ simulation can generate realistic transient light curves from various spectrophotometric models of transients. In the DES SN simulations, we include SNe Ia (Sec.~\ref{sec:sim_ia}) and various classes of SN contaminants (Sec.~\ref{sec:sim_cc}).

\subsection{Simulation of SNe Ia}
\label{sec:sim_ia}

SNe Ia are simulated using the SALT3 framework. The SALT3 parameters (redshift, day of peak $B$-band brightness, stretch and color) are simulated as following.
Redshifts are generated using SN Ia volumetric rates from \citet{2019MNRAS.486.2308F} and simulated $t_{\rm peak}$ are uniformly distributed between August 2013 and March 2018.
The distribution of $x_1$ and the dependency of $x_1$ with host galaxy properties is empirically determined using the method presented by \citet{2021ApJ...913...49P}. 

While the SALT3 light curve fitting includes 4 SN-related parameters per event, the simulation includes additional parameters to describe intrinsic scatter and the populations for stretch and color. We assume that SN Ia intrinsic scatter, color distribution and color-luminosity correlations are well described by the formalism presented by \citetalias{BS20}, but with updated parameters following \citet{P21_dust2dust}. In the formalism introduced by \citetalias{BS20}, the distribution of SN colors is modelled as the sum of an \textit{intrinsic color} Gaussian component (described by mean $c_{\rm int}$ and standard deviation $\sigma_{c_{\rm int}}$) and a reddening tail due to \textit{dust} (described as an exponentially decreasing function with the exponent scaled by $\tau_{E}$). SN luminosity color corrections are modelled as $\beta_{\rm int} c_{\rm int} + R_V E(B-V)_{\rm dust}$. Assuming that the average $R_V$ values in high and low-mass galaxies differ by approximately 1.25 reproduces the mass-step across different SN colors.

The distributions of intrinsic color $c_{\rm int}$, intrinsic $\beta_{\rm int}$, $R_V$ and $E(B-V)_{\rm dust}$ in high and low mass galaxies are determined using the \lq Dust2Dust\rq\ fitting code presented by \citet{P21_dust2dust}. For different combinations of color/dust parameters, \lq Dust2Dust\rq\ generates synthetic SNANA SN simulations and fits them with BBC. The best fit color/dust parameters are determined by iteratively comparing the SNANA simulated Hubble diagrams and the observed Hubble diagram (see Fig.~\ref{fig:data_sim_comparison_DES}). \lq Dust2Dust\rq\ in particular uses the simulated and observed Hubble residuals calculated \textit{without} applying bias corrections (i.e.\ the so-called \lq BBC0D\rq\ approach) as bias corrections are estimated making strong assumptions on SN colour/dust distribution. In Fig.~\ref{fig:mu_vs_c_des}, we present the comparison for our \lq Nominal\rq\ (also referred to \lq P21($M_{\star}$)\rq) simulation, built using the \citetalias{BS20} formalism but with the \lq Dust2Dust\rq\ best fit parameters, which are summarized in Table \ref{tab:dust} and Fig.~\ref{fig:corner_D2D}. In our baseline dust-modeling approach, we do not include a mass-step or color step (we set $\gamma=0$ in Eq.~\ref{eq:std_gamma}), however in Fig.~\ref{fig:mu_vs_c_des} we notice a difference in the residuals for $c<0$ SNe that is not captured by simulations (but it is captured by fitting for $\gamma$, see Sec.~\ref{sec:nuisance}).  

\subsubsection{SN Ia and host galaxy colour}
\label{sec:sim_ia_plus_host}
Following the findings of \citet{2023MNRAS.519.3046K} and analogous studies \citep[e.g.][]{2022A&A...657A..22B, 2023MNRAS.520.6214W}, we develop an alternative set of SN Ia simulations that use host galaxy rest-frame $u-r$ colour (instead of host galaxy \textit{stellar mass}) as the galaxy proxy to model SN-host correlations.

First, we adapt the method presented by \citet{2021ApJ...913...49P} to reproduce the (steeper) correlation between $x_1$ and host $u-r$ color (see Fig.~\ref{fig:data_sim_comparison_DES_corr}). Second, we run the \lq Dust2Dust\rq\ fitting code splitting SNe by host galaxy $u-r$ rest-frame color instead of host stellar mass and model dust parameters for intrinsically red and intrinsically blue galaxies (\lq P21($u-r$)\rq\ simulation, see Table \ref{tab:dust}). For this model, we do not simulate any additional \lq color-step\rq\ (see Eq.~\ref{eq:std_gamma}). In Sec.~\ref{syst:intr_scatter}, we discuss how the two simulations are implemented in the analysis.

\begin{deluxetable}{lccc}
\centering
\tablecolumns{4}
\tablewidth{20pc}
\tablecaption{Dust parameters used to model the SNe Ia population and estimate systematic uncertainties. We present here three sets of dust parmeters: \lq P21($M_{\star}$) \rq\ (best fit parameters found using the algorithms by \citetalias{P21_dust2dust} and splitting the SN sample on stellar mass), \lq BS20 \rq\ (using the original \textit{fudged} dust parameters presented in the \citetalias{BS20} paper), \lq P21($u-r$) \rq\ (best fit parameters found using the algorithms by \citetalias{P21_dust2dust} and splitting the SN sample on $u-r$ rest-frame color).}
\tablehead { 
\colhead {Parameter}  &
\colhead {P21($M_{\star}$)}  &
\colhead {\citetalias{BS20} $^{(\dag)}$} &
\colhead {P21($u-r$)}}
\startdata
$c_{\rm int}$ & $-0.07$ & $-0.084$ & $-0.07$ \\
$\sigma_{c_{\rm int}}$ & 0.053 & 0.042 & 0.035 \\
$\beta_{\rm int}$ & 2.07 & 1.98 & 1.86 \\
$\sigma_{\beta_{\rm int}}$ & 0.22 & 0.35 & 0.21 \\
$R_V$ highM/red $^*$ hosts & 1.66 & 1.25 & 1.5  \\
$\sigma_{R_V}$ highM/red hosts & 0.95 & 1.3  & 1.0  \\
$R_V$ lowM/blue hosts & 3.25 & 2.75 & 3.05  \\
$\sigma_{R_V}$ lowM/blue hosts & 0.93 & 1.3  & 1.0  \\
DES $\tau_E$ highM/red hosts & 0.15 & 0.15 & 0.13 \\
DES $\tau_E$ lowM/blue hosts & 0.12 & 0.12 & 0.10 \\
Low-$z$ $\tau_E$ highM/red hosts & 0.11 & 0.19 & 0.13 \\
Low-$z$ $\tau_E$ lowM/blue hosts & 0.14 & 0.10 & 0.10 \\
\enddata
    \label{tab:dust}
    \begin{tablenotes}\footnotesize
    \item $^{\dag}$ \textit{Original} \citetalias{BS20} dust parameters. \citetalias{BS20} did not use the dust parameter optimizisation code  by \citet{P21_dust2dust};
    \item $^*$ \lq highM\rq\ refers to high Mass galaxies ($>10^{10} M_{\odot}$), \lq lowM\rq\ refers to low Mass galaxies ($<10^{10} M_{\odot}$), \lq red\rq\ refers to intrinsically red galaxies ($u$-$r>$1) and \lq blue\rq\ refers to intrinsically blue galaxies ($u$-$r<$1).
    \end{tablenotes}\footnotesize
\end{deluxetable}

\subsection{Simulation of SN contaminants}
\label{sec:sim_cc}
In our simulations, we include four classes of SN contaminants: two types of peculiar SN Ia (SN Iax and 91bg-like SNe) and two types of core-collapse SNe (stripped-envelope and hydrogen-rich SNe). SN Iax and 91bg are simulated using the templates and assumptions presented by \citet{2019PASP..131i4501K}, with the revisions presented by \citet{Vincenzi_2020}. Core collapse SNe are generated using templates by \citet{Vincenzi_2019}, using the rates by \citet{Strolger_2015} and \citet{Shivvers_2017} and luminosity functions in \citet{Li_2011} \citep[with revisions by][]{Vincenzi_2020}. A detailed description of the core-collapse simulations used for the DES analysis is presented by \citet{Vincenzi_2020}. 

\subsection{Simulation of host galaxies and modeling survey selection effects}
\label{sec:sim_host}
We simulate host galaxies from the galaxy catalog presented by \citet{Qu_hostMismatch}. This catalog includes all galaxies detected in the coadded images of the DES-SN fields. For each galaxy, photometric redshifts (when spectroscopic redshifts are not available) are measured using the Self-Organizing Map (SOM) algorithm described in \citet{Qu_hostMismatch}, and galaxy properties are estimated from $griz$ DES photometry using the same galaxy SED fitting code used for the DES-SN hosts \citep{2010MNRAS.406..782S}.
SNe Ia are assigned to galaxies following the SN rates presented by \citet{DES_rates}. For peculiar SNe Ia and core-collapse SNe, we follow the same approach presented by \citet{Vincenzi_2020}.

\subsection{Simulations of the low redshift samples}
To simulate external low-$z$ SN samples, we use the same inputs and modeling assumptions presented by \citet{scolnic2018, Jones_2017_I, Jones_Foundation}, with minor adjustments mainly related to the modeling of host galaxy properties, as host galaxy properties have been remeasured for this analysis (see Sec.~\ref{sec:data_host}).

Intrinsic color distributions and dust properties for the low-$z$ SN samples are the same as for the DES-SN sample (the dust fitting code by \citet{P21_dust2dust} is simultaneously run on the DES and low-$z$ samples combined), and only the $\tau_E$ dust parameters are differentiated between high and low-$z$ samples (see Table~\ref{tab:dust}).

\section{The Hubble Diagram}
\label{sec:HD}

In Fig.~\ref{fig:HD}, we present the Hubble diagram of the DES-SN5YR analysis. This includes \numdeshd\ SNe from DES and \numlowz\ SNe from external low-$z$ samples. The selection cuts applied are summarized in Table \ref{tab:selection} and discussed below.

\subsection{Sample selection}
\label{sec:sample_selection}
\begin{deluxetable*}{ccccc}
\centering
\tablecolumns{5}
\tablewidth{20pc}
\tablecaption{\# SN After Iteratively Applied Cuts}
\tablehead {
\colhead {}  &
\colhead {} &
\multicolumn{2}{c}{\textbf{DES-SN}} & \colhead { } \vspace{-1mm}\\ \cline{3-4}
\colhead {Requirement} &
\colhead {\textbf{Low-$z$} \hspace{10mm}} &
\colhead {\# all}   &
\colhead { $P_{\rm Ia}>0.5$ $^{(1)}$}  &
\colhead {\hspace{4mm} \textbf{Total}\hspace{4mm} } } 
\startdata
Spec-$z$ available, SALT3 fit converged and $z>0.025$ & 247 & 3621 & 2200 [60\%] & 3868 \\
`Normal~SNIa' ($|x_1|<3$~\&~$|c|<0.3$) & 238 & 2449 & 2052 [83\%] & 2687 \\
`Well~constrained' ($\sigma_{x_1}<$1,~$\sigma_{t_{\rm peak}}<2$) & 238 & 1917 & 1639 [85\%] & 2155 \\
Fit probability (fitprob$>0.001$) & 221 & 1835 & 1627 [88\%] & 2056 \\
Detected host galaxy & 211 & 1806 & 1602 [88\%] & 2017 \\
Spec-$z$ from the host galaxy emission lines (not SN spectrum) $^{(2)}$ & 211 & 1765 & 1563 [88\%] & 1976 \\
Chauvenets criterion & 209 & 1757 & 1557 [88\%] & 1966 \\
Valid bias correction & 204 & 1694 & 1541 [90\%] & 1898 \\
Sub-sample of common CIDs across all systematic variants $^{(3)}$ & 194 & 1635 & 1499 [91\%] & 1829 \\
\hline \vspace{-1mm}\\
Cosmological Sample & \textbf{194} & \textbf{1635} & \textbf{1499 [91\%]} & \textbf{1829} \vspace{1mm} \\
\enddata
    \label{tab:selection}
    \begin{tablenotes}\footnotesize
    \item $^{(1)}$ Probabilities are from SNN classifier trained on \citetalias{Vincenzi_2019}. In parenthesis, we report the percentage of likely SN Ia for each given cut.
    \item $^{(2)}$ We exclude SNe for which a host was not detected and/or redshift information is from SN spectroscopic data, not from host galaxy emission lines.
    \item $^{(3)}$ In order to build the systematic covariance matrix, we require to have the same SNe across all systematic variants.
    \end{tablenotes}\footnotesize
\end{deluxetable*}

\begin{figure*}\centering
    \includegraphics[width=0.95\linewidth]{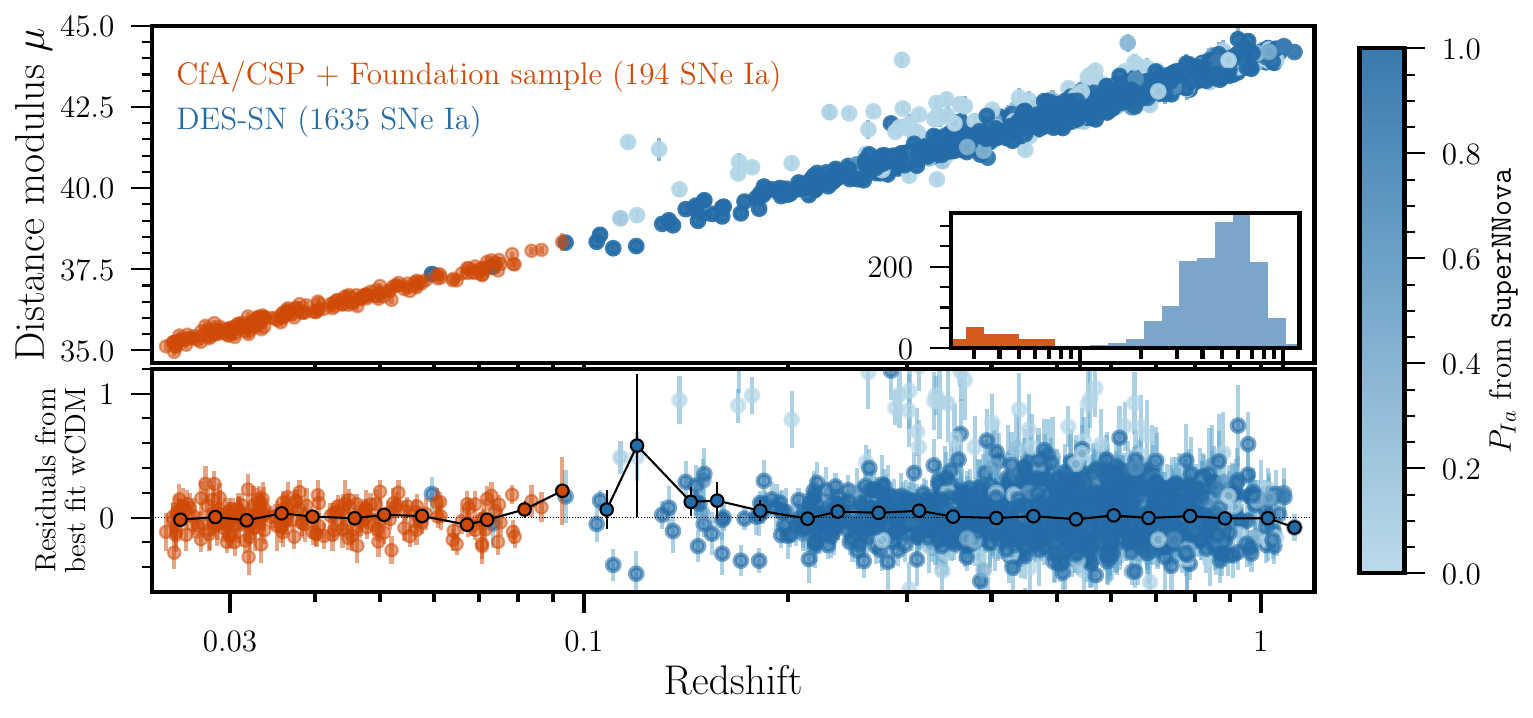}
    \caption{Hubble diagram (upper panel) and Hubble residuals (lower panel) combining the DES sample (blue) and external low-$z$ SN samples (orange, see Section~\ref{sec:lowz_samples}). For each SN event in the DES, we present classification probabilities, $P_{\rm Ia}$, estimated using the \snn\ algorithm (see Section~\ref{sec:classifiers}). Distance uncertainties are calculated using Eq.~\ref{eq:sig1} and do not include the renormalization term included in Eq.~\ref{eq:sig_renorm}.}
    \label{fig:HD}
\end{figure*}

\begin{deluxetable*}{cccccccccl}
\centering
\tablecolumns{9}
\tablewidth{18pc}
\tablecaption{Nuisance parameters and cosmological fit when combining DES with different low-$z$ external samples and when using DES alone}
\tablehead {
\colhead {Sample}  &
\colhead {$N_{\mathrm{SNe}}$} &
\colhead {$\alpha$}   &
\colhead {$\beta$}   &
\colhead {$\gamma$}   &
\colhead {$\sigma_{\rm gray}$} &
\colhead {RMS $^{*}$} &
} 
\startdata
DES-SN + low-$z$ &  1829 & 0.161(1) & 3.12(3) & 0.038(7) & 0.04 & 0.168 \vspace{2mm}\\
\hline
DES-SN only &  1678 & 0.170(4) & 3.14(3) & 0.046(9) & 0.04 & 0.177  \\
DES-SN + Foundation &  1796 & 0.166(3) & 3.13(0) & 0.042(8) & 0.04 & 0.173 \\
DES-SN + CfA/CSP &  1760 & 0.167(3) & 3.12(4) & 0.043(9) & 0.04 & 0.175 \\
Foundation+ CfA/CSP only &  204 & 0.137(8) & 2.90(10) & 0.019(19) & 0.06 & 0.118 \vspace{2mm}\\
\enddata
\tablenotetext{$*$}{RMS is measured applying a cut of $P_{\mathrm{Ia}}>0.5$ on the DES SN sample.}
\label{tab:nuisance_baseline}
\end{deluxetable*}

\begin{figure}\centering
    \includegraphics[width=0.95\linewidth]{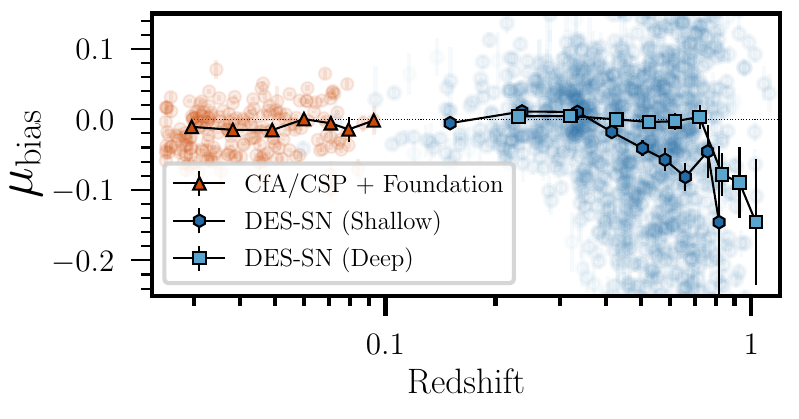}
    \caption{Bias corrections on $\mu$ for the DES sample (squares for deep fields, hexagon for shallow fields) and external low-$z$ SN samples (triangles). Depending on the depth of the survey, corrections for selection effects start increasing at different redshifts.}
    \label{fig:HD_biascorMu}
\end{figure}

\begin{figure*}\centering
    \includegraphics[width=0.95\linewidth]{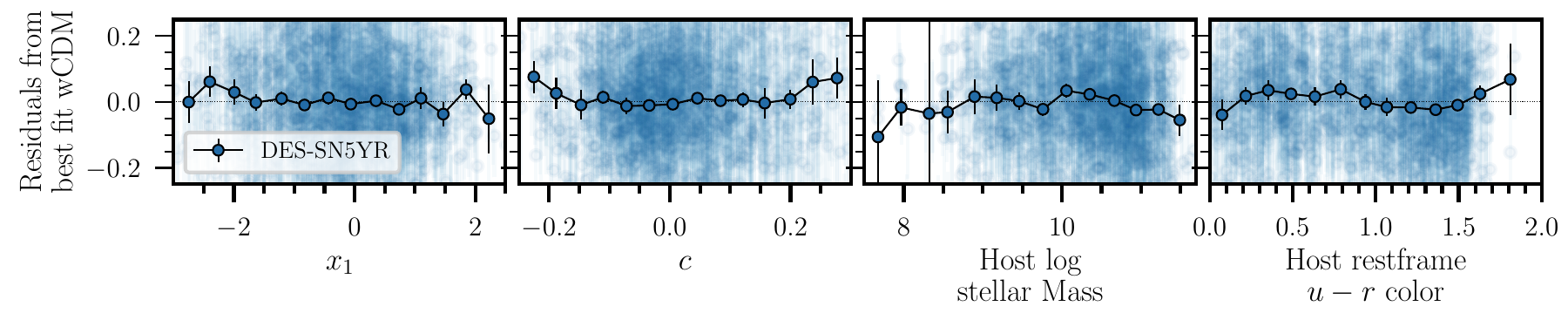}
    \includegraphics[width=0.95\linewidth]{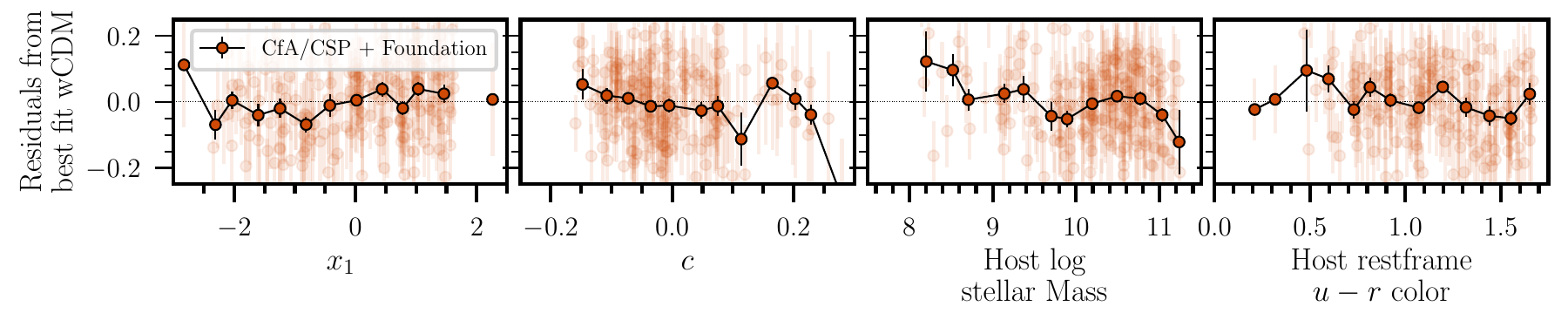}
    \caption{Hubble residuals from the DES sample (top row) and external low-$z$ SN samples (bottom row) as a function of (from left to right) SALT3 stretch $x_1$, SALT3 color $c$, host galaxy stellar mass and host galaxy rest-frame $u-r$ color. Note that the residuals shown include bias corrections and the `mass step correction' following Eq.~\ref{eq:tripp}.}
    \label{fig:HD_res}
\end{figure*}
First, we consider only DES SNe with a host galaxy spectroscopic redshift. We do not include DES SNe for which the host galaxy was not detected and the redshift information could only be inferred from the SN spectrum \citep[the sample selection function of this sample is different, see][for more details]{Vincenzi_2020}. For this reason, the DES SN sample presented in this analysis does not include all the DES SNe Ia presented in the DES-SN3YR analysis \citep[in this analysis, spectroscopically followed-up DES SNe were selected, regardless of host spectroscopic redshift information, see][]{DES_SMP, DES_abbott}.
To ensure good convergence of the SALT3 fit, we apply several cuts based on the quality of the light-curve. We select DES SNe with two bands that each have at least one detection with SNR$>5$. In line with previous SN cosmological analyses, we require at least one observation before phase +5 days after $B$-band peak (we do not require observations before light-curve peak). In Appendix~\ref{sec:require_SNPeak}, we discuss how the size of our sample and the final cosmological results change when requiring at least one detection before SN peak brightness.

We also apply SALT3-based selection cuts in both stretch and color ($-3<x_1<+3$
and $-0.3<c<+0.3$) and $\sigma_{x_1}<1$ and $\sigma_{t_{\rm peak}}<2$ days. These cuts are commonly applied in SN analyses to select \lq normal\rq\ SNe Ia; they also significantly reduce contamination from peculiar SNe Ia and core-collapse SNe \citep{Vincenzi_2020} (see Table~\ref{tab:selection}).

In addition, applying bias corrections (as described in Sec.~\ref{sec:bias_corrections}) constitute a sample selection cut in itself. 
In BBC, for a small fraction of SNe it is not possible to robustly determine bias corrections because the simulations of SN Ia used to calculate bias corrections do not have enough events in some regions of the SALT3 parameter space.
As discussed in \citet{Vincenzi_2020}, this \lq valid bias correction\rq\ requirement implicitly reduces contamination because SN contaminants generally populate regions of the SALT3 parameter space that are atypical for SNe Ia.
Moreover, applying Chauvenet's criterion, we iteratively apply a $4\sigma$ cut on the Hubble diagram residuals.

Finally, in order to build the unbinned systematics covariance matrix (see Sec.~\ref{sec:cosmology_fitting}), we require each analysis variant to have the same set of events as the nominal model. This results in an additional 3.6 percent loss of SNe. In total, there are \numdeshd\ DES SNe in the Cosmological Sample (see Table \ref{tab:selection}) for which \numdesia\ (91\%) are classified as likely Type Ia ($P_{\mathrm{Ia}}>0.5$). When combining with the external low-$z$ samples, the total number of SNe on our Hubble diagram is \numtot. This sample is smaller than the photometric DES SN sample presented by \citet{2022MNRAS.514.5159M}, as it is built applying more stringent selection criteria.

In Fig.~\ref{fig:HD}, we present the weighted mean of Hubble residuals as a function of redshift and do not observe any significant residual trend in our data. In Fig.~\ref{fig:HD_biascorMu}, we present the weighted mean of bias corrections $\mu_{\rm bias}$ (see Eq.~\ref{eq:bias_corrections}) as a function of redshift. These corrections become increasingly significant ($\sim 0.1$ mag) at higher redshifts (redshift 0.5 for the SNe in the shallow DES SN fields, 0.8 for the deep DES SN fields), where selection effects have a more significant effect. Bias corrections have also strong dependency on SN color and SN stretch.

\subsection{Nuisance parameters}
\label{sec:nuisance}
In our baseline analysis, we fit for the nuisance parameters $\alpha$, $\beta$, $\gamma$ and $\sigma_{\rm gray}$, and we present the fitted values in Table \ref{tab:nuisance_baseline}. We do not fix $\gamma$ to zero because we want to test for any residual brightness step that is not explained by our dust model, and might be related to intrinsic SN astrophysics.

When combining the DES sample with the external low-$z$ samples, we find $\alpha=0.161\pm0.001$, $\beta=3.12\pm0.03$. After accounting for dust law variation in the reported distances and uncertainties, we find $\gamma=0.038\pm0.007$ ($>5\sigma$ residual mass step) and a residual intrinsic scatter, $\sigma_{\rm gray}$, of 0.04.
From the DES sample alone, we find consistent $\beta$ ($\beta=3.14\pm0.03$), slightly higher $\alpha$ ($\alpha=0.170\pm0.004$) and still significant mass-step ($\gamma=0.046\pm0.009$), with a residual intrinsic scatter of 0.04.
We also compare nuisance parameters estimated from the DES-SN combined with the different low-$z$ samples. We find that most nuisance parameters are consistent between the different sample combinations considered. The most significant discrepancy is in the fitted $\alpha$ for low-$z$ samples alone (we find a significantly lower $\alpha$ compared to the value found when including the DES-SN sample). We discuss these discrepancies in Sec.~\ref{disc:nuisance_evolution}. 

\subsection{Hubble residuals}
In Fig.~\ref{fig:HD} and Fig.~\ref{fig:HD_res}, we present Hubble residuals of DES-SNe and low-$z$ SNe as a function of redshift and other relevant SN and host parameters. These are the residuals estimated after applying the BBC4D approach described in Sec.~\ref{sec:bias_corrections}.

In Fig.~\ref{fig:HD} and Fig.~\ref{fig:HD_res}, we do not observe any significant residual trend in our data. In general, we highlight that the BBC4D approach (with the additional grey residual step $\gamma$) fully corrects the color trends highlighted in Fig.~\ref{fig:mu_vs_c_des}. Hubble residuals as a function of host stellar mass and host restframe $u-r$ color present discontinuities at $10^{10} M_{\odot}$ and restframe $u-r$ color $\sim0.75$ respectively, which might suggest that our modeling of discontinuous dust properties in high/low mass host galaxies might be too simplistic.

\section{Systematic uncertainties}
\label{sec:syst_overview}
In this section, we describe the various sources of systematic uncertainties considered in the analysis. These are also summarized in Table~\ref{tab:syst_description}.

\begin{deluxetable*}{p{8.8cm}p{0.5cm}p{5cm}p{3cm}}
\tablecolumns{4}
\tablecaption{Sources of Uncertainty}
\tablehead {
\colhead {Baseline}  &
\colhead {Size\tablenotemark{a}}  &
\colhead {Systematic} & \colhead {Label} }\vspace{2mm}
\startdata
\multicolumn{4}{l}{\textbf{Calibration and Light-curve Modeling (Section \ref{syst:calibration})}}\\
\hspace{2mm} SALT3 surfaces $\&$ ZP & 1/10 & 10 covariance realizations  & \lq SALT3+Calibration\rq \\
\hspace{2mm} HST Calspec 2020 Update & 1 & 5 mmag/7000\AA\ & \lq HST Calspec\rq  \vspace{2mm}\\
\multicolumn{4}{l}{\textbf{SN Ia properties and astrophysics (Section \ref{syst:astro})}}\\
\hspace{2mm} Dust-based model \citet{P21_dust2dust} (\lq P21($M_{\star}$)\rq) & 1/3 & 3 realizations from MCMC dust model fitting code & \lq P21 dust pop 1/2/3\rq \\
\hspace{2mm} & 1 & Original \citetalias{BS20} dust parameters & \lq BS21\rq \\
\hspace{2mm} & 1 & Splitting on $u-r$ & \lq P21($u-r$)\rq \\
\hspace{2mm} Empirical modeling of $x_1$-M$_{\star}$ correlations & 1 & Modeling SN age following \citet{W22_x1age} & \lq Model SN age\rq\\
\hspace{2mm} No $\alpha$ evolution & 1 & $\alpha(z) = \alpha_{0} + \alpha_{1}\times z$ & \lq $\alpha$ Evolution\rq \\
\hspace{2mm} No $\beta$ evolution & 1 & $\beta(z) = \beta_{0} + \beta_{1}\times z$ & \lq $\beta$ Evolution\rq \\
\hspace{2mm} No $\gamma$ evolution & 1 & $\gamma(z) = \gamma_{0} + \gamma_{1}\times z$ & \lq $\gamma$ Evolution\rq \\
\hspace{2mm} Mass step location at $10^{10} M_{\odot}$& 1 & $10^{10.3} M_{\odot}$ & \lq Mass Location\rq \\
\hspace{2mm} $\sigma_{\rm int}$ modeling with scaling+additive scatter terms (eq.~\ref{eq:sig_floor}) & 1 & Scaling term only & \lq $\sigma_{\rm int}$ modeling\rq \vspace{2mm}\\
\multicolumn{4}{l}{\textbf{Milky Way extinction (Section \ref{syst:MW})}}\\
\hspace{2mm} MW scaling \citet{Schlafly11} & 1 & 5\% scaling & \lq MW scaling\rq \\
\hspace{2mm} MW color law $R_V$=3.1 and F99 & 1/3 & $R_V$=3.0 and CCM &\lq MW color law\rq\vspace{2mm}\\
\multicolumn{4}{l}{\textbf{Host and survey modeling (Section \ref{syst:survey})}}\\
\hspace{2mm} SN host catalog by \citet{Qu_hostMismatch} & 1 & SN host catalog using DES-SVA galaxy catalog &\lq DES SV catalog\rq \\
\hspace{2mm} Efficiency $\epsilon_{z}^{\mathrm{spec}}$ presented by \citetalias{Vincenzi_2020}  & 1 & Shift of $\pm$0.2 mag in the efficiency curves &\lq Shift in host spec eff\rq \vspace{2mm}\\
\multicolumn{4}{l}{\textbf{Contamination and photometric classifiers (Section \ref{syst:contamination})}}\\
\hspace{2mm} Classification using \snn & 1 &  SCONE, SNIRF \\
\hspace{2mm} Classifier training sample simulated using \citetalias{Vincenzi_2019} templates & 1 & \citetalias{Jones_2017_I} templates, DES CC templates (\lq \texttt{SuperNNova} training\rq )\\
\hspace{2mm} Core-collapse SN prior using \citetalias{Vincenzi_2019} simulation & 1 & Polynomial fit as in \citet{Hlozek_2012} &\lq CC SN prior\rq \vspace{2mm}\\
\multicolumn{4}{l}{\textbf{Redshift (Section \ref{syst:redshift_vpec})}}\\
\hspace{2mm} Peculiar velocities using 2M$++$ & 1 & 2M$++$(Line-of-sight integration) or 2MRS &\lq Pec Velocities\rq \\
\hspace{2mm} No redshift shift & 1/6 &  $\Delta z = 4 \times 10^{-5}$ &\lq Redshift shift\rq \\
\enddata
\tablenotetext{a}{Weighting adopted for each source of systematic uncertainty when building the systematic covariance matrix (see also Eq.~\ref{eq:cov_matrix}). In Sec.~\ref{sec:syst_overview}, we provide an explanation for the weights that are different from 1.}
\label{tab:syst_description}
\end{deluxetable*}

\subsection{Calibration and light-curve modeling}
\label{syst:calibration}
In this section, we discuss all the sources of systematics uncertainties related to the calibration of the DES SN Ia fluxes and of the samples of SNe Ia that are used in the training of the SALT3 light-curve model.

The photometric systems of DES, Foundation and the other low-$z$ SN samples are cross-calibrated using the large and uniform sky coverage of the public Pan-STARRS stellar photometry catalog \citep{Fragilistic}. In this cross-calibration approach, the filter zero point and mean wavelength in all systems are fitted simultaneously in order to produce a calibration uncertainty covariance matrix\footnote{\href{https://github.com/PantheonPlusSH0ES/DataRelease/tree/main/Pantheon+_Data/2_CALIBRATION/FRAGILISTIC_COVARIANCE.npz}{https://github.com/PantheonPlusSH0ES/DataRelease/tree/
\\
main/Pantheon+\_Data/2\_CALIBRATION/
\\
FRAGILISTIC\_COVARIANCE.npz}} that can be used in cosmological-model constraints. The calibration uncertainty covariance matrix is used to randomly draw ten mock realizations of zero-point calibration offsets and effective mean wavelength shifts. These correlated shifts are applied to re-calibrate the SALT3 training sample and produce ten perturbations of the SALT3 model.
Following this approach, calibration uncertainties and light-curve modeling uncertainties are propagated \textit{simultaneously} to the light curve fitting (and not decoupled as in most previous SN cosmological analyses).
The calibration uncertainty covariance matrix implemented in our analysis is presented by \citet{Fragilistic} and the relative set of SALT3 surfaces is presented by \citet{2023MNRAS.520.5209T}.

In addition, we consider uncertainties associated to the fundamental flux calibration of the HST CALSPEC standards. These uncertainties are estimated to be of 5~mmag/7000~\AA\ \citep{2014PASP..126..711B}.

\subsection{SN Ia properties and astrophysics}
\label{syst:astro}
In this section, we discuss sources of systematics uncertainties related to the astrophysics of SN Ia and their host galaxies. Assumptions on SN Ia intrinsic properties, their correlations with host galaxy properties, and their evolution with redshift primarily affect the bias corrections.

\subsubsection{Intrinsic scatter model}
\label{syst:intr_scatter}
As discussed in Sec.~\ref{sec:sim_ia}, SNe Ia intrinsic scatter is modelled using dust-based formalism introduced by \citetalias{BS20}.  
For our nominal analysis, we use the best-fit dust parameters determined using the MCMC fitting code \lq Dust2Dust\rq\ by \citetalias{P21_dust2dust} and splitting SNe by their host galaxy stellar mass (see values summarized in Table \ref{tab:dust}). In addition to the baseline approach (also referred to as \lq P21($M_{\star}$)\rq), we consider the following variations of dust-based intrinsic scatter models:
\begin{itemize}
    \item We randomly draw three sets of dust parameters from the MCMC chains produced by \lq Dust2Dust\rq. The three realizations are presented in Fig.~\ref{fig:mu_vs_c_des}. We refer to these models as \lq P21 population 1\rq, \lq P21 population 2\rq and \lq P21 population 3\rq ;
    \item We use the dust parameters \textit{originally} presented in \citetalias{BS20} (see Table \ref{tab:dust});
    \item The \lq Dust2Dust\rq\ best fits when splitting SNe by host galaxy $u-r$ rest-frame color instead of host stellar mass. We refer to this model as \lq P21($u-r$)\rq. For this model, we measure the color-step $\gamma_{u-r}$ (see Sec.~\ref{sec:measure_distances}).
\end{itemize}

In Fig.~\ref{fig:mu_vs_c_des}, we show how the different models listed above reproduce the observed correlations between Hubble residuals and SN color, both for SNe in high and low mass galaxies and for SNe in red and blue galaxies. 

Historically, most SN cosmological analyses have included the two following intrinsic scatter models:
\begin{itemize}
    \item The model presented by \citet{G10} (generally referred as \lq G10\rq ), according to which the SN luminosity dispersion is mostly (70\%) wavelength-independent and 25\% chromatic.
    \item The model presented by \citet{C11} (referred as \lq C11\rq ) according to which the SN luminosity dispersion is mostly chromatic dependent (70\%). 
\end{itemize}
We do not include these models in our analysis as they are highly disfavoured by both publicly released \citep{PantheonP_cosmo} and the DES-SN5YR data in this work (see discussion in Sec.~\ref{disc:intrinsic_scatt_models}).

\subsubsection{Modeling of residual intrinsic scatter and distance uncertainties}
In Eq.~\ref{eq:sig1} and Eq.~\ref{eq:sig_floor}, we present how uncertainties and residual intrinsic scatter floor ($\sigma_{\rm gray}$) are modelled in our analysis. The scaling and additive terms in Eq.~\ref{eq:sig1} and Eq.~\ref{eq:sig_floor} are used to deflate and inflate SN distance uncertainties so that the reduced $\chi^2$ of the cosmological fitting is close to unity across different regions of the redshift/color/host stellar mass parameter space. The additive term encapsulates any unaccounted for SN intrinsic scatter. 

We test the alternative approach of fitting the unexplained intrinsic scatter as a constant floor and using the scaling term only to inflate/deflate uncertainties when necessary.
The two approaches should be fundamentally identical (and, in fact, when testing the two methods on 25 simulations, we recover the input cosmology in both cases). However, the approach used in our nominal analysis allows us to directly test whether our simulations reproduce SN Ia intrinsic scatter by testing whether $\sigma_{\rm gray}$ is 0.

\subsubsection{Modeling host galaxies and SN-host galaxy correlations}
\label{syst:W22}
In our nominal analysis, we model correlations between SN stretch $x_1$ and SN host stellar mass $M_{\star}$ following the empirical approach presented by \citet{2021ApJ...913...49P}. However, we incorporate in our systematic error budget an alternative `galaxy-driven' approach that models $x_1-M_{\star}$ correlations starting from our current knowledge of the underlying astrophysics causing this correlation.

The galaxy-driven model used in this analysis is presented by \citet[hereafter W22]{W22_x1age}. This model is based on the SN rates and SN delay time distributions presented by \citet{DES_rates}. 

The model uses galaxy evolution models to generate mock catalogs of galaxies and their properties, e.g., stellar population age, stellar mass, star formation rate and observed optical photometry.
For each galaxy, the distribution of SN progenitor ages is determined by convolving the SN delay time distribution by \citet{DES_rates} with the galaxies’ star-formation histories and stellar populations. 

\begin{figure}\centering
    \includegraphics[width=0.95\linewidth]{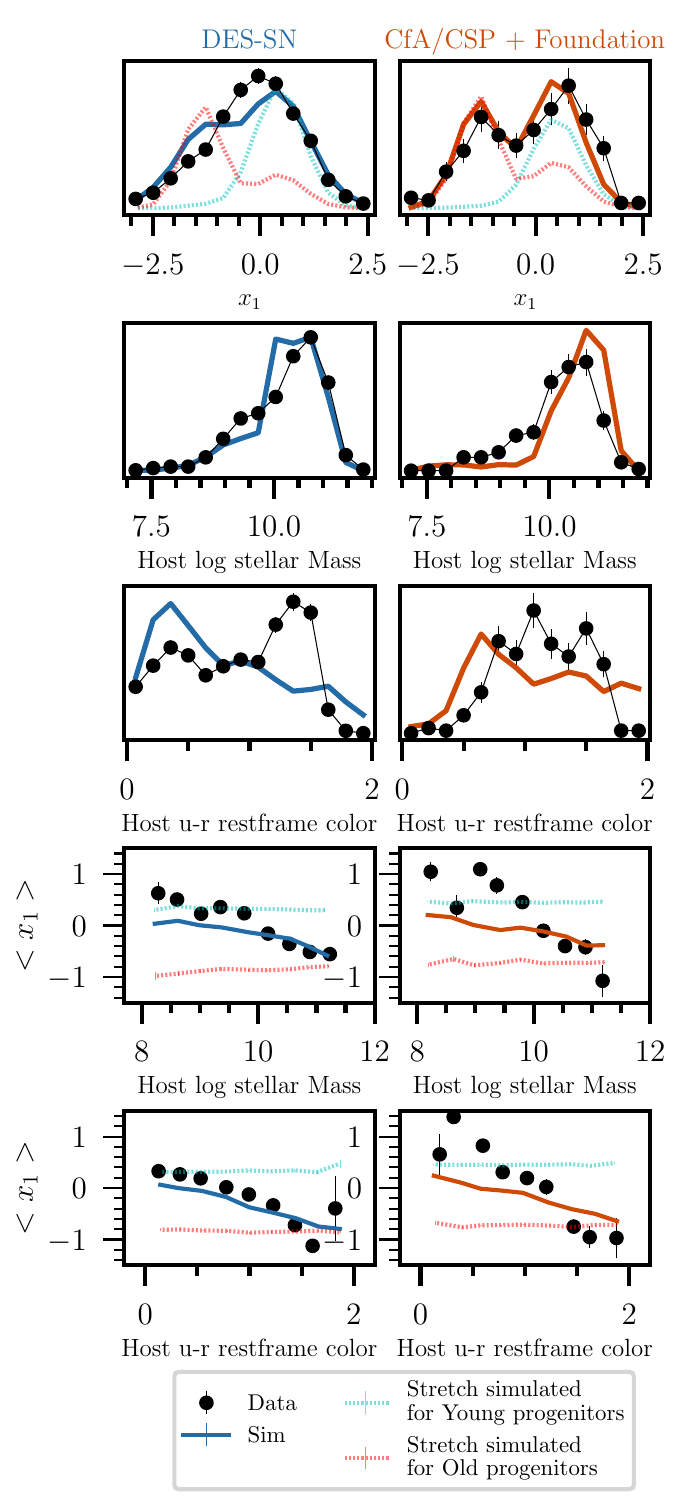}
    \caption{Comparison between data and alternative simulations generated using the galaxy-driven model by \citet{W22_x1age}. We compare distributions of (from top to bottom) SN $x_1$, SN host stellar mass, SN host $u-r$ restframe color, and correlations between $x_1$ and host $M_{\star}$ and host $u-r$ restframe color. We present our results for DES-SN (left) and low-$z$ samples (right). We also present the underlying distributions of young and old progenitors from which SN $x_1$ values are inferred. A similar comparison using our baseline simulations is presented in Fig.~\ref{fig:data_sim_comparison_DES}.}
    \label{fig:age}
\end{figure}

Given a galaxy and its associated SN age estimate, the stretch parameter $x_1$ is assigned following the prescription presented by \citet{Nicolas21} (old and young SN progenitors are associated to two separate distributions determined from external nearby SN sample and represented in Fig.~\ref{fig:age}). 
In this galaxy-driven approach, $x_1-M_{\star}$ correlations are the result of a physically motivated modeling of correlations between SN age and SN host galaxy mass. 

In Fig.~\ref{fig:age}, we present the distribution of SN stretch and SN host galaxy properties simulated using this galaxy-driven approach and compare it to the observed properties for both DES-SN and external low-$z$ samples. We find good agreement between simulated and measured SN stretch and SN host stellar mass distributions, with the exception of low-mass galaxies. However, we find that the simulations built using this alternative model significantly underestimate the number of intrinsically red host galaxies (host rest-frame $u$-$r>1$). This is likely to be caused by an oversimplified approach to modeling galaxy quenching in the initial galaxy catalog. Moreover, correlations between $x_1$ and host properties observed in the data are steeper than what is reproduced by this alternative simulation, and this suggests that a revision of the modeling proposed by \citet{Nicolas21} is required. 

Despite the discrepancies between observations and simulations, we include this model in the systematic error budget as it provides an astrophysically-motivated method to model SN-host galaxy correlations.

Finally, we include as an additional systematic uncertainty shifting the splitting point to measure the mass step from $10^{10} M_{\odot}$ to $10^{10.3} M_{\odot}$, since the typical uncertainty on our stellar mass estimates is 0.3 dex.

\subsubsection{Standardization parameter evolution}
\label{syst:nuisance_evol}
Given the wide redshift range probed by the DES-SN sample (see Fig.~\ref{fig:redshift_distr}), we test for evolution of the standardization parameters $\alpha$, $\beta$ and $\gamma$, as well as the inferred $\sigma_{\mathrm{floor}}$ (see Eqs.~\ref{eq:sig1} and \ref{eq:sig_floor}). We define $\alpha(z) = \alpha_0 +
\alpha_1 \times z$, and similarly for $\beta(z)$ and $\gamma(z)$. We present and discuss our results in Sec.~\ref{disc:nuisance_evolution}.

\subsection{Milky Way extinction corrections}
\label{syst:MW}
Inaccurate Milky Way extinction corrections can introduce biases in cosmology, especially because Milky Way extinction affects low and high redshift SNe differently (the average Milky Way reddening in low redshift SNe is twice the average Milky Way reddening in DES SNe, see Sec.~\ref{sec:MW}). 

For our systematic analysis, we test the effect of a global 5\% scaling on Milky Way corrections, following the reanalysis and uncertainties presented by \citet{Schlafly11}. As discussed in \citet{Schlafly10}, the Milky Way reddening law favoured by the data is a \citet{Fitzpatrick99} reddening law with $R_V=3.1$. However, we conservatively include a systematic uncertainty in the Milky Way reddening law and analyze the data using a \citet{CCM89} color law. The \citet{CCM89} color law has the second lowest $\chi^2$ when compared to the extinction derived from star photometry \cite[twice the $\chi^2$ associated to the \citet{Fitzpatrick99} reddening law, see][section 5.1.3]{Schlafly10} therefore we weight the Milky Way color law systematic by a factor of 1/3 ($W_S^2=1/3$, see eq.~\ref{eq:cov_matrix}).

\subsection{Host association and survey modeling}
\label{syst:survey}

In this section, we discuss systematic uncertainties related to the modeling of host association and host-related survey selection effects.

\subsubsection{Host galaxy mismatch}
Host galaxy mis-association can occur for various reasons: the true host and/or the true host outskirts are too faint to be detected and a brighter, apparently closer (in terms of DLR) galaxy is identified as the likely host instead; or the SN is far from the true host, and fails our DLR cut. In the first case, deeper images can reduce host mis-association.

After upgrading from the shallower DES Science Verification (SV) images to the deeper co-added images, \citet{DES_deepstacks} finds that less than 1.1 per cent of the DES SN candidates change host galaxies, thus providing a preliminary estimate of the fraction of host mis-associations expected in the DES sample.

\citet{Qu_hostMismatch} provide a more robust assessment of host mis-association in DES-SN and estimated the percentage of misidentified hosts to be 1.7 per cent. This prediction is based on high quality simulations, built from a galaxy catalog that accurately models galaxy light profiles. These simulations reproduce the observed distributions of SN-galaxy separations and DLRs for DES-SN (Fig. 5 and 6 of \citealt{Qu_hostMismatch}). 

When comparing the two deep SN catalogs by \citet{DES_deepstacks} and \citet{Qu_hostMismatch}, we find that 7 of 1635 DES SNe ($0.5$ per cent) have uncertain assigned hosts (i.e., they are assigned to different hosts depending on the catalog used). Therefore, we assume this source of systematic is negligible for our analysis.

\begin{figure}\centering
    \includegraphics[width=0.9\linewidth]{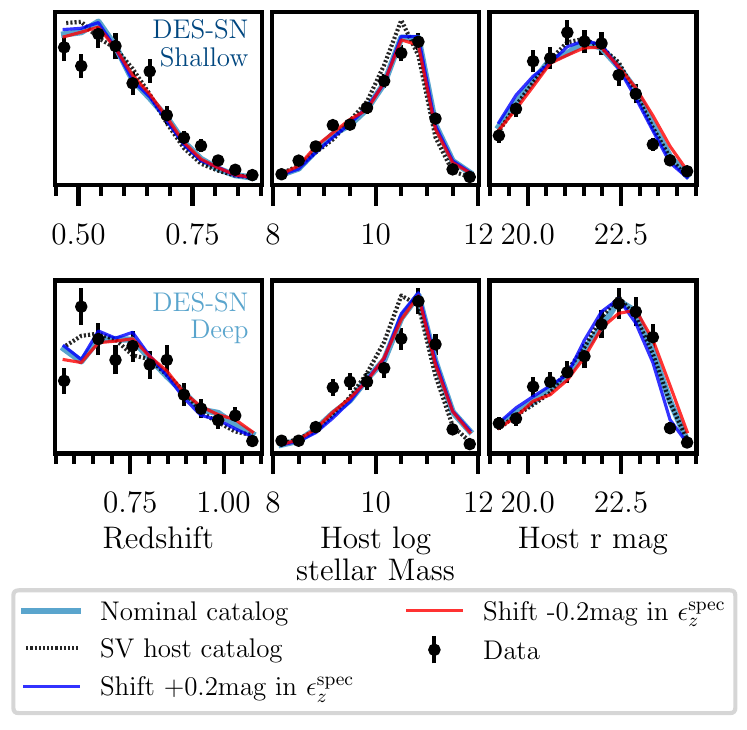}
    \caption{Comparison between observed and simulated distributions of SN redshifts (zoomed at higher redshifts, left panel), host galaxy mass (center panel) and host galaxy $r$-band magnitudes (the Kron-like \texttt{MAG\char`_AUTO} magnitudes measured with \textsc{SExtractor} \citep{1996A&AS..117..393B} from the $r$-band deep coadds, right panel) for our nominal simulation and two variants: using the galaxy catalog from DES Science Verification images \citep[instead of catalog by][]{Qu_hostMismatch} and using a shifted spectroscopic redshift efficiency instead of the nominal efficiency by \citet{Vincenzi_2020}.}
    \label{fig:sim_deep}
\end{figure}
\begin{figure}\centering
    \includegraphics[width=0.8\linewidth]{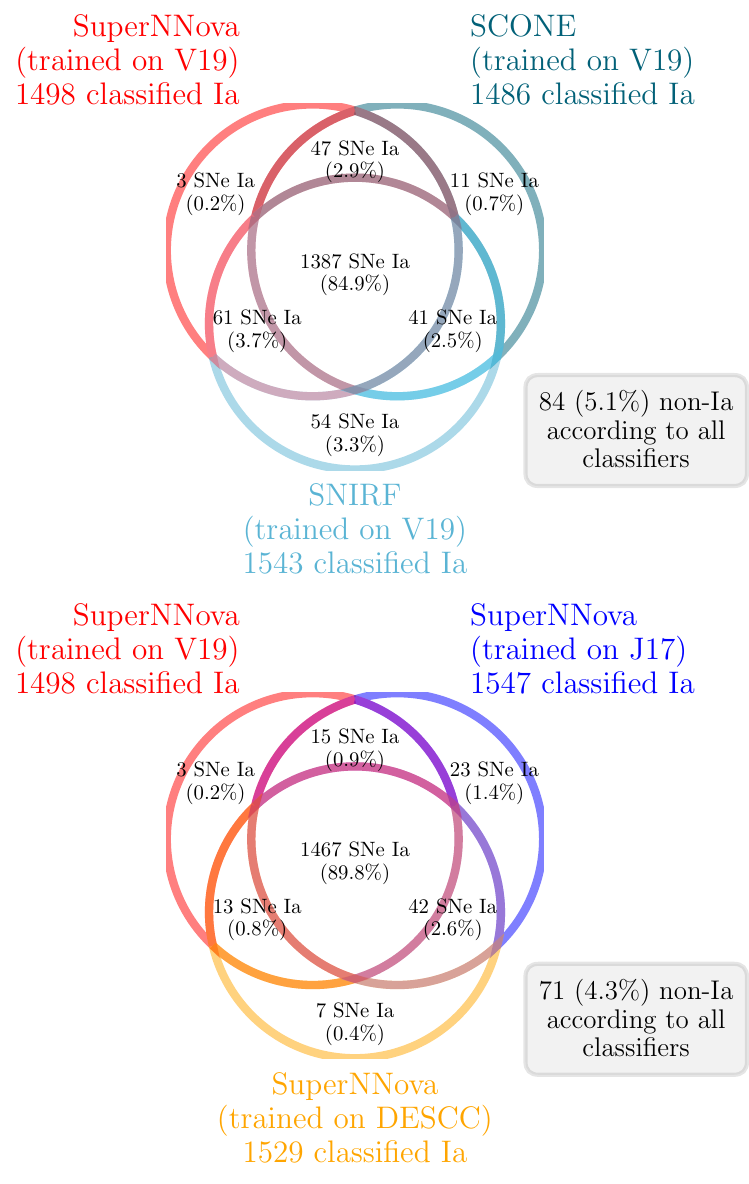}
    \caption{Likely DES SNe Ia (i.e., $P_{\mathrm{Ia}}>0.5$) according to the different classification models used in our analysis. \textit{Upper diagram}: we compare three classification tools: SuperNNova (red), SCONE (dark teal) and the \lq Supernova Identification with Random Forest\rq\ (SNIRF, teal). \textit{Lower diagram}: we compare three different SuperNNova models: one trained using core-collapse templates by \citet{Vincenzi_2019} (V19, red), one trained using core-collapse templates by \citet{Jones_2017_I} (J17, blue) and one trained using core-collapse templates from DES data (DES-CC, orange).}
    \label{fig:classification}
\end{figure}

\subsubsection{Galaxy catalog for SN host simulations}
In our analysis, we test two different galaxy catalogs for simulations of DES-SN hosts. The first is generated using deep DES coadded images and is described in Sec.~\ref{sec:sim_host} and presented in detail by \citet{Qu_hostMismatch}. The second is compiled using galaxies detected in the (shallower) DES SV images and is presented by \citet{DES_massstep}. The DES SV catalog includes  photo-$z$'s from template fitting techniques and galaxy properties measured using the galaxy SED fitting code by \citet{Sullivan_2006}. This second galaxy catalog is inferior in depth compared to our nominal catalog; however, we decide to include it in our systematic error budget because it has been used in many previous DES-SN studies \citep{DES_massstep, DES_biascor, Vincenzi_2020, Vincenzi_2021, DES_rates}. 
A comparison between simulations generated using the two galaxy catalogs is presented in Fig.~\ref{fig:sim_deep}.

\subsubsection{OzDES selection effects}
One of the most important selection effects in the DES SN sample is the requirement of a spectroscopic redshift from the SN host galaxy. The spectroscopic redshift efficiency for the DES SN sample is presented by \citet{Vincenzi_2020} and modelled as a function of the host galaxies' total brightness (\texttt{MAG$\_$AUTO} in Source Extractor).
We apply a shift of 0.2 mag to the efficiency curves presented by \citet{Vincenzi_2020} and include this in the systematic error budget.
A comparison between simulations generated using the alternative spectroscopic redshift efficiency is presented in Fig.~\ref{fig:sim_deep}.

\subsection{Contamination and photometric classification}
\label{syst:contamination}

To correct for core-collapse SN contamination, we test different photometric classification algorithms and different training methods. We include the different classification variants in the systematic error budget.

For the baseline analysis, we use the algorithm \snn\ by \citet{2020MNRAS.491.4277M}. \citet{Vincenzi_2021} present a detailed analysis of the training and performances of \snn\ in the context of the DES SN cosmological analysis.
For our baseline analysis, we train \snn\ using the simulations presented in Sec.~\ref{sec:simulations}. For our systematic analysis, we train \snn\ using two alternative and independent libraries of core-collapse SN templates: the one presented by \citet[][hereafter J17]{Jones_2017_I} and the one built from core-collapse SNe observed in DES (Hounsell et al. in prep., hereafter DES-CC).

As an alternative to \snn\, we consider two additional classification algorithms: the classifier SCONE by \citet{SCONE_Qu} and the \lq Supernova Identification with Random Forest\rq\ (SNIRF) algorithm.\footnote{\url{https://github.com/evevkovacs/ML-SN-Classifier}} 
We train SCONE and SNIRF on the same set of simulations used to train the \snn\ baseline model. We compare results from the different classifiers in Fig.~\ref{fig:classification} and discuss the results in Sec.~\ref{disc:contamination}. 

Finally, we test different approaches of modeling the contamination likelihood term in eq.~\ref{eq:likelihoods}. While the baseline approach uses simulations based on \citet{Vincenzi_2020}, we test the approach of using a polynomial fitting as in \citet{Hlozek_2012} \citep[see][for a detailed comparison of the two methods]{Vincenzi_2021}.

\begin{figure}\centering
    \includegraphics[width=\linewidth]{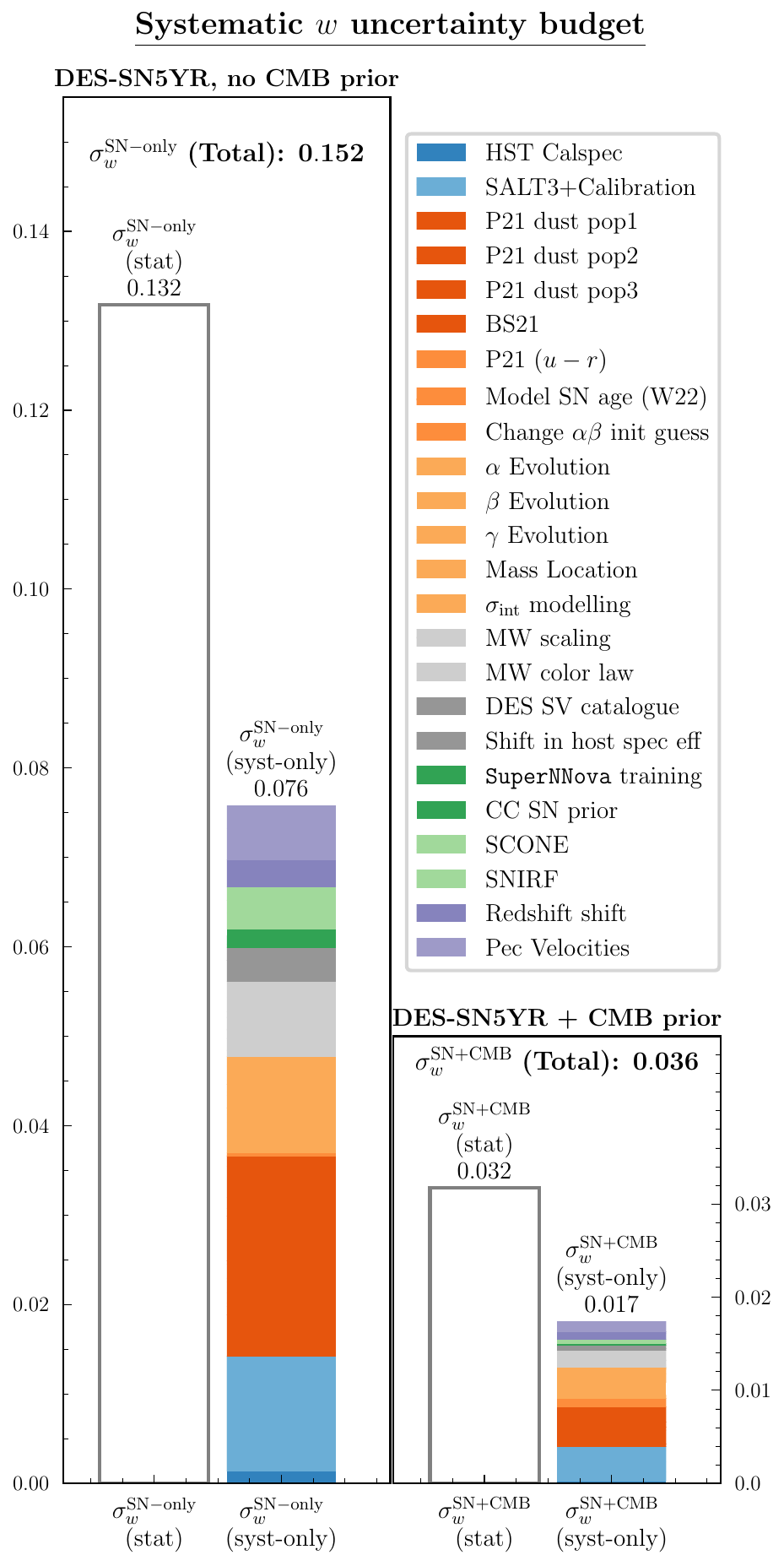}
    \caption{Systematic error budget on $w$ and comparison with statistical uncertainty on $w$. We present results both with and without including a CMB prior (left and right plots respectively, note the different y-axis scale in the two plots). The different sources of systematic uncertainty considered in this analysis are presented in Table~\ref{tab:syst_description} and \ref{tab:syst_size} and described in detail in Section~\ref{sec:syst_overview}.}
    \label{fig:error_budg}
\end{figure}

\begin{deluxetable}{lccc}
\tablecolumns{4}
\tablewidth{10pc}
\tablecaption{Size of Systematic uncertainty (SN-only, no CMB prior). A detailed description of the different sources of systematics and the labelling conventions are presented in Sec.~\ref{sec:syst_overview} and Table \ref{tab:syst_description}. }
\tablehead {
\colhead {Systematic}  &
\colhead {$\sigma_{w,\mathrm{syst}}$ $^*$}  &
\colhead {$\% \sigma_{\mathrm{tot}}$} &
\colhead {$\delta w_{\mathrm{syst}}^{\dag}$}} \vspace{2mm}
\startdata
\textbf{Total Stat+Syst} & 0.152 & 100 & $-$0.032 \\
\textbf{Total Statistical} & 0.132 & 87 & 0.000  \\
\textbf{Total Systematic ($\mathcal{C_{\mathrm{unbin}}}$)} & 0.076 & 50 & $-$0.032  \\
\hline \vspace{-2mm}\\
\textbf{Calibration $\&$ LC model} & 0.057 &15  & \\
\hspace{2mm} SALT3+Calibration & 0.052 & 34 & $-$0.036  \\
\hspace{2mm} HST Calspec & 0.006 & 4 & 0.002  \vspace{2mm}\\
\textbf{SN Ia astrophysics} & 0.133 &35  & \\
\hspace{2mm} P21 dust pop 1 &  0.019 & 12 & $-$0.010  \\
\hspace{2mm} P21 dust pop 2 &  0.024 & 16 & 0.003  \\
\hspace{2mm} P21 dust pop 3 &  0.020 & 13 & $-$0.004  \\
\hspace{2mm} P21($u-r$) &  0.000 & 0 & 0.048  \\
\hspace{2mm} Dust model as in \citetalias{BS20}  &  0.027 & 18 & $-$0.006  \\
\hspace{2mm} Model SN age (Sec. \ref{syst:W22}) &  0.000 & 0 & 0.048  \\
\hspace{2mm} Change $\alpha \beta$ initial estimate &  0.002 & 1 & 0.000  \\
\hspace{2mm} $\alpha$ Evolution &  0.020 & 13 & $-$0.008  \\
\hspace{2mm} $\beta$ Evolution &  0.000 & 0 & $-$0.007  \\
\hspace{2mm} $\gamma$ Evolution &  0.011 & 7 & $-$0.001 \\
\hspace{2mm} Mass step location &  0.000 & 0 & $-$0.002 \\
\hspace{2mm} $\sigma_{\mathrm{int}}$ modeling &  0.013 & 8 & $-$0.002  \vspace{2mm}\\
\textbf{Milky Way extinction} & 0.034 &9 & \\
\hspace{2mm} MW 5$\%$ scaling  & 0.020 & 13 & $-$0.011  \\
\hspace{2mm} MW colour law CCM  & 0.014 & 9 & $-$0.003  \vspace{2mm}\\
\textbf{Survey modeling } & 0.015 &4  & \\
\hspace{2mm} DES SV catalog & 0.009 & 6 & 0.002  \\
\hspace{2mm} Shift $\epsilon_{z}^{\mathrm{spec}}$ & 0.005 & 4 & 0.002 \vspace{2mm}\\
\textbf{Contamination} & 0.028 &7  & \\
\hspace{2mm} Classifier SCONE  & 0.006 & 4 & $-$0.000  \\
\hspace{2mm} Classifier SNIRF & 0.013 & 9 & $-$0.003  \\
\hspace{2mm} SuperNNova different training  & 0.006 & 4 & $-$0.000  \\
\hspace{2mm} Core-collapse SN prior   & 0.003 & 2 & $-$0.000  \vspace{2mm}\\
\textbf{Redshift} & 0.037 &10 & \\
\hspace{2mm} Redshift shift & 0.012 & 8 & 0.002  \\
\hspace{2mm} Peculiar velocities & 0.025 & 16 & $-$0.012\\
\enddata
\tablenotetext{$\dag$}{Shift in $w$ when including ONLY this systematic;}
\tablenotetext{$*$}{The quadrature sum of systematic uncertainties is larger than the total systematic uncertainty. Internal correlations in the sample cause the effects of some systematics to partially cancel out when considering the full covariance matrix.}
\label{tab:syst_size}
\end{deluxetable}

\subsection{Redshift and Peculiar velocity corrections}
\label{syst:redshift_vpec}
All SN redshifts are corrected for peculiar velocities and converted to CMB frame.
For the nominal analysis, we measure peculiar velocity corrections from 2M++. For systematics, we test two alternative approaches for the correction of peculiar velocities, both discussed in \citet{2022ApJ...938..112P}. The first approach uses the 2M++ corrections integrating over the line-of-sight relation between distance and the measured redshift. The second approach is to use the 2MRS \citep{2021MNRAS.507.1557L} peculiar velocity map. The two approaches both have $W_S=0.7$ (see Eq.~\ref{eq:cov_matrix}) so that their sum in quadrature results in an effective contribution of 1. The details of how peculiar velocity uncertainties are incorporated into the systematic covariance matrix $\mathcal{C}_{\rm syst}$ are presented in \citet[][, sec. 3.1.3]{PantheonP_cosmo}.

Additionally, we account for potential biases due to a local void or other systematic redshift error and apply a systematic redshift shift of $4 \times 10^{-5}$ \citep{CalcinoDavis}.

\begin{figure}[ht!]
\centering
    \includegraphics[width=\linewidth]{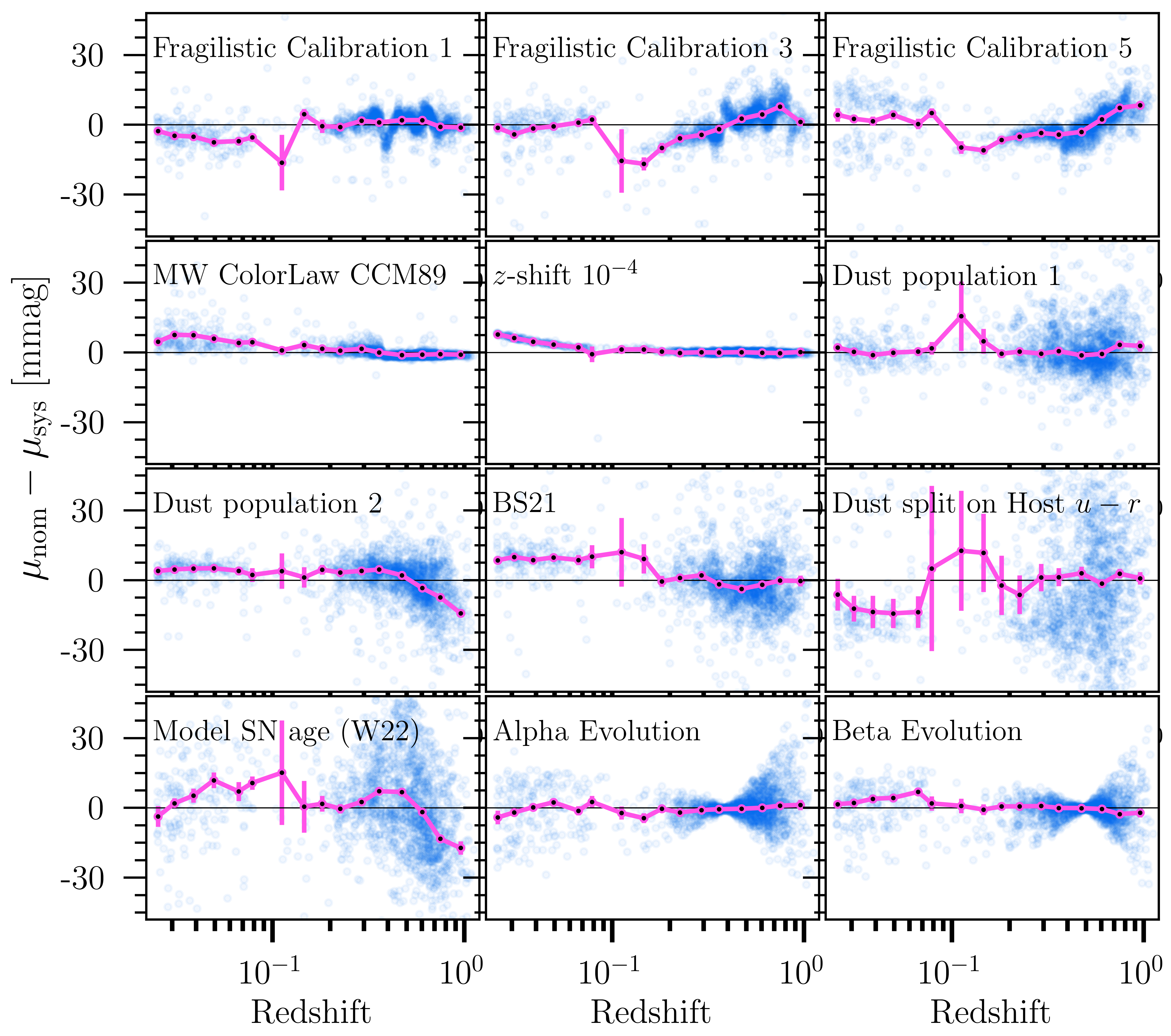}
    \includegraphics[width=\linewidth]{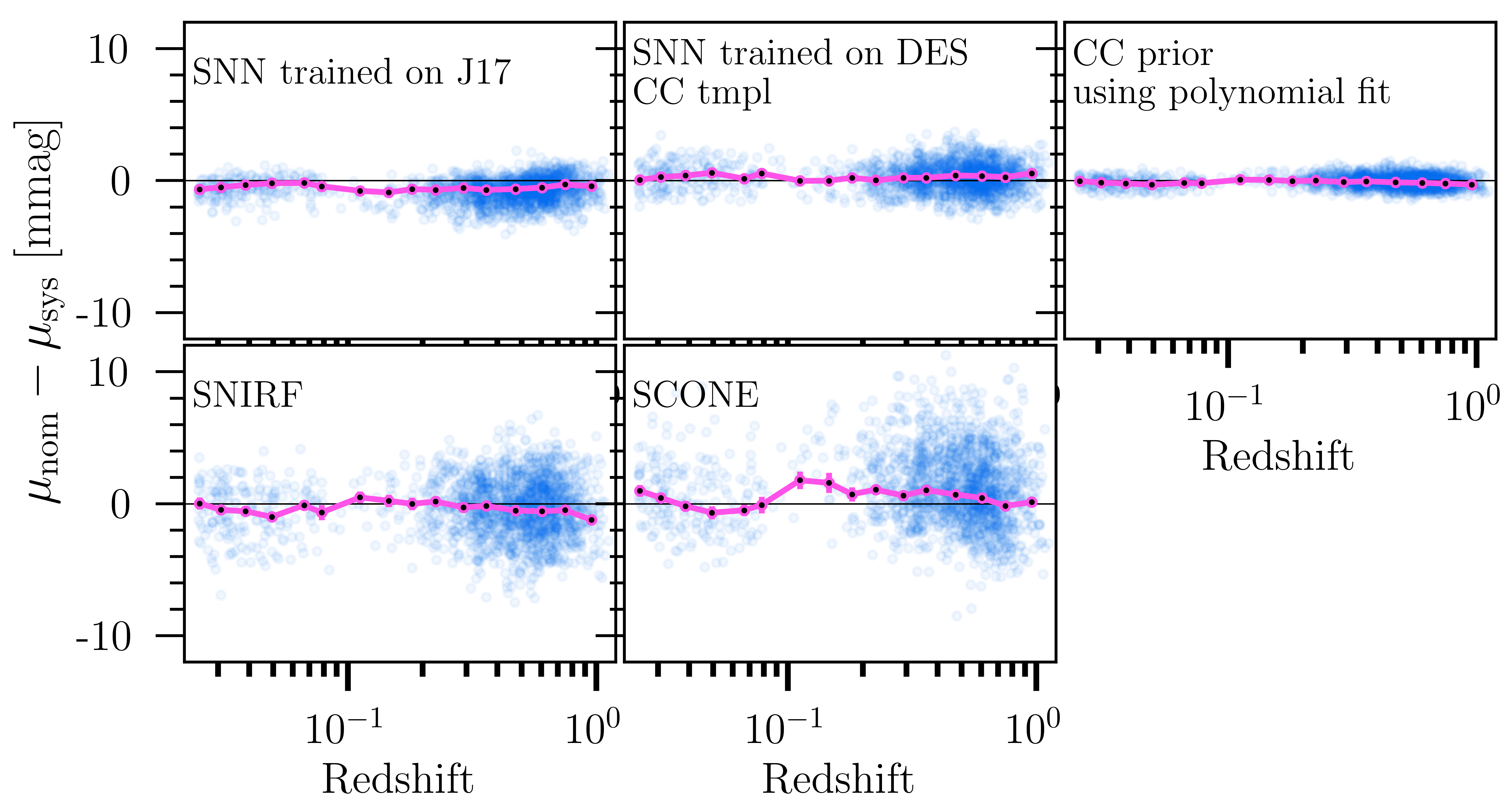}
    \caption{ Effects of different sources of systematic uncertainties on SN distances. \textit{Upper panel:} Systematic uncertainties presented in Table~\ref{tab:syst_description}, excluding systematics related to contamination and classification. \textit{Lower panel:} Contamination-related systematics. We note that the range of the y-axis is a fifth compared to the upper panel.}
    \label{fig:SYST}
    \end{figure}

\begin{deluxetable*}{@{\extracolsep{3pt}}lccccccc}
\tablecolumns{8}
\tablewidth{10pc}
\tablecaption{BBC-fitted nuisance parameters, and cosmology-fitted $\chi_S^2$, Hubble residuals RMS and shift in best fit $w$ for different analysis variants. The $\chi_S^2$ and RMS are measured after applying a cut of $P_{\mathrm{Ia}}>0.5$. In the $\chi_S^2$ calculations (see eq.~\ref{eq:chi2_S}), residuals are measured w.r.t. the best fit cosmology estimated from each analysis variant, and always using residual uncertainties from the Nominal analysis. A detailed description of the different sources of systematics and the labelling conventions are presented in Sec.~\ref{sec:syst_overview} and Table \ref{tab:syst_description}.}
\label{tab:intr_scatter_model}
\tablehead {
\colhead {}  &
\multicolumn{4}{c}{BBC-fit} & 
\multicolumn{3}{c}{Cosmology-fit (\footnotesize{SN-only, no CMB prior})}\\
\cline{2-5} \cline{6-8}
\colhead {Model}  &
\colhead {$\alpha$ $^{(c)}$}  &
\colhead {$\beta$ $^{(c)}$} & 
\colhead {$\gamma$ $^{(c)}$} & 
\colhead {$\sigma_{\rm gray}$ } & 
\colhead {$\Delta \chi_S^2$$^{(d)}$} & 
\colhead {RMS } & 
\colhead {$\Delta w_{\rm stat}$ $^{(e)}$}}
\startdata
& \multicolumn{7}{c}{\textbf{DES-SN+low-$z$ data}} \\
\textbf{Nominal$^{(a)}$} & 0.161(1) & 3.12(3) & 0.038(7) & 0.04 & - & 0.168 & - \\
BiasCor sim: \textbf{P21 dust pop 1} & 0.162(0) & 3.17(0) & 0.043(7) & 0.06 & $-$1 & 0.168 & $-$0.038\\
BiasCor sim: \textbf{P21 dust pop 2} & 0.160(3) & 3.06(3) & 0.033(8) & 0.04 & +5 & 0.169 & 0.079 \\
BiasCor sim: \textbf{P21 dust pop 3} & 0.161(0) & 3.09(0) & 0.040(8) & 0.05 & +18 & 0.17 & 0.039 \\
BiasCor sim: \textbf{Original \citetalias{BS20} param.} & 0.161(4) & 3.18(1) & 0.026(8) & 0.06 & +9 & 0.169 & $-$0.110 \\
BiasCor sim: \textbf{P21($u-r$) (fit $\gamma_{u-r}$)$^{(b)}$} & 0.158(0) & 3.11(0) & 0.033(7) & 0.06 & +16 & 0.169 & 0.066 \\
BiasCor sim: \textbf{model SN age \citepalias{W22_x1age}} & 0.148(3) & 3.06(3) & 0.017(8) & 0.05 & +1 & 0.168 & 0.113 \\
\textbf{Change $\alpha \beta$ init estimate} & 0.171(3) & 3.34(0) & 0.037(8) & 0.05 & +0 & 0.168 & 0.002 \\
\textbf{$\alpha$ evolution} & 0.149(0) & 3.12(0) & 0.036(8) & 0.04 & $-$5 & 0.168 & $-$0.007 \\
\textbf{$\beta$ evolution} & 0.161(3) & 2.99(0) & 0.037(8) & 0.04 & +0 & 0.168 & $-$0.009 \\
\textbf{$\gamma$ evolution} & 0.161(0) & 3.12(0) & 0.046(14) & 0.04 & +0 & 0.168 & $-$0.006 \\
\markgrey{BiasCor sim: \textbf{no $x_1$-$M_{\star}$ correlations}} & 0.152(2) & 3.07(2) & 0.019(8) & 0.04 & +6 & 0.167 & 0.006 \\
\markgrey{BiasCor sim: \textbf{G10} (no dust model)} & 0.157(6) & 3.20(6) & 0.055(9) & 0.11 & +102 & 0.172 & $-$0.120 \vspace{2mm}\\
\hline
& \multicolumn{7}{c}{\textbf{DES-SN+low-$z$ simulations} (average of 25 independent simulations) $^{(f)}$}\\
\textbf{Nominal$^{(a)}$} & 0.144(2) & 2.83(4) & 0.002(8) & 0.00 & - & 0.160(4) & - \\
BiasCor sim: \textbf{P21 dust pop 1} & 0.146(3) & 2.90(5) & 0.006(7) & 0.00 & $-$1\textbf{(4)} & 0.161(4) & 0.008 \\
BiasCor sim: \textbf{P21 dust pop 2} & 0.143(2) & 2.77(4) & $-$0.002(8) & 0.00 & 5\textbf{(4)} & 0.161(4) & 0.015 \\
BiasCor sim: \textbf{P21 dust pop 3} & 0.146(3) & 2.80(4) & 0.005(8) & 0.00 & 0\textbf{(8)} & 0.161(4) & 0.113 \\
BiasCor sim: \textbf{Original \citetalias{BS20} param.}  & 0.146(2) & 2.89(5) & $-$0.009(8) & 0.00 & 1\textbf{(5)} & 0.161(4) & $-$0.042 \\
BiasCor sim: \textbf{P21($u-r$) (fit $\gamma_{u-r}$)$^{(b)}$} & 0.141(3) & 2.87(4) & $-$0.020(8) & 0.00 & +35\textbf{(11) }& 0.163(4) & 0.051 \\
BiasCor sim: \textbf{model SN age \citepalias{W22_x1age}}  & 0.137(3) & 2.79(4) & $-$0.009(8) & 0.00 & 19\textbf{(12)} & 0.162(4) & 0.110 \\
\textbf{Change $\alpha \beta$ init estimate} & 0.154(2) & 3.05(4) & 0.002(7) & 0.00 & 1\textbf{(0)} & 0.160(4) & 0.004 \\
\textbf{$\alpha$ evolution} & 0.148(6) & 2.83(4) & 0.002(8) & 0.00 & $-$1\textbf{(2)} & 0.160(4) & 0.006 \\
\textbf{$\beta$ evolution }& 0.144(2) & 2.88(7) & 0.002(8) & 0.00 & $-$2\textbf{(2)} & 0.160(4) & 0.007 \\
\textbf{$\gamma$ evolution }& 0.144(2) & 2.83(4) & $-$0.003(15) & 0.00 & $-$1\textbf{(2)} & 0.160(4) & 0.005 \\
\markgrey{BiasCor sim: \textbf{no $x_1$-$M_{\star}$ correlations}}  & 0.135(3) & 2.77(4) & $-$0.011(7) & 0.00(0) & +7\textbf{(6)} & 0.150(5) & 0.083 \\
\markgrey{BiasCor sim: \textbf{G10} (no dust model)} & 0.148(4) & 2.87(9) & 0.030(8) & 0.07(0) & +26\textbf{(10)} & 0.151(5) & $-$0.069 \vspace{2mm}\\
\enddata
\tablenotetext{$(a)$}{\hspace{7pt}Nominal is P21($M_{\star}$) and the step is measured splitting on host galaxy stellar mass ($\gamma_{M_{\star}}$).}
\tablenotetext{$(b)$}{\hspace{7pt}For the alternative model P21($u-r$), the dust is modelled for blue/red galaxies and the step is defined and measured as a \lq color\rq\ step, $\gamma_{u-r}$(i.e., difference in brightness between SNe find in blue/red galaxies).}
\tablenotetext{$(c)$}{\hspace{7pt}For data, reported uncertainties are the uncertainties from the BBC fit. For simulations, uncertainties are estimated as the standard deviations from 25 simulations.}
\tablenotetext{$(d)$}{\hspace{7pt}This is the increase (or decrease) in $\chi_S^2$ (see eq.~\ref{eq:chi2_S}) compared to nominal (i.e., P21($M_{\star}$)).}
\tablenotetext{$(e)$}{\hspace{7pt} $w$-shift from different intrinsic scatter model in biasCor sim 
  ($C_{\rm syst}$ is not used), and without CMB prior in cosmology fit.} 
\tablenotetext{$(f)$}{\hspace{7pt}Reported uncertainties are estimated as the standard deviations from 25 simulations. The 25 simulations are all generated using the Nominal simulation approach.}
\end{deluxetable*}

\begin{figure}
\includegraphics[width=\linewidth]{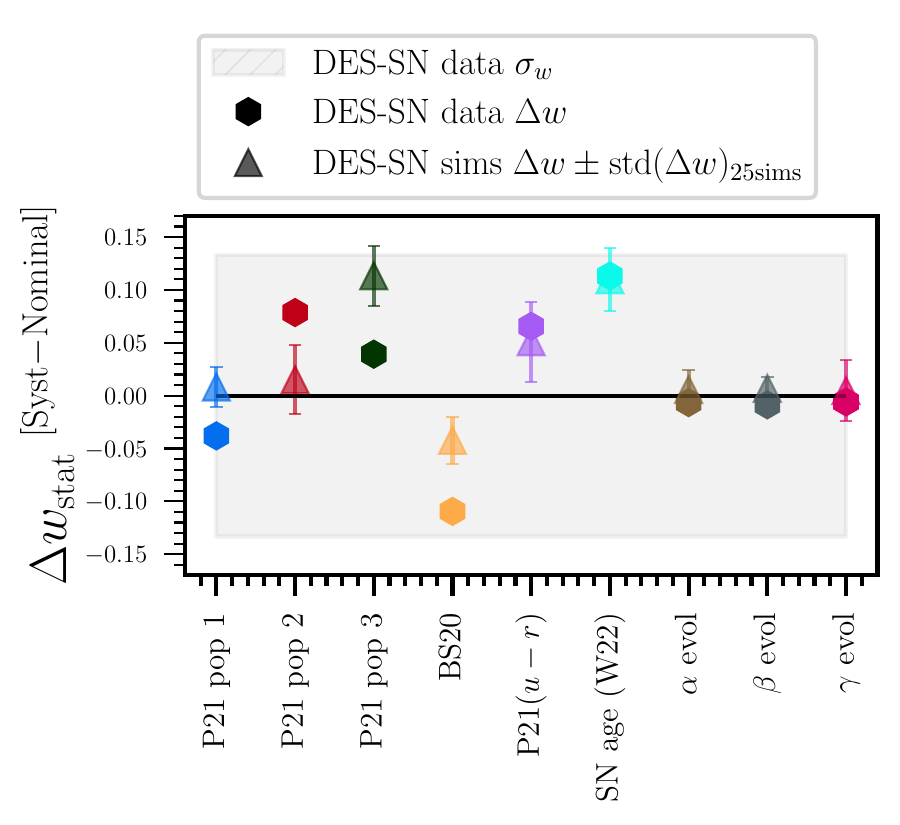}
\caption{Shifts in $w$ ($\Delta w$) for different intrinsic scatter models and $\alpha$, $\beta$, $\gamma$ evolution (see Sec. \ref{syst:intr_scatter}). $\Delta w$ measured from the data (hexagons) are compared to $\Delta w$ measured from averaging 25 DES+Low-$z$ simulations (triangles). For simulations, each errorbar is the standard deviation of the $\Delta w$ from the 25 simulated samples. Shifts are measured w.r.t. the nominal intrinsic scatter model., i.e. P21 measuring the mass step at $10^{10} M_{\odot}$ (first data point on the left). The statistical uncertainty on $w$ from the DES data is represented by the grey shaded area. }  
 \label{fig:scat_models}
 \end{figure}

\section{Discussion}
\label{sec:discussion}
\subsection{Systematic uncertainty budget on $w$}

The sources of systematic uncertainties described in Sec.~\ref{sec:syst_overview} are used to construct the covariance matrix $\mathcal{C}$ (Sec.~\ref{sec:cosmology_fitting}). The unbinned Hubble diagram and $\mathcal{C}$ are used in a cosmological fit to determine constraints on cosmological parameters $w$ and $\Omega_M$ using a flat $w$CDM model.
Here we present our sensitivity to cosmological parameters; the final unblinded results are presented in the DES key paper \citep{DES_final_cosmo}. The total uncertainty budget on $w$ is presented in Fig.~\ref{fig:error_budg} and Table \ref{tab:syst_size}.

In Fig.~\ref{fig:error_budg}, we compare statistical-only and systematic-only uncertainties on $w$. We present uncertainties both when measuring cosmological constraints from SN only (DES and low-$z$ external SNe), and when combining SNe with CMB (see inset in Fig.~\ref{fig:error_budg}). The contribution from systematic uncertainties only is evaluated as $\sqrt{\sigma_{w,\mathrm{tot}}^2-\sigma_{w,\mathrm{stat}}^2}$, where $\sigma_{w, \mathrm{tot}}$ is the total uncertainty on $w$ measured including the covariance matrix $\mathcal{C}_{\rm syst}$ described in Eq.~\ref{eq:cov_matrix}.

For a Flat$w$CDM model considering SNe alone, we find that statistical uncertainties on $w$ dominate at $\sigma_{w,\mathrm{stat}}^{w{\rm CDM}}=$\deswfwcdmstatonly\, which increases slightly to $\sigma_{w,\mathrm{stat+syst}}^{w{\rm CDM}}=$\deswfwcdm\ when including both systematic and statistical uncertainties.  When adding a CMB prior to the cosmological fit, we find $\sigma_{w,\mathrm{stat}}^{w{\rm CDM}}=$\desplanckwfwcdmstatonly\ and $\sigma_{w,\mathrm{stat+syst}}^{w{\rm CDM}}=$\desplanckwfwcdm. Both with and without the prior the statistical uncertainties dominate over the systematic ones. 
CMB measurements are highly complementary to SN measurements and have an orthogonal direction of degeneracy on the $\Omega_M-w$ plane for a Flat$w$CDM model. For this reason, combining SNe and CMB significantly reduces uncertainties on $w$, and also reduces the impact of systematics for those that primarily move the SN contours along the direction of SN degeneracy. 

To separately evaluate the contribution from each source of systematic
uncertainty, we evaluate the $w$-uncertainty using $C_{\mathrm{stat+syst}}$ with a single systematic and compare to the stat-only uncertainty. In Fig.~\ref{fig:error_budg}, we show each systematic uncertainty contribution with a separate color/pattern.

In Table \ref{tab:syst_size}, we summarize the size of systematic uncertainties visually presented in Fig.~\ref{fig:error_budg} (in the table, we only focus on results determined using SN only, without a CMB prior) and additionally, we present the observed shifts on best-fit $w$ when including statistical+systematic covariance matrix compared to statistical only.

For the purpose of interpreting Table \ref{tab:syst_size}, we make two important caveats.
First, the quadrature sum of all systematic uncertainties presented in Table \ref{tab:syst_size} is larger than the total systematic uncertainty on $w$. The difference is due to internal correlations in the sample that cause the effects of some systematics to partly cancel out when considering the full covariance matrix.\footnote{For the purpose of Fig.~\ref{fig:SYST}, every systematic contribution is rescaled so that their sum is the total systematic uncertainty, 0.076 (or 0.019 when including the CMB prior).}  For the same reason, the $\delta w$'s measured for each systematic separately do not necessarily sum to the overall $\delta w$. 

Second, the size of a single systematic, $\sigma_{w, \mathrm{syst}}$, and size of the related $\delta w$, are not necessarily correlated and some systematics might have a small impact on $\sigma_{w, \mathrm{syst}}$ but cause a large $\delta w$ (i.e., model SN age using \citetalias{W22_x1age} or using the P21($u-r$) intrinsic scatter model), or viceversa (i.e., including $\gamma$ evolution). This difference between $\sigma_{w, \mathrm{syst}}$ and $\delta w$ can arise if a systematic results in significantly less/more scatter in the Hubble residuals or in a significantly better/worse maximum likelihood. 

In order to understand why some sources of systematics have larger $\sigma_{w, \mathrm{syst}}$ and/or $\delta w$, it is useful to look at changes in SN distances and fitting $\chi^2$ when running the nominal analysis and the analysis run changing the systematic $S$. 
For a subsample of systematics, we present differences in SN distances in Fig.~\ref{fig:SYST}, while in Table~\ref{tab:intr_scatter_model} we report differences in fitting chi-squared, fitted nuisance parameters and cosmological best fits. We compute the fitting chi-squared for each systematic variant as 
\begin{equation}
\chi_{S}^2 = \sum_{i=1}^{N_{\mathrm{likely-Ia}}} \frac{(\mu^i_{\mathrm{obs, }S}-\mu_{\mathrm{best-fit, }S}(z_i))^2}{\sigma_{\mu,\mathrm{ Nominal}}^2},
\label{eq:chi2_S}
\end{equation}
where $\mu_{\mathrm{obs, }S}$ are SN distances determined when changing the systematic $S$, $\mu_{\mathrm{best-fit, }S}$ are the theoretical distances given the best fit cosmology when adopting systematic $S$ in the analysis, and $\sigma_{\mu,\mathrm{Nominal}}$ are the uncertainties on $\mu_{\mathrm{Ia}}$ determined for the Nominal analysis. As described in Sec.~\ref{sec:dist_uncertainties}, SN distance uncertainties are rescaled and inflated using the terms $f(z_i, c_i, M_{\star,i})$ and $\sigma_{\mathrm{floor}}(z_i, c_i, M_{\star,i})$ (see eq.~\ref{eq:sig1}). These terms are estimated from simulations and can vary from systematic to systematic. In the $\chi_{S}^2$ calculations, we fix the distance uncertainties $\sigma_{\mu,\mathrm{ Nominal}}$ for all systematics to ensure that changes in $\chi_{S}^2$ ($\Delta \chi_{S}^2$) are not driven by inflated/reduced uncertainties, but \textit{by an effective change in the modeling of Hubble residuals}. The $\chi_S^2$ are measured using \lq likely SNe Ia\rq, i.e., SNe with a $P_{\mathrm{Ia}}>0.5$.

To estimate the significance the observed changes in $\chi_{S}^2$ and best-fit $w$, we generate a set of 25 simulations of the DES and Low-$z$ SN samples and propagate the effects of the same systematics considered for the data (simulations are generated following on our Nominal modeling approach, i.e., P21$(M_{\star}$)). 
We perform a full analysis on each simulated data sample and in Table~\ref{tab:intr_scatter_model}, we report the mean $\Delta \chi_S^2$ and its standard deviation (highlighted in bold), BBC-fitted nuisance parameters and best fit $w$ determined from the simulations. Standard deviations measured from the 25 independent realizations of our SN sample provide a robust estimate of uncertainties on the observed shifts.

Our evaluations of $\Delta \chi_{S}^2$ (and relative uncertainties) enable us to quantify which models or analysis variants are favoured by the data: $\chi_{S}^2>\chi_{\mathrm{Nominal}}^2$ or $\Delta \chi_{S}^2>0$ suggest that the analysis variant introduced with the systematic $S$ provides a worse modeling of the Hubble residuals, 
 while $\Delta \chi_{S}^2<0$ suggests a better modeling of Hubble residuals intrinsic scatter. The more a systematic is favoured by the data (i.e., more negative $\Delta \chi_{S}^2$), the larger its impact on the final error budget will be (especially if it significantly changes the best-fit cosmology).
 For these reasons, the results presented in Table~\ref{tab:intr_scatter_model} and Fig.~\ref{fig:SYST} are useful to interpret the error budget presented in Table~\ref{tab:syst_size} and Fig.~\ref{fig:error_budg}. 
 In the next sections, we discuss in detail each sub-group of systematic uncertainties considered in the error budget. 

\subsubsection{Systematic: SN Ia intrinsic scatter}
\label{disc:intrinsic_scatt_models}
Uncertainties related to the modeling of SN intrinsic scatter are the largest source of systematic uncertainties in our analysis. 
The different intrinsic scatter models included in our analysis are all equally favoured by the data, i.e., have $\chi_S^2$ comparable to the nominal analysis. This is shown in Table~\ref{tab:intr_scatter_model}; the three P21 realizations (P21 population 1, P21 population 2 and P21 population 3), the \citetalias{BS20} model, the dust model implemented splitting on $u-r$ color (P21($u-r$)) and the SN age model by \citetalias{W22_x1age} all have $\Delta\chi_S^2$ between $0$ and $+15$. These fluctuations are expected considering the $\Delta\chi_S^2$ scatter over 25 simulations, see bottom section of Table~\ref{tab:intr_scatter_model}).
The only systematic that is significantly favoured by the data is $\alpha$ evolution ($\Delta\chi_S^2=-7 \pm 2$). We discuss the effects of nuisance parameter redshift evolution in Sec.~\ref{disc:nuisance_evolution}.
For comparison, we present in Table~\ref{tab:intr_scatter_model} the nuisance parameters and $\Delta\chi_S^2$ for the intrinsic scatter model by \citet{Guy_2010} and \citet{2013ApJ...764...48K} (historically referred to as \lq G10\rq\ model). This model is significantly disfavoured by the data ($\Delta\chi_S^2=102$) and therefore is not included in our systematic error budget. We make a similar test and reach the same conclusion when testing the model by \citet{C11} (historically referred to as \lq C11\rq\ model)

While the different intrinsic scatter models considered in Table~\ref{tab:intr_scatter_model} are almost equally favoured by the data, the relative best-fit cosmologies can change. In Fig.~\ref{fig:scat_models}, we present the shifts in the best fit $w$ determined for each variant in Table~\ref{tab:intr_scatter_model}. We present our results both for the data and for the simulations.
The size and direction of the $w$-shifts observed in the data are well reproduced by the simulations (which are all generated assuming our nominal intrinsic scatter model i.e., P21($M_{\star}$)). 

\subsubsection{Systematic: Residual mass/color steps}
The most interesting result in Table~\ref{tab:intr_scatter_model} is related to the recovered nuisance parameters. As already noted in Sec.~\ref{sec:nuisance}, the recovered mass step for the data is $\gamma_{M_{\star}}\sim0.039$ mag ($5\sigma$ significance) across most analysis variants, and regardless of what SN sub-sample is considered (DES-SN only or DES-SN + Low-$z$, see Table~\ref{tab:nuisance_baseline}).
When fixing $\beta=3.14$ and considering either only blue SNe Ia ($c<0$) or only reddened SNe Ia ($c>0$), we find $\gamma_{M_{\star}}=0.039\pm0.012$ and $\gamma_{M_{\star}}=0.028\pm0.014$, which shows no significant color dependence.

The mass step decreases when considering the \citetalias{BS20} model ($\gamma_{M_{\star}}=0.026$) and the modeling approach by \citetalias{W22_x1age} ($\gamma_{M_{\star}}=0.017$). In the first case, \citetalias{BS20} assume the intrinsic colour distribution to be bluer ($c_{\rm int}=-0.084$) and dust to have longer tails (i.e., larger $\tau_E$ values) compared to our nominal analysis model. 
In the second case, the reduced mass step is likely a consequence of the fact that the SN age modeling approach by \citetalias{W22_x1age} does not reproduce $x_1-M_{\star}$ correlations well (see Fig.~\ref{fig:age}). In the BBC approach, if the simulations used to determine bias corrections underestimate $x_1-M_{\star}$ correlations, the mass step can be partially absorbed into the the bias correction term $\mu_{\rm bias}$ \citep[this has been discussed and demonstrated by][]{DES_massstep, 2021ApJ...913...49P}. As a test, we generate a large simulation following the same approach used for our Nominal simulations, but removing $x_1-M_{\star}$ correlations. We use this alternative simulation to bias-correct our SN sample and measure the mass step. We recover a mass step of 0.019 mag, half of the mass step found in the nominal analysis (see Table~\ref{tab:intr_scatter_model}).

The color-step measured for the P21($u-r$) model is also non negligible ($\gamma_{u-r}=0.033 \pm 0.007$ mag, $\sim5\sigma$ significance) but slightly smaller than the mass-step, which suggests that $u-r$ colour is a better proxy to describe SN-host correlations (even though P21($u-r$) model is not significantly favoured by the data).

In simulations, we recover a mass step consistent with zero (average uncertainty on $\gamma_{M_{\star}}$ is less than 0.01) and lower values of fitted $\beta$. While the zero mass step is expected (by construction, simulations have only a dust-based mass step), it is interesting to note that the \lq effective\rq\ $\beta$ measured from the simulations and the data is different. This is unexpected since simulations are generated using a dust-based model specifically built to reproduce the intrinsic ($\beta_{\rm int}$) and extrinsic ($R_V$) color-luminosity corrections.  
The $4\sigma$ discrepancies in the $\beta$ and $\gamma$ recovered in data and simulations suggest that either \textit{(i)} dust-parameter errors from \lq Dust2Dust\rq\ fit are larger than the statistical uncertainties because of our incomplete understanding of SN Ia intrinsic properties and correlations, or \textit{(ii)} the mass step is not fully explained by dust.

Despite the discrepancies observed between data and simulations and our likely still incomplete modeling of SNe Ia intrinsic scatter, our overall modeling of bias corrections is significantly improved over previous models \citep[e.g., \lq G10\rq\ from][]{2013ApJ...764...48K}. The associated systematics from our current discrepancies are discussed in Sec.~\ref{disc:variants}.

\subsubsection{Systematic: Nuisance parameter evolution}
\label{disc:nuisance_evolution}

\begin{deluxetable}{lcc}
\tablecolumns{3}
\tablewidth{8pc}
\tablecaption{Nuisance parameter evolution}
\label{tab:nuisance_evolution}
\tablehead {\colhead {} & \colhead {$\alpha_0$} & \colhead {$\alpha_1$}}\vspace{2mm}
\startdata
DES-SN+low-$z$ &  0.146(6) & 0.033(14) \\
DES-SN only &  0.176(10) & $-$0.013(20) \\
DES-SN+low-$z$ (25 sims $^{\S}$) &   0.147(5) & 0.003(12) \vspace{2mm}\\
 &  $\beta_0$ & $\beta_1$ \\
\hline
DES-SN+low-$z$ &  3.05(0) & 0.17(6) \\
DES-SN only &  3.13(0) & 0.04(6) \\
DES+low-$z$ (25 sims $^{\S}$) &   2.88(8) & $-$0.03(14) \vspace{2mm}\\
 &  $\gamma_0$ & $\gamma_1$ \\
\hline
DES-SN+low-$z$ &   0.051(15) & $-$0.033(34) \\
DES-SN only &  0.068(24) & $-$0.045(49)\\
DES+low-$z$ (25 sims $^{\S}$) &  $-$0.006(12) & 0.010(34) \\
\enddata
\tablenotetext{$\S$}{Reported uncertainties are estimated as the standard deviations from 25 simulations.}
\end{deluxetable}

\begin{figure}[ht!]
\centering
    \includegraphics[width=\linewidth]{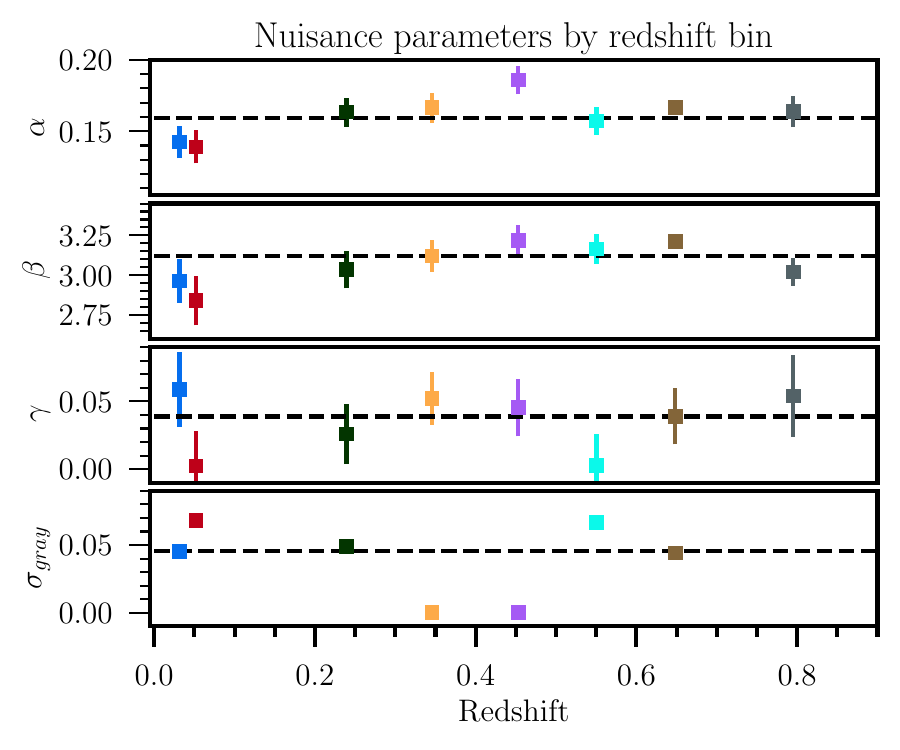}
    \caption{Values of $\alpha$, $\beta$, $\gamma$ and $\sigma_{\rm gray}$ for discrete redshift bins using the DES-SN + Low-$z$ samples. The Low-$z$ sample is divide into two redshift bins ($0.025<z<0.04$ and $0.04<z<0.1$, $\sim80$ SNe in each bin), while the DES-SN sample is divided into equally spaced bins of $\Delta z =0.1$. The baseline fit (assuming no redshift evolution of nuisance parameters) is shown in black for each panel. Fitted $\alpha$, $\beta$, $\gamma$ evolution are presented in Table~\ref{tab:nuisance_evolution}.}
    \label{fig:nuisance_param}
    \end{figure}
    
When considering the combined DES and Low-$z$ samples, we detect a significant ($> 2\sigma$) redshift evolution in the $\alpha$ and $\beta$ nuisance parameters. 
The best fit parameters for the redshift evolution of $\alpha$, $\beta$ and $\gamma$ are summarized in Table~\ref{tab:nuisance_evolution}. The fitted parameters suggest that $\alpha$ and $\beta$ increase with redshift ($\alpha<0.15$ and $\beta<3.05$ at lower redshifts and $\alpha>0.16$ and $\beta>3.2$ at higher redshifts).
The effects of redshift evolving $\alpha$ and $\beta$ on SN distances are presented in Fig.~\ref{fig:SYST}. For both $\alpha$ and $\beta$ evolution, there is no evidence of a net trend in the binned SN distance differences and, as a result, the differences between the best fit $w$ with and without $\alpha$ and $\beta$ evolution is consistent with zero (see \lq $\alpha$ evolution\rq\ and \lq $\beta$ evolution\rq\ in Table~\ref{tab:intr_scatter_model} and Fig.~\ref{fig:scat_models}).

While we do not detect significant $\gamma$ evolution in our sample, we conservatively allow for the possibility of a redshift evolution in $\gamma$ and include this as a source of systematic uncertainty. The contribution to the systematic error budget from nuisance parameter evolution ($\alpha$, $\beta$ and $\gamma$) is presented in Fig.~\ref{fig:error_budg} and Table~\ref{tab:syst_size}.

When measuring $\alpha$, $\beta$ and $\gamma$ evolution on the 25 simulations of DES-SN+Low-$z$, we successfully recover zero evolution for all nuisance parameters (see Table~\ref{tab:nuisance_evolution}), which indicates that evolution is not an artifact of our analysis.

Finally, we note that evolution of $\alpha$ and $\beta$ are negligible when considering the DES-SN sample only. This is particularly clear when looking at the values of $\alpha$, $\beta$ and  $\gamma$ evaluated in discrete redshift bins, as shown in Fig.~\ref{fig:nuisance_param}.
The parameters $\alpha$, $\beta$ and  $\gamma$ are consistent between redshift 0.2 and 1 (the redshift range covered by the DES-SN sample) and present more significant discrepancies at low redshift. This was also highlighted in Sec.~\ref{sec:nuisance}, where we presented the best-fit nuisance parameters for DES alone and low-$z$ alone, and note significant $\alpha$ discrepancies.

If the underlying cause of the observed evolution and nuisance parameters discrepancies between DES and low-$z$ (Table~\ref{tab:nuisance_baseline}) had an astrophysical origin (e.g., SN progenitor physics), it would be challenging to explain how this could produce such large fluctuations in nuisance parameters at $z\sim0.1$. 
A likely explanation of the observed evolution is an incomplete understanding of the properties and selection effects of the low-$z$ SN samples. The modeling of the low-$z$ samples is challenging for two reasons. First, the selection functions of the low-$z$ SN samples are poorly understood. Second, the small statistics in the low-$z$ samples makes \textit{(i)} dust modeling more uncertain, and \textit{(ii)} empirical modeling of $x_1-M_{\star}$ or $x_1-(u-r)$ correlations more difficult to model (see Fig.~\ref{fig:data_sim_compare_lowz} and Fig.~\ref{fig:mu_vs_c_des_lowz}).

\subsubsection{Systematic: Photometry and Calibration}
\label{disc:phot_calibration}
Historically, calibration uncertainties --- especially between low and high redshift samples --- have been the largest source of systematic uncertainties in SN analyses. In our analysis, systematic uncertainty associated to calibration is smaller than systematic uncertainties related to SN Ia intrinsic properties.

The reduced impact of calibration on our systematic error budget is due to various factors:
\begin{itemize}
    \item The FGCM method presented by \citet{2018AJ....155...41B} improve the accuracy of the DES internal calibration to $\sim5$ mmag in $griz$ bands;
    \item In our analysis, we combine samples from different surveys, with different filter systems and calibrations. The work presented by \citet{Supercal} and extended by \citet{Fragilistic} significantly improved cross-calibration between the different SN surveys used in this work. In particular, it improved the calibration of older low-$z$ SN samples;
    \item Calibration affects our ability to train light-curve fitting models like SALT3. For this analysis, we train the SALT3 model on a larger and more accurate training sample compared to the \citet{Betoule_2014} SALT2 model of previous SN cosmology analyses, which makes our SALT3 model less sensitive to calibration uncertainties. As shown in \citet{2023MNRAS.520.5209T}, scatter between SALT3 surfaces considered for systematic uncertainties is significantly reduced compared to previous SALT2 models;
    \item The approach introduced by \citet{Fragilistic} to propagate calibration uncertainties to cosmology has significantly reduced the impact of this source of systematics. Following \citet{Fragilistic}, we propagate calibration uncertainties \textit{simultaneously} on the SALT3 training and on the SN light-curves, thus accounting for calibration correlations for data used in both the training and cosmology analysis. \footnote{This approach was implemented in Joint-Lightcurve-Analysis \citep{Betoule_2014}, but it was not applied in subsequent SN cosmological analyses due to the unavailability of the SALT2 training code.}
\end{itemize}

\subsubsection{Systematic: Non-Ia Contamination}
\label{disc:contamination}
One of the most remarkable results of our analysis is that systematic uncertainties associated with contamination are only 9 per cent of the total systematic error budget and they bias SN distances by only a few mmag (see Fig.~\ref{fig:SYST}, lower panel). 

As discussed by \citet{Vincenzi_2021} and highlighted in Table~\ref{tab:selection}, even before photometric classifiers are applied, contamination is reduced to 9 per cent.
This reduction is due to fact that the SALT3 fitting and the BBC fitting serve as classifiers themselves:
\begin{enumerate}
\item The SALT3 model fitting (and associated quality cuts) already eliminate a significant fraction of non-Ia SNe (from $\sim 40$ per cent to $\sim11$ per cent, see Table~\ref{tab:selection});
\item The requirement of a valid bias correction further adds the constraint that the fitted SALT3 parameters of each SN are representative of what is found in the 3 dimensional parameter space populated by large SN Ia simulations used for bias corrections;
\item Chauvenet's criterion iteratively applied in BBC provides a cosmological model independent outlier rejection.
\end{enumerate}

Classifiers contribute to further reduce the weight  of contaminants in the cosmological fit. 
When testing classifiers on simulations, algorithms like \snn\ and SCONE can achieve levels of purity and efficiency $>98\%$ \citep[see results reported in ][]{2020MNRAS.491.4277M, SCONE_Qu, Vincenzi_2021}. In Fig.~\ref{fig:classification}, we compare classification of DES data using the different algorithms tested in our analysis, and we note some discrepancies. The three classification algorithms (\snn, SCONE and SNIRF, all trained on \citetalias{Vincenzi_2019}) have consistent classification on 90.0 per cent of DES-SNe (84.9 per cent are classified as Ia by all three classifiers and 5.1 per cent are classified as non-Ia by all three classifiers), while the three \snn\ models trained on the \citetalias{Vincenzi_2019}, \citetalias{Jones_2017_I} and \citetalias{DES-CC} templates have consistent classification for 94.2 per cent of SNe (89.8 per cent Ia and 4.3 per cent non-Ia). In particular, we note that the baseline classifier \snn\ (trained on \citetalias{Vincenzi_2019}) and SCONE are the more \lq conservative\rq\ classifiers (smaller number of SNe are classified as SNe Ia). 

When testing the classifiers on the sample of 207 spectroscopically classified SNe Ia from the DES-SN3YR analysis \citep[][]{DES_abbott, DES_spec} and on the sample of 43 spectroscopically classified non-Ia from the first three years of the DES survey \citepalias[see][]{DES-CC}, we find that \snn\ trained on \citetalias{Vincenzi_2019} templates have the highest accuracy (correctly classifying as Ia 203 out of 207 SNe Ia, and misidentify only 2 out of 43 non-Ia), \snn\ trained on \citetalias{Jones_2017_I} has the lowest purity (6 out of 43 non-Ia are misclassified as SNe Ia) and SNIRF has the lowest efficiency (192 out of 207 SNe classified as Ia). 
These results confirm that our baseline classifier (\snn) provides excellent results not only on simulations but also data.

In general, it is not surprising that classification accuracy is lower on real data rather than on simulations, as data often present defects and outliers that are not fully modelled in simulations \citep[for the DES data, these artefacts are significantly reduced by SMP, and this contributes to improve classification, see][]{DES5YR_SMP}. Despite the differences between the classification algorithms used in our analysis and despite the additional challenges that classification on real data presents, we find that SN distances measured from using different classification algorithms vary by 10 mmag at most (see Fig.~\ref{fig:SYST}) and differences in the estimated $w$ are not significant (see Table~\ref{tab:contamination_BEAMS}).

In order to further validate that contamination and classification are consistent between simulations and data, we consider an additional metric. The true number of contaminants in the DES data is unknown, however the BEAMS contaminants likelihoods, $\sum \mathcal{L}_{\rm CC}^i$(see eq.~\ref{eq:likelihoods}) provides an estimate of how many SNe are associated with the contaminants population. We determine the sample contamination as the fraction of the contaminants total likelihood divided by the total sample likelihood $\sum ( \mathrm{log} (\mathcal{L}_{\rm Ia}^i+\mathcal{L}_{\rm CC}^i$)). This fraction provides a measurement of the fraction of contaminants estimated in the Hubble diagram during the BBC cosmological fit. This quantity is also defined in \citet{KV_2022} as the sum of the \lq  BEAMS probabilities\rq\ \citep[see equation 9 in][]{KV_2022}.

In Table~\ref{tab:contamination_BEAMS}, we present the contaminants likelihood fractions estimated on data and simulations when implementing different classifiers. For every classification method, the contamination predicted in the simulations is consistent with the contamination observed in data, and it is $\sim6-7$ per cent. This is the first SN analysis that demonstrates such a close agreement between contamination estimated in the data and in simulations. The \textit{true} percentage of contaminants to total likelihood estimated from the simulations 5.3$\pm$0.4 and the simulated value of $w=-1$ is fully recovered by our pipeline.

\begin{deluxetable*}{@{\extracolsep{3pt}}lcccc}
\tablecolumns{5}
\tablewidth{12pc}
\tablecaption{Fraction of contaminants likelihood to total likelihood.}
\label{tab:contamination_BEAMS}
\tablehead{ &
\multicolumn{2}{c}{$\sum\mathcal{L}_{\rm CC}^i / \sum \mathcal{L}_{\rm tot}^i$ $^{*}$} & \multicolumn{2}{c}{$\Delta w_{\rm stat} ^{\ddag}$}\\
\cline{2-3} \cline{4-5}
Classification systematic & data & 25 sims$^{\dag}$ & data & 25 sims$^{\dag}$ } \vspace{1mm}
\startdata
\textbf{Nominal} \citep[SuperNNova trained on ][]{Vincenzi_2019} & 0.065 & 0.066 $\pm$ 0.008 & 0.000 & 0.000 $\pm$ 0.024 \\
SuperNNova trained on J17 \citep{Jones_2017_I} & 0.074 & 0.067 $\pm$ 0.009 & $-$0.005 & 0.016 $\pm$ 0.024 \\
SuperNNova trained on DESCC \citep{DES-CC}  & 0.069 & 0.066 $\pm$ 0.009 & 0.015 & 0.013 $\pm$ 0.024 \\
SNIRF classifier& 0.059 & 0.063 $\pm$ 0.009 & 0.029 & 0.037 $\pm$ 0.022 \\
Replace sim core-collapse SN prior with fitted polynomial prior $\S$ & 0.069 & 0.068 $\pm$ 0.011 & 0.002 & 0.003 $\pm$ 0.025 \\
\enddata
\tablenotetext{$*$}{$\sum \mathcal{L}_{\rm tot}^i$ is the total likelihood, $\sum (\mathcal{L}_{\rm Ia}^i + \mathcal{L}_{\rm CC}^i)$ described in eq.~\ref{eq:SUM}. See Eq.~\ref{eq:likelihoods} and \ref{eq:likelihoodsIa} in Sec.~\ref{sec:BEAMS} for the definition of the SN Ia and contaminants likelihood terms.}
\tablenotetext{$\dag$}{Mean and standard deviation of the fraction of contaminants to total likelihood \textit{measured} over 25 simulations. The \textit{true} fraction of contaminants to total likelihood is 0.053$\pm$0.004 per cent.}
\tablenotetext{$\ddag$}{Shifts in $w$ when considering different classification methods (no systematic covariance matrix is used to measure these shifts). 
$w$ is constrained using SN only data (no additional CMB prior).}
\tablenotetext{$\S$}{See \citet{Hlozek_2012}.}
\end{deluxetable*}

\subsection{Validation of the BBC fitting approach and final cosmological contours}
In SN cosmological analyses, one of the most critical aspects is to model and correct for sample selection biases. For the majority of cases, analytical modeling of selection effects in SN experiments is an intractable problem. For this reason, we have to rely on complex simulations like the ones described in Sec.~\ref{sec:simulations}. An important limiting factor of using simulations is that they require assumptions of the input cosmology (as it is computationally prohibitive to generate a new simulation for every step of the cosmological fit).
For this reason, it is important to \textit{(i)} quantify the size of the biases on $w$ when assuming the wrong input cosmology in the bias correction simulations, and \textit{(ii)} validate the Bayesian cosmological contours determined from SN distances bias-corrected assuming a specific input cosmology.

The first aspect has been discussed in section 6.1 in \citet{Kessler_2017} and further tested by \citet{Camilleri}.
\citet{Kessler_2017} use a set of simulated SN samples and demonstrate that the effects on $w$  of assuming the wrong input cosmology in the bias correction simulations is small (one seventh of the statistical uncertainties) when including a strong $\Omega_M$ prior.

\citet{Camilleri} reproduce an analogous test without including the $\Omega_M$ prior and show that the direction of the bias is always \textit{along} the SN contour degeneracy (hence the small biases when including a $\Omega_M$ prior, which is orthogonal to the SN contour degeneracy). 

The validation of the Bayesian cosmological contours is addressed by  \citet{Validation_PA}. \citet{Validation_PA} validate the Bayesian cosmological contours produced using BBC Hubble diagram by generating 150 realizations of the DES-SN sample and making use of approximate Neyman confidence intervals.
This work demonstrates that the size of the cosmological contours produced with our pipeline and the Neyman confidence intervals agree at the $>95\%$ level.  

\subsection{Analysis variants and their impact on $w$}
\label{disc:variants}
In this section, we present different analysis variants and their impact on $w$. These analysis variants aim to answer different questions: \textit{(a)} Do our cosmological results vary significantly when considering different groups of DES fields and measuring cosmology along different directions in the sky? \textit{(b)} Do our cosmological results vary significantly when considering sub-samples of SNe found in specific host environments? \textit{(c)} Do our cosmological results vary significantly when considering different SALT3 wavelength ranges? \textit{(d)} What is the impact on cosmological results of incorrect implementation of e.g., MW corrections? 
The analysis variants implemented to answer \textit{(a)} and \textit{(b)} and \textit{(c)} are not included in our systematic error budget because they effectively select only specific sub-sample of the data-set. The analysis variants implemented to answer \textit{(d)} are also not included in the main analysis because they are incorrect implementations of the cosmological analysis.

The tested analysis variants and the associated $w$ shifts are shown in Table \ref{tab:shifts_w}. We present not only the shifts in $w$ estimated from DES SNe (no external priors) and but also the standard deviation of the $w$ shifts measured from 25 simulations (for a robust estimation of the significance of the measured $w$ shifts). When considering different DES-SN sub-fields \textit{(a)} or host-dependent sub-samples \textit{(b)}, we do not see significant ($>2.5\sigma$) deviation from the baseline $w$. When we incorrectly implement the cosmological analysis \textit{(d)}, we note that not including bias corrections (i.e., using the BBC0D approach), ignoring Milky Way extinction corrections or modeling bias corrections for deep and shallow fields together would have potentially produced the most significant biases on $w$.

\begin{deluxetable}{lrr}
\tablecolumns{3}
\tablewidth{8pc}
\tablecaption{Miscellaneous $w$-shifts (no CMB prior included) when implementing analysis variants described in Sec.~\ref{disc:variants} and not included in the analysis error budget.}
\tablehead {
\colhead {Analysis variant}  &
\colhead {$\Delta w_{\mathrm{stat}}$ $^{*}$} &
\colhead {std$(\Delta w)$}\vspace{-2mm}\\
 &  & [25 sims]}
\startdata
\multicolumn{3}{l}{\textit{(a)} \textbf{Different sets of DES SN fields}}  \\
DES-SN(C3)+Low-$z$ & -0.249 & 0.126 \\
DES-SN(X3)+Low-$z$ & -0.196 & 0.180 \\
DES-SN(X1,X2)+Low-$z$ & -0.008 & 0.207 \\
DES-SN(C1,C2)+Low-$z$ & -0.535 & 0.239 \\
DES-SN(S1,S2)+Low-$z$ & 0.023 & 0.218 \\
DES-SN(E1,E2)+Low-$z$ & 0.016 & 0.219 \\
DES-SN(Deep)+Low-$z$ & -0.166 & 0.098 \\
DES-SN(Shallow)+Low-$z$ & -0.069 & 0.135\vspace{2mm}\\
\multicolumn{3}{l}{\textit{(b)} \textbf{Host prop sub-samples}}  \\
Only SNe in $M_{\star}>10^{10} M_{\odot}$ & -0.071 & 0.081 \\
Only SNe in $M_{\star}<10^{10} M_{\odot}$ & 0.138 & 0.213 \\
Only SNe in $u$-$r>$1 hosts & -0.000 & 0.041 \\
Only SNe in $u$-$r<$1 hosts & -0.220 & 0.149\vspace{2mm}\\
\multicolumn{3}{l}{\textit{(c)} \textbf{SALT3 wavelength coverage}}  \\
SALT3 using 3500-7000\AA $^{\dag}$ & -0.027 & 0.024 \\
SALT3 using 4000-8000\AA $^{\dag}$ & -0.105 & 0.128 \vspace{2mm}\\
\multicolumn{3}{l}{\textit{(d)} \textbf{Incorrect implementation}}  \\
No BBC $^{\ddag}$ & -1.024 & 0.116 \\
No BEAMS, only $P_{\rm Ia}>0.5$ cut $^{\S}$& -0.033 & 0.034 \\
No Milky Way Ext corrections & 0.185 & 0.035 \\
Biascor Deep/Shall together & -0.136 & 0.040 \\
Ignore SB anomaly in data & -0.012 & 0.000 \\
Force $\gamma=0$ in BBC fit & -0.113 & 0.007 \\
\enddata
\tablenotetext{$\dag$}{In the nominal analysis, the wavelength range used for the SALT3 model is 3500-8000\AA.}
\tablenotetext{$*$}{Shift in $w$ when implementing each analysis variance (no CMB prior included). $w$ is measured using SN only (DES and low-$z$) and without including any systematic covariance matrix (statistical uncertainties only).}
\tablenotetext{$\ddag$}{No BEAMS and no bias-corrections.}
\tablenotetext{$\S$}{We apply a $P_{\mathrm{Ia}}$-based cut and assume every SN is a type Ia. We apply bias corrections but we do not implement the full BEAMS approach and we do not incorporate probabilities in the BEAMS framework.}
\label{tab:shifts_w}
\end{deluxetable}

\subsection{Systematic budget on $w_0 w_a$ Figure of Merit}
\label{disc:wowa_error_budget}
\begin{figure}\centering
    \includegraphics[width=\linewidth]{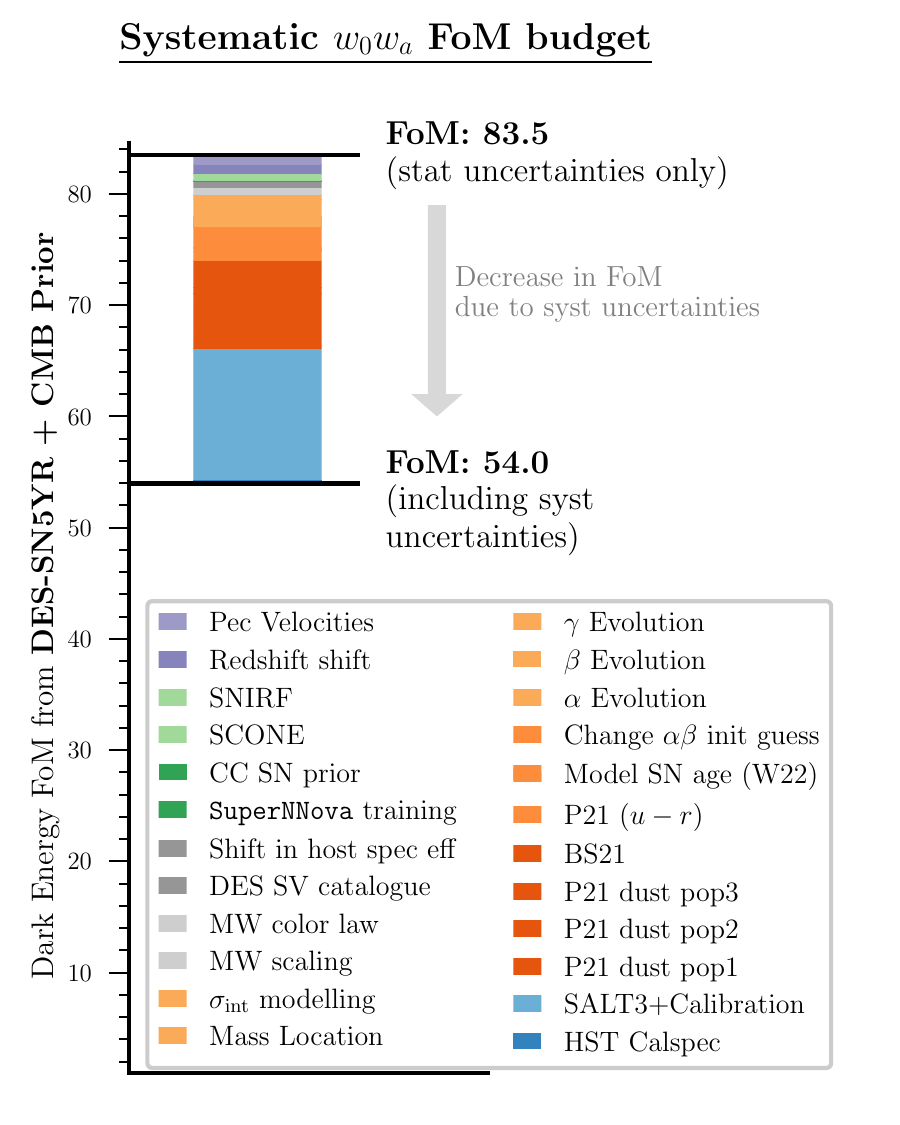}
    \caption{Decrease in the DES-SN5YR Figure of Merit (FoM) when including systematic uncertainties. Similarly to the systematic error budget presented in Fig.~\ref{fig:error_budg}, we highlight the sources of systematic uncertainties that degrade the FoM the most.}
    \label{fig:fom}
\end{figure}

The unbinned Hubble diagram and $\mathcal{C}$ can also be used to determine constraints on cosmological parameters $w_0$ and $\Omega_M$ using a flat $w_0 w_a$CDM model. The evaluation of present and forthcoming dark energy experiments involves assessing their capacity to enhance the Dark Energy Task Force Figure of Merit (DETF-FoM, Albrecht et al., 2006), which is determined as the reciprocal of the region enclosed within the $w_0 w_a$ contours. In this section, we present the $w_0 w_a$ constraining power of DES-SN5YR, before and after including systematic uncertainties. 
In Fig.~\ref{fig:fom}, we present the $w_0 w_a$ FoM with and without including systematic uncertainties (similar to the systematic $\sigma_w$ budget presented in Fig.~\ref{fig:error_budg}). We estimate the FoM including DES-SN5YR and CMB-like prior by \citet{collaboration2018planck}. We find $w_0^{\rm stat}=0.098$ and $w_a^{\rm stat}=0.49$ and $w_0^{\rm stat+syst}=0.12$ and $w_a^{\rm stat+syst}=0.59$. This corresponds to a $\sim$35 per cent decrease in FoM (from 83.5 when including statistical uncertainties only to 54 when including statistical and systematic uncertainties) and we find that calibration, light-curve modeling and modeling of dust properties are the dominating sources of systematics.

\section{Conclusions and future}
\label{sec:conclusions}
We present the DES-SN5YR cosmological analysis using the 5-year photometric SN sample for the DES-SN program. This new independent sample constitutes the largest and deepest SN sample from a single telescope to date and provides constraints on the dark energy EoS competitive with the best existing compilation of all high-$z$ SNe Ia previously discovered \citep{PantheonP_cosmo}. Additional DES-SN cosmological analyses using photo-$z$ information only (Chen et al. in prep.) and using additional DES SN spectroscopic data are currently ongoing.

In the following paragraphs, we summarise the main conclusions of the analysis presented in this paper, and outline key areas for future research, particularly in light of upcoming SN Ia surveys such as LSST and Roman.

Historically, the limiting systematic for SN analyses has been photometric calibration. We are entering a new phase in SN cosmology where intrinsic SN properties are becoming the dominating source of systematic (a third of the systematic $w$-uncertainty budget, and half of the $w_0 w_a$ FoM budget). For the future of SN cosmology, it will be crucial to significantly improve our understanding of SN intrinsic scatter and correlations with hosts. Alternatively, our way of doing SN Ia cosmology will likely be revised, for example, by selecting only specific sub-samples of SNe Ia \citep[e.g., only blue SNe Ia in low mass or blue environments as suggested by][]{2023MNRAS.519.3046K}.

For the modeling of SN Ia intrinsic scatter, the dust-driven model proposed by \citetalias{BS20} is the most successful currently available, however it requires further improvements. This is highlighted by the several discrepancies observed between the simulations and data. In our analysis, we build simulations using \lq Dust2Dust\rq\ \citepalias{P21_dust2dust}, a software designed to infer the dust properties of any SN Ia sample, operating under the assumption that the \citetalias{BS20} model accurately describes SN dust and its correlations with SN host properties.
When comparing simulations and data, we find interesting discrepancies in the inferred nuisance parameters (especially $\beta$ and $\gamma$). In particular, we find a non-negligible residual mass step, which might indicate the need of including additional astrophysics in the \citetalias{BS20} model.

Moreover, contamination has been one of the most important challenges for the DES-SN5YR analysis compared to the DES-SN3YR analysis. We demonstrated that contamination from non-Ia SNe is not the dominating source of systematic uncertainties for photometric SN samples. Our analysis is also the first SN cosmological analysis to present simulations that can accurately model and reproduce the amount of core-collapse contamination observed in the data (Table \ref{tab:contamination_BEAMS}).  Our results mark a transition point for future SN experiments such as LSST, for which spectroscopic classification of SNe will be extremely limited (see Frohmaier et al. in prep.).

Finally, in the next era of high-redshift SN experiments (e.g., Rubin Observatory's Legacy Survey of Space and Time, and Nancy Grace Roman Space Telescope), it will be essential to obtain low-$z$ SN samples with well defined selection functions and accurate calibration. Experiments like the Young Supernova Experiment \citep{2021ApJ...908..143J, 2023ApJS..266....9A} and Zwicky Transient Factory \citep[][Smith et al. in prep.]{2022MNRAS.510.2228D} and DEBASS (PI: Dillon Brout) will provide the next generation of low-$z$ samples and significantly improve cosmological constraints from SNe Ia.

\section*{Acknowledgements}

Funding for the DES Projects has been provided by the U.S. Department of Energy, the U.S. National Science Foundation, the Ministry of Science and Education of Spain, 
the Science and Technology Facilities Council of the United Kingdom, the Higher Education Funding Council for England, the National Center for Supercomputing  Applications at the University of Illinois at Urbana-Champaign, the Kavli Institute of Cosmological Physics at the University of Chicago, 
the Center for Cosmology and Astro-Particle Physics at the Ohio State University, the Mitchell Institute for Fundamental Physics and Astronomy at Texas A\&M University, Financiadora de Estudos e Projetos, 
Funda{\c c}{\~a}o Carlos Chagas Filho de Amparo {\`a} Pesquisa do Estado do Rio de Janeiro, Conselho Nacional de Desenvolvimento Cient{\'i}fico e Tecnol{\'o}gico and 
the Minist{\'e}rio da Ci{\^e}ncia, Tecnologia e Inova{\c c}{\~a}o, the Deutsche Forschungsgemeinschaft and the Collaborating Institutions in the Dark Energy Survey. 

M.V.\ was partly supported by NASA through the NASA Hubble Fellowship grant HST-HF2-51546.001-A awarded by the Space Telescope Science Institute, which is operated by the Association of Universities for Research in Astronomy, Incorporated, under NASA contract NAS5-26555. R.K. is supported by DOE grant DE-SC0009924.  T.M.D.\ is the recipient of an Australian Research Council Laureate Fellowship (FL180100168) funded by the Australian Government. 
L.K.\ thanks the UKRI Future Leaders Fellowship for support through the grant MR/T01881X/1. A.M.\ is supported by the ARC Discovery Early Career Researcher Award (DECRA) project number DE230100055. L.G.\ acknowledges financial support from the Spanish Ministerio de Ciencia e Innovaci\'on (MCIN), the Agencia Estatal de Investigaci\'on (AEI) 10.13039/501100011033, and the European Social Fund (ESF) "Investing in your future" under the 2019 Ram\'on y Cajal program RYC2019-027683-I and the PID2020-115253GA-I00 HOSTFLOWS project, from Centro Superior de Investigaciones Cient\'ificas (CSIC) under the PIE project 20215AT016, and the program Unidad de Excelencia Mar\'ia de Maeztu CEX2020-001058-M, and from the Departament de Recerca i Universitats de la Generalitat de Catalunya through the 2021-SGR-01270 grant. We acknowledge the University of Chicago’s Research Computing Center for their support of this work.

The Collaborating Institutions are Argonne National Laboratory, the University of California at Santa Cruz, the University of Cambridge, Centro de Investigaciones Energ{\'e}ticas, 
Medioambientales y Tecnol{\'o}gicas-Madrid, the University of Chicago, University College London, the DES-Brazil Consortium, the University of Edinburgh, 
the Eidgen{\"o}ssische Technische Hochschule (ETH) Z{\"u}rich, 
Fermi National Accelerator Laboratory, the University of Illinois at Urbana-Champaign, the Institut de Ci{\`e}ncies de l'Espai (IEEC/CSIC), 
the Institut de F{\'i}sica d'Altes Energies, Lawrence Berkeley National Laboratory, the Ludwig-Maximilians Universit{\"a}t M{\"u}nchen and the associated Excellence Cluster Universe, 
the University of Michigan, NSF's NOIRLab, the University of Nottingham, The Ohio State University, the University of Pennsylvania, the University of Portsmouth, 
SLAC National Accelerator Laboratory, Stanford University, the University of Sussex, Texas A\&M University, and the OzDES Membership Consortium.

Based in part on observations at Cerro Tololo Inter-American Observatory at NSF's NOIRLab (NOIRLab Prop. ID 2012B-0001; PI: J. Frieman), which is managed by the Association of Universities for Research in Astronomy (AURA) under a cooperative agreement with the National Science Foundation. Based in part on data acquired at the Anglo-Australian Telescope. We acknowledge the traditional custodians of the land on which the AAT stands, the Gamilaraay people, and pay our respects to elders past and present.

The DES data management system is supported by the National Science Foundation under Grant Numbers AST-1138766 and AST-1536171.
The DES participants from Spanish institutions are partially supported by MICINN under grants ESP2017-89838, PGC2018-094773, PGC2018-102021, SEV-2016-0588, SEV-2016-0597, and MDM-2015-0509, some of which include ERDF funds from the European Union. IFAE is partially funded by the CERCA program of the Generalitat de Catalunya.
Research leading to these results has received funding from the European Research
Council under the European Union's Seventh Framework Program (FP7/2007-2013) including ERC grant agreements 240672, 291329, and 306478.
We  acknowledge support from the Brazilian Instituto Nacional de Ci\^encia
e Tecnologia (INCT) do e-Universo (CNPq grant 465376/2014-2).

This manuscript has been authored by Fermi Research Alliance, LLC under Contract No. DE-AC02-07CH11359 with the U.S. Department of Energy, Office of Science, Office of High Energy Physics.

\newpage
\appendix
\input{Appendix_lowz}

\input{Appendix_TMINcut.tex}
\input{Affiliations}

\bibliography{bibliography}
\bibliographystyle{yahapj_twoauthor_amp}

\end{document}

%% file: Appendix_lowz.tex
\section{Modelling the external low-$z$ samples and DES-SN spectroscopic sample}
\label{AppendixLowz}

\begin{figure*}
    \includegraphics[width=\linewidth]{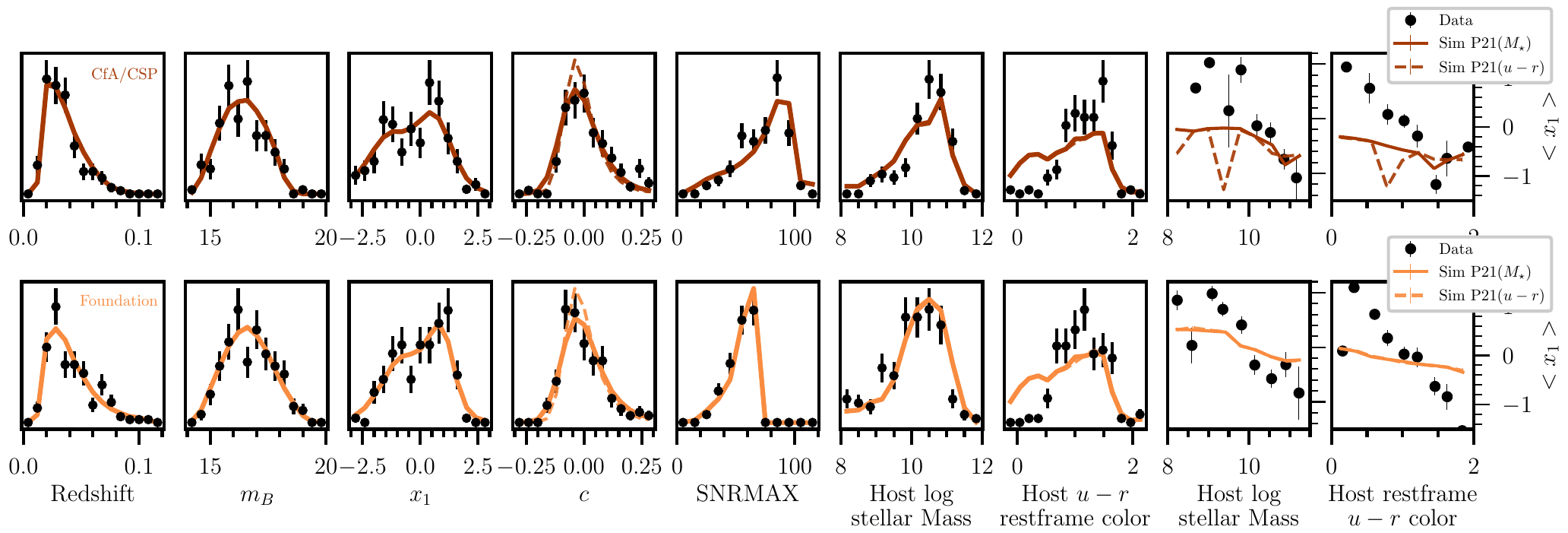}
    \caption{Same as Fig.~\ref{fig:data_sim_comparison_DES} but for the low redshift samples included in this analysis.}
   \label{fig:data_sim_compare_lowz}
\end{figure*}

In this Appendix, we focus on modelling of low-$z$ external samples (CfA, CSP and Foundation SN surveys). In Fig.~\ref{fig:data_sim_compare_lowz}, we present a comparison between observed and simulated distributions of various SN and host galaxy properties (e.g., SN stretch, SN colour, SN host stellar masses) and correlations between $x_1$ and host stellar mass and host rest-frame $u-r$ colour.

\begin{figure}\centering
    \includegraphics[width=0.49\linewidth]{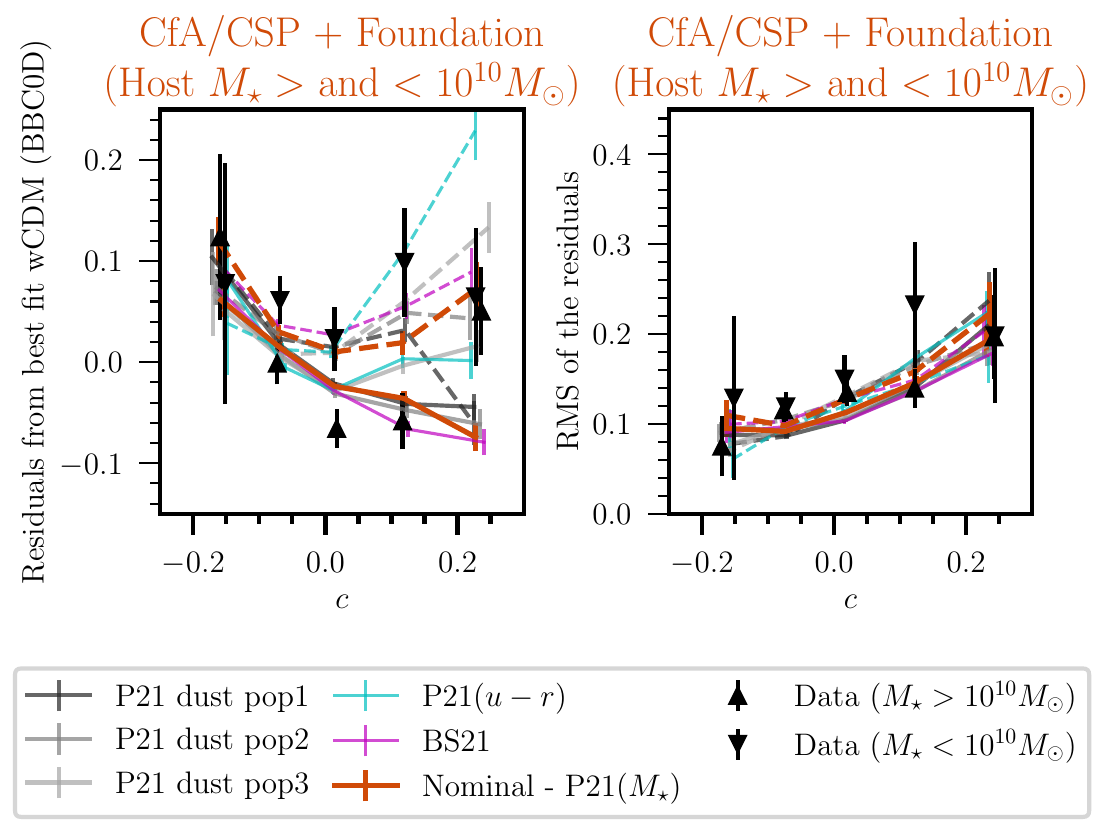}
\includegraphics[width=0.49\linewidth]{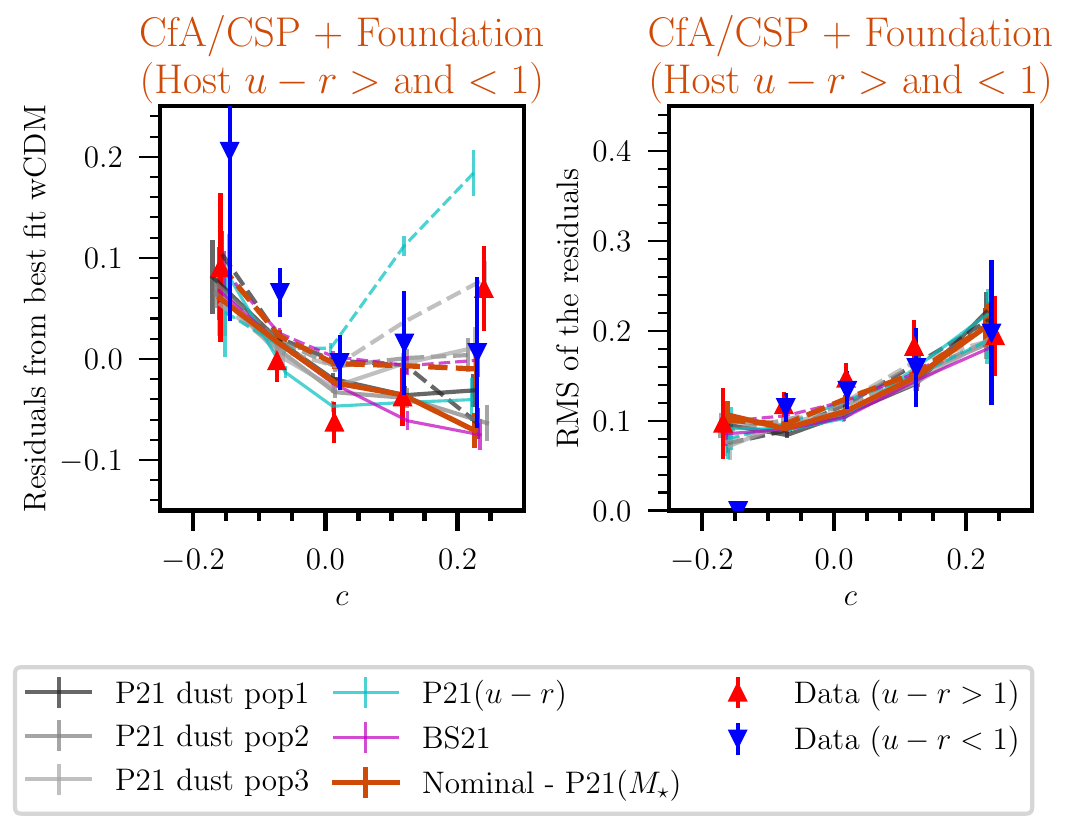}
    \caption{Same as \ref{fig:mu_vs_c_des}, but for the external low-$z$ samples (all combined).}
    \label{fig:mu_vs_c_des_lowz}
\end{figure}

Dust properties and correlations between $x_1$ and host properties are modelled following the same techniques used for the DES SN sample. In Fig.~\ref{fig:mu_vs_c_des_lowz}, we present a comparison between observed and simulated Hubble residuals measured using the BBC0D approach (same as Fig.~\ref{fig:mu_vs_c_des}, but for external low-$z$ samples). The dust models implemented in our analysis generally reproduce the observed trends, however the low statistics makes it challenging to infer SN dust properties in these samples.

Moreover, we note that the selection functions of the low-$z$ SN samples are not as well understood as for the DES SN sample.
Early SN surveys such as CfA were designed as targeted surveys, specifically focused on the discovery and follow-up of the brightest SNe Ia in high mass nearby galaxies. Therefore, the selection function in the low-$z$ samples can only be inferred fudging simulations to match the data.

\begin{figure*}
    \includegraphics[width=\linewidth]{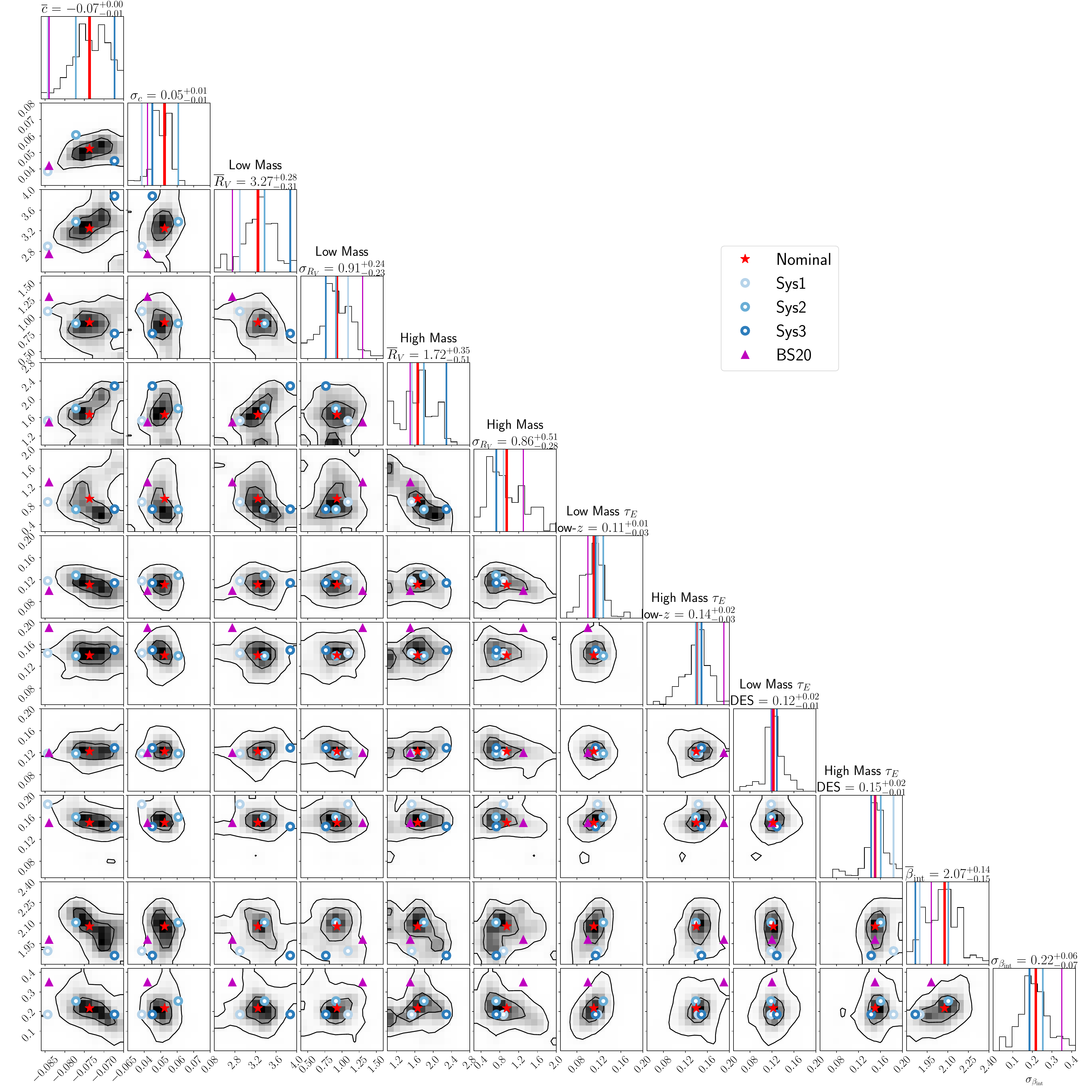}
    \caption{Best fit from Dust fitting code \citep{P21_dust2dust} and three realizations included in the systematic uncertainties (see Sec.~\ref{syst:intr_scatter}).}
   \label{fig:corner_D2D}
\end{figure*}

%% file: Appendix_TMINcut.tex
\section{Selecting only SNe detected before SN peak brightness}
\label{sec:require_SNPeak}
The lack of pre-peak SN observations can significantly impact the accuracy of SN stretch and SN color estimates, and therefore estimates of SN distances. Conversely, selecting only SNe detected before peak brightness can significantly reduce the size of a SN sample, especially for the oldest low-$z$ targeted surveys. For the DES SN sample we have that only 1599 out of 1635 (2\%) have the first detection after $-$2 days from peak brightness. In the low-$z$ sample, the fraction increases to 80 over 194 (41\%).

In order to test the effects of this cut, we run our cosmological analysis selecting only SNe with pre-peak data (at least a detection before $-$2 days from peak brightness) and compare the results with the nominal analysis (that only requires at least a detection before $+$2 days from peak brightness).
We find that the fitted nuisance parameters are consistent within uncertainties (see Table~\ref{tab:nuisance_Tpeak}) and that the cosmological parameter $w$ shifts by 0.013. To test the significance of this $w$-shift, we perform the same test on a set of 25 DES-SN5YR simulated samples. We find an average shift in $w$ of 0.013, with standard deviation of 0.051. Therefore, the shift observed in the data is not significant.

\begin{deluxetable*}{cccccccccl}
\centering
\tablecolumns{9}
\tablewidth{18pc}
\tablecaption{Nuisance parameters and cosmological fit when combining DES with different low-$z$ external samples and when using DES alone.}
\tablehead {
\colhead {Sample}  &
\colhead {$N_{\mathrm{SNe}}$} &
\colhead {$\alpha$}   &
\colhead {$\beta$}   &
\colhead {$\gamma$}   &
\colhead {$\sigma_{\rm gray}$} &
\colhead {RMS $^{*}$} &
\colhead {$\Delta w_{\rm stat}$ $^{\dag}$}
} 
\startdata
DES-SN + low-$z$ &  1829 & 0.161(1) & 3.12(3) & 0.038(7) & 0.04 & 0.168 & 0.000$\pm$0.133 \\
DES-SN + low-$z$ (detection before $-$2 days from peak) &  1713 & 0.161(3) & 3.13(3) & 0.036(8) & 0.04 & 0.17 & 0.013$\pm$0.139 \\
\enddata
\tablenotetext{$*$}{RMS is measured applying a cut of $P_{\mathrm{Ia}}>0.5$ on the DES SN sample.}
\label{tab:nuisance_Tpeak}
\end{deluxetable*}

%% file: Affiliations.tex
\section{Affiliations}
$^{1}$ Department of Physics, Duke University Durham, NC 27708, USA\\
$^{2}$ Einstein Fellow\\
$^{3}$ Department of Astronomy, Boston University, 725 Commonwealth Ave., Boston, MA 02215, USA\\
$^{4}$ Department of Physics, Boston University, 590 Commonwealth Ave., Boston, MA 02215, USA\\
$^{5}$ Center for Astrophysics $\vert$ Harvard \& Smithsonian, 60 Garden Street, Cambridge, MA 02138, USA\\
$^{6}$ The Research School of Astronomy and Astrophysics, Australian National University, ACT 2601, Australia\\
$^{7}$ Univ Lyon, Univ Claude Bernard Lyon 1, CNRS, IP2I Lyon / IN2P3, IMR 5822, F-69622, Villeurbanne, France\\
$^{8}$ Department of Physics and Astronomy, University of Pennsylvania, Philadelphia, PA 19104, USA\\
$^{9}$ School of Mathematics and Physics, University of Queensland,  Brisbane, QLD 4072, Australia\\
$^{10}$ Institute of Cosmology and Gravitation, University of Portsmouth, Portsmouth, PO1 3FX, UK\\
$^{11}$ Centre for Gravitational Astrophysics, College of Science, The Australian National University, ACT 2601, Australia\\
$^{12}$ Department of Astronomy and Astrophysics, University of Chicago, Chicago, IL 60637, USA\\
$^{13}$ Kavli Institute for Cosmological Physics, University of Chicago, Chicago, IL 60637, USA\\
$^{14}$ Centre for Astrophysics \& Supercomputing, Swinburne University of Technology, Victoria 3122, Australia\\
$^{15}$ Centre de Physique des Particules de Marseille, 163 Av. de Luminy, CEDEX 09, Marseille, France\\
$^{16}$ School of Physics and Astronomy, University of Southampton,  Southampton, SO17 1BJ, UK\\
$^{17}$ Departamento de Física Teórica and Instituto de Física de Partículas y del Cosmos (IPARCOS-UCM), Universidad Complutense de Madrid, 28040 Madrid, Spain\\
$^{18}$ African Institute for Mathematical Sciences, 6 Melrose Road, Muizenberg, 7945, South Africa\\
$^{19}$ South African Astronomical Observatory, P.O.Box 9, Observatory 7935, South Africa\\
$^{20}$ INAF-Osservatorio Astronomico di Trieste, via G. B. Tiepolo 11, I-34143 Trieste, Italy\\
$^{21}$ Santa Cruz Institute for Particle Physics, Santa Cruz, CA 95064, USA\\
$^{22}$ Institute of Space Sciences (ICE, CSIC),  Campus UAB, Carrer de Can Magrans, s/n,  08193 Barcelona, Spain\\
$^{23}$ Institut d'Estudis Espacials de Catalunya (IEEC), 08034 Barcelona, Spain\\
$^{24}$ Argonne National Laboratory, 9700 South Cass Avenue, Lemont, IL 60439, USA\\
$^{25}$ Australian Astronomical Optics, Macquarie University, North Ryde, NSW 2113, Australia\\
$^{26}$ Lowell Observatory, 1400 Mars Hill Rd, Flagstaff, AZ 86001, USA\\
$^{27}$ School of Mathematics and Physics, University of Surrey, Guildford, Surrey, GU2 7XH, UK\\
$^{28}$ Department of Physics, Baylor University, One Bear Place \#97316, Waco, TX 76798-7316, USA\\
$^{29}$ Fermi National Accelerator Laboratory, P. O. Box 500, Batavia, IL 60510, USA\\
$^{30}$ Cerro Tololo Inter-American Observatory, NSF's National Optical-Infrared Astronomy Research Laboratory, Casilla 603, La Serena, Chile\\
$^{31}$ Laborat\'orio Interinstitucional de e-Astronomia - LIneA, Rua Gal. Jos\'e Cristino 77, Rio de Janeiro, RJ - 20921-400, Brazil\\
$^{32}$ Department of Physics, University of Michigan, Ann Arbor, MI 48109, USA\\
$^{33}$ Physics Department, 2320 Chamberlin Hall, University of Wisconsin-Madison, 1150 University Avenue Madison, WI  53706-1390\\
$^{34}$ Department of Physics \& Astronomy, University College London, Gower Street, London, WC1E 6BT, UK\\
$^{35}$ Kavli Institute for Particle Astrophysics \& Cosmology, P. O. Box 2450, Stanford University, Stanford, CA 94305, USA\\
$^{36}$ SLAC National Accelerator Laboratory, Menlo Park, CA 94025, USA\\
$^{37}$ Instituto de Astrofisica de Canarias, E-38205 La Laguna, Tenerife, Spain\\
$^{38}$ Universidad de La Laguna, Dpto. Astrofísica, E-38206 La Laguna, Tenerife, Spain\\
$^{39}$ Institut de F\'{\i}sica d'Altes Energies (IFAE), The Barcelona Institute of Science and Technology, Campus UAB, 08193 Bellaterra (Barcelona) Spain\\
$^{40}$ Jodrell Bank Center for Astrophysics, School of Physics and Astronomy, University of Manchester, Oxford Road, Manchester, M13 9PL, UK\\
$^{41}$ University of Nottingham, School of Physics and Astronomy, Nottingham NG7 2RD, UK\\
$^{42}$ Hamburger Sternwarte, Universit\"{a}t Hamburg, Gojenbergsweg 112, 21029 Hamburg, Germany\\
$^{43}$ Department of Physics, IIT Hyderabad, Kandi, Telangana 502285, India\\
$^{44}$ Institute of Theoretical Astrophysics, University of Oslo. P.O. Box 1029 Blindern, NO-0315 Oslo, Norway\\
$^{45}$ Center for Astrophysical Surveys, National Center for Supercomputing Applications, 1205 West Clark St., Urbana, IL 61801, USA\\
$^{46}$ Instituto de Fisica Teorica UAM/CSIC, Universidad Autonoma de Madrid, 28049 Madrid, Spain\\
$^{47}$ University Observatory, Faculty of Physics, Ludwig-Maximilians-Universit\"at, Scheinerstr. 1, 81679 Munich, Germany\\
$^{48}$ Department of Astronomy, University of Illinois at Urbana-Champaign, 1002 W. Green Street, Urbana, IL 61801, USA\\
$^{49}$ Center for Cosmology and Astro-Particle Physics, The Ohio State University, Columbus, OH 43210, USA\\
$^{50}$ Department of Physics, The Ohio State University, Columbus, OH 43210, USA\\
$^{51}$ Jet Propulsion Laboratory, California Institute of Technology, 4800 Oak Grove Dr., Pasadena, CA 91109, USA\\
$^{52}$ George P. and Cynthia Woods Mitchell Institute for Fundamental Physics and Astronomy, and Department of Physics and Astronomy, Texas A\&M University, College Station, TX 77843,  USA\\
$^{53}$ LPSC Grenoble - 53, Avenue des Martyrs 38026 Grenoble, France\\
$^{54}$ Instituci\'o Catalana de Recerca i Estudis Avan\c{c}ats, E-08010 Barcelona, Spain\\
$^{55}$ Department of Physics, Carnegie Mellon University, Pittsburgh, Pennsylvania 15312, USA\\
$^{56}$ Observat\'orio Nacional, Rua Gal. Jos\'e Cristino 77, Rio de Janeiro, RJ - 20921-400, Brazil\\
$^{57}$ Ruhr University Bochum, Faculty of Physics and Astronomy, Astronomical Institute, German Centre for Cosmological Lensing, 44780 Bochum, Germany\\
$^{58}$ Department of Physics and Astronomy, Pevensey Building, University of Sussex, Brighton, BN1 9QH, UK\\
$^{59}$ Centro de Investigaciones Energ\'eticas, Medioambientales y Tecnol\'ogicas (CIEMAT), Madrid, Spain\\
$^{60}$ Computer Science and Mathematics Division, Oak Ridge National Laboratory, Oak Ridge, TN 37831\\
$^{61}$ Lawrence Berkeley National Laboratory, 1 Cyclotron Road, Berkeley, CA 94720, USA\\